\pdfoutput=1
\documentclass[11pt, a4paper]{article}
\usepackage{relsize}
\usepackage{array}
\usepackage{a4wide}
\usepackage[hidelinks]{hyperref}
\usepackage{latexsym}
\usepackage{cite}
\usepackage{mathrsfs}
\usepackage{amssymb}
\usepackage{xcolor}
\usepackage{amsmath}
\usepackage{amsfonts}
\usepackage{amsthm}
\usepackage{graphicx}
\usepackage{enumerate}
\usepackage[dvipsnames]{xcolor}
\usepackage{todonotes}
\usepackage{bm}
\usepackage{comment}
\usepackage{cancel}
\usepackage{float}
\usepackage{verbatim}
\usepackage{mathtools}
\usepackage{subcaption}
\usepackage{xcolor}

\usepackage{tikz}\usepackage[compat=1.1.0]{tikz-feynman}
\usetikzlibrary{calc}
\usetikzlibrary{fit}
\usetikzlibrary{calc,arrows.meta,shapes.misc,decorations.pathreplacing}
\definecolor{outeryellow}{RGB}{224,185,60}
\definecolor{innerred}{RGB}{207,103,70}
\definecolor{innergreen}{RGB}{116,176,96}
\definecolor{softviolet}{RGB}{156,142,234}

\graphicspath{ {Images/} }
\usepackage{pgfplots}
\pgfplotsset{compat=1.18}

\renewcommand{\vec}{\bs}  % Use bold for 3d vectors is better?
\newcommand{\bs}[1]{\boldsymbol{#1}}

\newcommand{\DiagramWidth}{0.95\textwidth}
\usepackage{etoolbox}
\apptocmd{\sloppy}{\hbadness 10000\relax}{}{}

\newcommand{\R}{\mathbb{R}}
\renewcommand{\P}{\mathbb{P}}

\newcommand{\p}{\partial}

\renewcommand{\P}{\mathbb{P}}

\newcommand{\be}{\begin{equation}\label}
\newcommand{\ee}{\end{equation}}
\newcommand{\bea}{\begin{eqnarray}\label}
\newcommand{\eea}{\end{eqnarray}}

\makeatletter
\newcommand*{\textoverline}[1]{$\overline{\hbox{#1}}\m@th$}
\newcommand*\bigcdot{\mathpalette\bigcdot@{.65}}
\newcommand*\bigcdot@[2]{\mathbin{\vcenter{\hbox{\scalebox{#2}{$\m@th#1\bullet$}}}}}
\makeatother

\definecolor{darkgreen}{rgb}{0.0, 0.4, 0.0}

\tikzfeynmanset{warn luatex=false}
\date{}

\begin{document}

\title{The Cosmological $\delta$-shift}
\author{Carolina Figueiredo$^a$\thanks{\texttt{cfigueiredo@fas.harvard.edu}}, Jiajie Mei$^b$\thanks{\texttt{j.mei@uva.nl}}, Yuyu Mo$^c$\thanks{\texttt{moyuyu@itp.ac.cn}}, Francisco Vazão$^{d,e}$\thanks{\texttt{fvvazao@ias.edu}} \vspace{7pt
} \\\normalsize
\textit{
${}^a$ Society of Fellows, Harvard University, Cambridge, MA 02138, USA}\\ \normalsize
\textit{
${}^b$ Institute of Physics, University of Amsterdam, Amsterdam, 1098 XH, The Netherlands}\\ 
\normalsize \textit{
${}^c$  New Cornerstone Laboratory, Institute of Theoretical Physics,}\\ 
\normalsize \textit{Chinese Academy of Sciences, Beijing 100190, China}
\\ 
\textit{
${}^d$ School of Natural Sciences, Institute for Advanced Study, Princeton, NJ 08540}
\\
\textit{
${}^e$
Center for Particle Cosmology, Department of Physics and Astronomy} \\ \textit{University of Pennsylvania, Philadelphia, PA 19104, USA}
}

{\let\newpage\relax\maketitle}
\vspace{1cm}
\begin{abstract}
We consider the extension of the $\delta$-shift connecting Tr$(\phi^3)$ and pion amplitudes to observables in cosmology. We start in flat-space and explain how one can extract the pion wavefunction and correlators from those of Tr$(\phi^3)$ theory, via a simple kinematic deformation. As opposed to the amplitude, the $\delta$-shift for the wavefunction is no longer a linear deformation on the kinematics since the wavefunction depends on magnitudes rather than the squares of momenta. We ``linearize'' this deformation by using the \textit{frequency representation} -- allowing us to recast the contribution from any graph as a sum over residues. We then show how this construction extends to de Sitter, and find that, remarkably, in the frequency representation, the pion de Sitter wavefunction and correlators can be formulated in precisely the same way as in flat-space. Amongst other things, this unified formulation reveals a simple universal soft-limit for flat-space and de Sitter, where, in the soft-limit, the wavefunction turns
into a sum of wavefunctions in a mixed theory of pions and scalars.
\end{abstract}
\vfill

\pagebreak
\setcounter{tocdepth}{2}
\tableofcontents

\section{Introduction}
The past two decades have seen the emergence of remarkable structures underlying some of the simplest observables in fundamental physics, beginning with scattering amplitudes but also starting to be seen in the context of observables in cosmology  \cite{Arkani-Hamed:2017fdk,Cosmohedron,Figueiredo:2025daa,KinFlow, GeometryCosmoCorr, Benincasa:2024lxe,Arkani-Hamed:2018kmz, Baumann:2019oyu,Goodhew:2020hob,Melville:2021lst,Baumann:2024ttn,DiffEq_CosmoCorr,De:2024zic,Glew:2025ypb,Cespedes:2023aal,korte_cosmological_2026,baumann_differential_2026,LoopInts}. Much of this progress has focused on a simple theory of conformally coupled scalars in FRW backgrounds. The simplicity of the cosmological wavefunction and correlators of these models relies on the fact that the action of conformally coupled scalars in FRW backgrounds is Weyl-equivalent to a flat-space action with time-dependent interactions. 

%This ultimately allows us to ``split'' the computation into two steps: 1. Computing the time-independent flat-space wave-function; 2. Introduce the time-dependence via the energy integrals. Purely at the level of the first step, in the last few years we learnt that, just like for amplitudes, there is an object which correctly captures the combinatorics of the singularities of the wavefunction -- the Cosmohedron \cite{Cosmohedron} -- and as for step 2. we have also seen that the time dependence is totally captured by a set of 

However, much less is understood when we go beyond conformally coupled scalars. In particular, if we hope to see this magic extend to observables of our actual universe -- for example to the correlations of the inflaton at the end of inflation -- we need to go beyond toy models. In the context of inflation, the fact that the inflaton is expected to be light naturally suggests describing it as a pseudo-Goldstone boson of a non-linear sigma model, with its shift symmetry weakly broken by the potential. In this paper, we take a first step towards systematically studying pseudo-goldstone inflationary correlators, studying the wavefunction and correlators of exact Goldstone bosons in de Sitter space.

One of the reasons why computing the wavefunction/correlator for Goldstone bosons/pions is challenging is because these are derivatively coupled, and so, already in flat-space, the resulting Feynman rules are non-trivial expressions of the spatial momenta; and even more so in de Sitter. In this note we propose a different route to extract the NLSM wavefunction (and correlator): we derive the generalization of the $\delta$-shift -- first introduced in \cite{Zeros,Arkani-Hamed:2024nhp} to connect amplitudes in Tr$(\phi^3)$ theory to those of the NLSM -- to the context of the observables relevant in cosmology and use it to compute pion wavefunctions and correlators in flat-space and de Sitter. 

At the level of the amplitude, the $\delta$-shift is a simple kinematic deformation of the planar variables, $X_{i,j} = (p_i+p_{i+1}+ \cdots p_{j-1})^2$, which takes the form: 
\begin{equation}
    X_{i,j} \to X_{i,j} + \delta_{i,j}^2, \quad \text{ with }\delta_{i,j}^2 = \begin{cases}
        +\delta^2, \, \text{ if }(i,j)\text{ are both even},\\
        -\delta^2, \, \text{ if }(i,j)\text{ are both odd},\\
        0,\, \text{ otherwise}\, .
    \end{cases} 
\end{equation}
In \cite{Arkani-Hamed:2024nhp} it is shown that the $2n$ Tr$(\phi^3)$ amplitude under this kinematic deformation, in the limit $\delta \to \infty$, yields NLSM amplitudes at leading order in $1/\delta$. The reason behind this connection is that the amplitudes in the shifted Tr$(\phi^3)$-theory can be interpreted as Goldstone amplitudes in a theory with spontaneous breaking of $\text{U}(N) \to \text{U}(k) \times \text{U}(N-k)  $, as we review in the section \ref{sec:NLSM_WF_FS}. This explanation then makes clear that the $\delta$-shift should have a generalization to the observables in cosmology. That is, starting from the wavefunction for massless and massive scalars (with $m^2 = \pm \delta^2$), we should be able to extract the pion wavefunction from the large mass $(\delta)$ limit.

Of course, this is simple to do in flat-space because the difference between a massive and massless scalar simply amounts to a redefinition of the scalar momenta, $k_i$; however the same is not true in de Sitter. It is well known that the propagation of massive fields in de Sitter leads to transcendental functions, which make the computation of the wavefunction significantly harder when compared to the flat-space case. However, since we are interested in the large mass limit, one might hope that the answer simplifies, and this is precisely the case. 

All the results of this paper hinge on a single, simple but remarkable observation: in the large-mass limit, there is indeed a unified formulation for the flat-space and de Sitter wavefunctions (as well as correlators). In this picture, we use the \textit{frequency representation} to write each graph in the cubic (massless and massive) theory in terms of a sum over frequency residues, whose contribution to the wavefunction is then: 
\begin{equation}
    \begin{gathered}
        \begin{tikzpicture}[scale=0.8,
    line/.style={line width=1.15pt},
    vertex/.style={circle, fill=black, inner sep=2.4pt},
    boundary/.style={gray!55, line width=0.9pt},
    every node/.style={font=\footnotesize}]

    \coordinate (X1) at (0,0);
    \coordinate (X2) at (1.3,0);
    \coordinate (X4) at (2.4,0);
    \coordinate (X3) at (3.6,0);

\def\boundaryY{1.}
    \foreach \x/\lab/\v in {-0.45/1/X1,0.45/2/X1,1.3/3/X2,2.4/4/X4,3.15/5/X3,4.05/6/X3} {
    \coordinate (E\lab) at (\x,\boundaryY);
    \draw[line] (E\lab) -- (\v);}

    \draw[line,color=Maroon] (X1) -- node[below=8pt] {} (X2);
    \draw[line,color=Maroon] (X2) -- node[below=8pt] {} (X3);

    \node[vertex] at (X1) {};
    \node[] at (-0.2, -0.4) {$C^{(t_1)}$};
    \node[vertex] at (X2) {};
    \node[] at (1.1, -0.4) {$C^{(t_2)}$};
    \node[vertex] at (X3) {};
    \node[] at (3.8, -0.4) {$C^{(t_4)}$};
    \node[vertex] at (X4) {};
    \node[] at (2.6, -0.4) {$C^{(t_3)}$};
    \end{tikzpicture}
    \end{gathered} \, \, \propto \, \, \int_\mathbb{R}  \prod_{(i,j) \in G} \frac{d \omega_{i,j}}{2 \pi i (w_{i,j}^2 - (k_{i,j}^2 +\delta^2_{i,j}))} \times \prod_{t_i \in G} C^{(t_i)}(\{k_i\},\{\omega_{i,j}\})\, \bigg \vert_{1/\delta^{2n-4}} ,
    \label{eq:FreqRep_GeneralStatement}
\end{equation}
where, analogous to what we get for amplitudes, we have a ($\delta$-shifted) denominator factor $(\omega_{i,j}^2-(k_{i,j}^2+\delta_{i,j}^2))$ for each propagator, with spatial momenta $k_{i,j} = |k_i +k_{i+1} + \cdots +k_{j-1}|$, in the graph; and for each vertex we associate a vertex factor given by a function $C^{(t_i)}$ which in general will depend on the $\omega_{i,j}$ and $k_l$ of the edges touching the vertex. In the large $\delta$ limit, the $n$-pion wavefunction is then the piece at order $1/\delta^{2n-4}$ (with n always even). 

Now what's remarkable is that \eqref{eq:FreqRep_GeneralStatement} is true both for flat-space and for de Sitter, with the only difference being the details of the vertex factors, $C^{(t_i)}$! Amongst other things, we find that this representation of the pion wavefunction reveals an especially  simple behavior under the soft limit. As a given momentum, say $k_1 \to 0$, the wavefunction turns into the sum of shifted $(n-1)$ wavefunctions, which can in turn be interpreted as objects in the mixed theory NLSM$\oplus \, \phi^3$ (analogous to what is found in \cite{Cachazo:2016njl,Splits}): 
\begin{equation}
    \Psi_n  \to  - \frac{i}{\delta^2} \left[  
   \Psi^{\text{NLSM} \oplus \, \phi^3}_{n-1}(2^{\text{odd}},3,\cdots,n) \bigg\vert_{k_2 \to \sqrt{k_2^2-\delta^2}}  -  \Psi^{\text{NLSM} \oplus \, \phi^3}_{n-1}(2^{\text{even}},3,\cdots,n) \bigg\vert_{k_n \to \sqrt{k_n^2+\delta^2}} \right],
   \label{eq:SoftStatement}
\end{equation}
where this formula again holds both in flat-space and in de Sitter.

Finally the closeness of the frequency integrand to a normal amplitude lets us write a simple recursive form for the pion wavefunction. Recall that for the amplitude we can construct the full answer at any $2n$, simply by knowing the effective $2k$-point vertices for all $2k\leq 2n$. In the $\delta$-shift picture, these $2k$-point contact terms come exclusively from the triangulations/diagrams of the $2k$-gon which contain exclusively propagators $k_{i,j}$ with $(i,j)$ both even or both odd. So similarly, if we define $\hat{\Psi}_{2k}$ as the contribution to $\Psi_{2k}$ from the $\delta$-shift of this type of diagrams, we can write the full flat-space/de Sitter wavefunctions exclusively in terms of these pieces as: 
\begin{equation}
    \Psi_{2n} = \sum_{\mathcal{C}} \int_{-\infty}^{+\infty} \prod_{(i,j) \in \mathcal{C}} \frac{\omega_{i,j} d \omega_{i,j}}{(2\pi i)(\text{Re}\psi_2[\omega_{i,j}]) [2(w_{i,j}^2-k_{i,j}^2)]} \ \prod_{P_{e} \subset \mathcal{C}} \hat{\Psi}_{P_e} + \hat{\Psi}_{2n}
\end{equation}
where we sum over all collections, $\mathcal{C}$, of $(\text{even},\text{odd})$ propagators which divide the graph into different even-point subgraphs -- $P_e$. Practically, this simple recursion plays a crucial role in letting us access higher $n$ wavefunctions in de Sitter.
\\ \\
The paper is organized as follows: In section \ref{sec:NLSM_WF_FS} we start by explaining how the $\delta$-shift can be applied to the flat-space wavefunction; then we explain the frequency representation in flat-space and use it to both derive the soft-limits and recursions above. In section \ref{sec:PionWf_DS}, we explain how all the flat-space results generalize to de Sitter and in particular how the frequency representation together with the large-mass expansion of the de Sitter time-integrals let us recover the universal formula \eqref{eq:FreqRep_GeneralStatement}. In section \ref{sec:Correlators} we explain how these results extend to correlators in flat-space and de Sitter.

We finish this section by reviewing the main aspects of the flat-space and de Sitter wavefunctions that will be important for the rest of the text.

\subsection{Flat-space wavefunction}
\label{sec:FlatSpacePsi}

The objects of interest when studying inflation are the wavefunction of the universe, and field correlation functions, see \cite{
Arkani-Hamed:2017fdk,Cosmohedron,Figueiredo:2025daa,KinFlow,
GeometryCosmoCorr,Benincasa:2024lxe,Arkani-Hamed:2018kmz,
Baumann:2019oyu,Goodhew:2020hob,Melville:2021lst,Baumann:2024ttn,
DiffEq_CosmoCorr,De:2024zic,Glew:2025ypb,Cespedes:2023aal,
korte_cosmological_2026,baumann_differential_2026,LoopInts,
Anninos:2014lwa,KinFlowLoop,CosmoReview,PajerUnitarityLocality,
GoodhewCutCosmoCorr,Benincasa:2025uie,CosmoSMatrix,Sachs2,
CosmoTreeTheo,Meltzer,Bittermann:2022nfh,CespedesScott,Sachs3,
HiddenZerosWF,Baumann:2025qjx,Glew:2025otn,Glew:2025ugf,
Benincasa:2019vqr,Fan:2024iek,Arkani-Hamed:2025mce,
Ardila-Mantilla:2026cbo
}, 
which are directly related to the correlations between temperature fluctuations we can measure in the cosmic microwave background. The vacuum wavefunction is simply the transition amplitude from the vacuum to some final state at some finite time, which we take to be $\eta_0 =0$, defined by the path integral:
\begin{align}
    \Psi [\Phi] &= \mathcal{N} \int\limits_{\phi(-\infty(1-i \epsilon))=0}^{\phi(0) = \Phi} \mathcal{D}\phi\: e^{-i  S[\phi]} \, ,
\end{align}
where the $i \epsilon$-prescription is such that we start in the vacuum at past infinity, which in the case of a de Sitter background is simply the Bunch-Davies vacuum \cite{Bunch:1978yq,Allen:1985ux}. 

Before proceeding to de Sitter, let us first consider a theory of massless colored scalars in a Minkowski background with cubic interactions, with action
\begin{equation}
    \mathcal{S}[\varphi]=\int_{-\infty}^{+\infty} d^d x\, \int_{-\infty}^{0} d\eta \, \frac{1}{2}\text{Tr}\left( \partial \phi\right)^2 - \frac{\lambda_3}{3}\text{Tr}\, \phi^3 \, ,
    \label{eq:action}
\end{equation}
where $\eta$ is the time coordinate. The wavefunction can be expanded around the classical solution to yield:
\begin{equation}
    \Psi = \exp\left\{ \sum_{n\geq 2} \int \prod_{i=1}^n\, d \vec{k}_{i} \, \Psi_n[k_{i} ]\prod_i\Phi(\vec{k}_i)\, \delta^d\left(\textstyle{\sum_i} \vec{k}_i\right)  \right\},
    \label{eq:Wf_Coeff}
\end{equation}
where $\Psi_n$ is the wavefunction coefficient for a field configuration with $n$ insertions, and $\vec{k}_i$ is the momentum of the external states, and we denote $k_i \equiv |\vec{k_i}|$. Picking a particular color-ordering of the
external states, after stripping out the color-factor Tr($\phi_1 \phi_2 \cdots \phi_n)$, we can take $\Psi_n$ as a sum over
all Feynman diagrams compatible with the ordering:
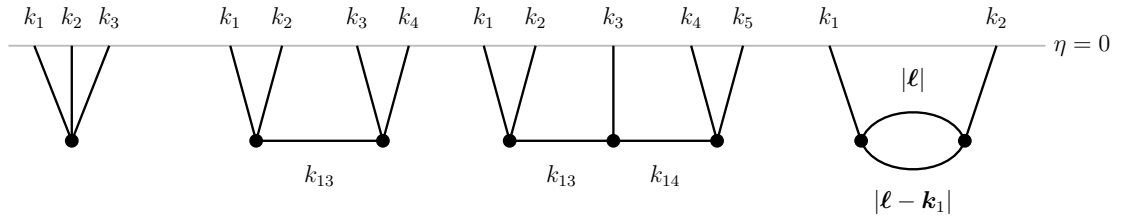
\begin{figure}[H]
    \centering
\resizebox{\DiagramWidth}{!}{%
\begin{tikzpicture}[
    line/.style={line width=1.15pt},
    vertex/.style={circle, fill=black, inner sep=2.4pt},
    boundary/.style={gray!55, line width=0.9pt},
    every node/.style={font=\large}
]

% Height of the eta=0 boundary
\def\boundaryY{1.65}

% Common boundary line
\draw[boundary] (-1.1,\boundaryY) -- (16.9,\boundaryY)
    node[right, black] {$\eta=0$};

% =====================================================
% 1. Three-point vertex
% =====================================================

\begin{scope}[shift={(0,0)}]
    \coordinate (X1) at (0,0);

    \foreach \x/\lab in {-0.65/1,0/2,0.65/3} {
        \coordinate (E\lab) at (\x,\boundaryY);
        \node[above=5pt] at (E\lab) {$k_{\lab}$};
        \draw[line] (E\lab) -- (X1);
    }

    \node[vertex] at (X1) {};
    %\node[below=7pt] at (X1) {$k_1 +k_2 +k_3$};
\end{scope}

% =====================================================
% 2. Four-point exchange
% =====================================================

\begin{scope}[shift={(3.2,0)}]
    \coordinate (X1) at (0,0);
    \coordinate (X2) at (2.2,0);

    \foreach \x/\lab in {-0.45/1,0.45/2} {
        \coordinate (E\lab) at (\x,\boundaryY);
        \node[above=5pt] at (E\lab) {$k_{\lab}$};
        \draw[line] (E\lab) -- (X1);
    }

    \foreach \x/\lab in {1.75/3,2.65/4} {
        \coordinate (E\lab) at (\x,\boundaryY);
        \node[above=5pt] at (E\lab) {$k_{\lab}$};
        \draw[line] (E\lab) -- (X2);
    }

    \draw[line] (X1) -- node[below=8pt] {$k_{13}$} (X2);

    \node[vertex] at (X1) {};
    \node[vertex] at (X2) {};

%    \node[below=7pt] at (X1) {$X_1$};
%    \node[below=7pt] at (X2) {$X_2$};
\end{scope}

% =====================================================
% 3. Five-point chain
% =====================================================

\begin{scope}[shift={(7.6,0)}]
    \coordinate (X1) at (0,0);
    \coordinate (X2) at (1.8,0);
    \coordinate (X3) at (3.6,0);

    \foreach \x/\lab/\v in {-0.45/1/X1,0.45/2/X1,1.8/3/X2,3.15/4/X3,4.05/5/X3} {
        \coordinate (E\lab) at (\x,\boundaryY);
        \node[above=5pt] at (E\lab) {$k_{\lab}$};
        \draw[line] (E\lab) -- (\v);
    }

    \draw[line] (X1) -- node[below=8pt] {$k_{13}$} (X2);
    \draw[line] (X2) -- node[below=8pt] {$k_{14}$} (X3);

    \node[vertex] at (X1) {};
    \node[vertex] at (X2) {};
    \node[vertex] at (X3) {};

%    \node[below=7pt] at (X1) {$X_1$};
%    \node[below=7pt] at (X2) {$X_2$};
%    \node[below=7pt] at (X3) {$X_3$};
\end{scope}

% =====================================================
% 4. One-loop bubble
% =====================================================

\begin{scope}[shift={(13.7,0)}]
    \coordinate (X1) at (0,0);
    \coordinate (X2) at (1.8,0);

    \coordinate (E1) at (-0.55,\boundaryY);
    \coordinate (E2) at (2.35,\boundaryY);

    \node[above=5pt] at (E1) {$k_1$};
    \node[above=5pt] at (E2) {$k_2$};

    \draw[line] (E1) -- (X1);
    \draw[line] (E2) -- (X2);

    \draw[line] (X1) to[out=70,in=110] node[above=6pt] {$|\vec{\ell}|$} (X2);
    \draw[line] (X1) to[out=-70,in=-110] node[below=6pt] {$|\vec{\ell}-\vec{k}_1|$} (X2);

    \node[vertex] at (X1) {};
    \node[vertex] at (X2) {};

%    \node[below=7pt] at (X1) {$X_1$};
%    \node[below=7pt] at (X2) {$X_2$};
\end{scope}

\end{tikzpicture}
}
\end{figure}
\noindent where the $k_{ij}$'s are simply the magnitudes of the momenta flowing through the internal edges, so $k_{12}=|\vec{k}_1+\vec{k}_2|$. Each graph, $G$, is associated with a time-integral which takes the form:
\begin{equation}
    \Psi \vert_G = \int_{-\infty(1-i \epsilon)}^{0}\prod_{v\in \mathcal{V}(G)} d\eta_v \; \prod_{i\in L(v)}\phi_{k_i}(\eta_v)\prod_{e\in \mathcal{E}(G)} G_{BB}(\eta_{v_e},\eta_{v_e^{\prime}}; k_e) \; \footnote{here we omitted the powers of $\lambda_3$ for simplicity.},
    \label{eq:time_int_rep}
\end{equation}
where $\mathcal{V}(G)$ is the set of vertices, $L(v)$ is the set of external legs attached to $v$, and $\mathcal{E}(G)$ the set of edges of graph $G$; $\phi_{k_v}(\eta_v)$ is the bulk-to-boundary propagator of an edge anchored at the boundary and attached to the vertex $v$ with energy $k_v$, it is given by the solution of the classical equation of motion which takes the form
\begin{equation}
    \phi_{k}(\eta) = e^{i k \eta} \, .
     \label{eq:FS_BulktoBdy}
\end{equation}
Finally $G_{BB}(\eta_{v_e},\eta_{v_e^{\prime}}; k_e)$ is the bulk-to-bulk propagator corresponding to the internal edge $e$ that connects the vertices $v_e$ and $v_e^{\prime}$ with internal energy $k_e$. It is given by the Green's function satisfying: 
\begin{equation}    
\left[\partial_{\eta}^2 + k^2\right]G_{BB}(\eta,\eta^\prime; k)=-i  \delta(\eta-\eta^{\prime})\, .
\label{eq:GreensFunc_FS}
\end{equation}
which together with the boundary condition, $G_{BB}(0,\eta_{v_e^{\prime}}; k_e) = G_{BB}(\eta_{v_e},0; k_e) =0$, yields
\begin{equation}
    G_{BB}(\eta,\eta^\prime; k) = \frac{1}{2\,k}\left[\phi_{k}(\eta) \phi^*_{k}(\eta^\prime) \Theta(\eta^\prime-\eta) + \phi_{k}(\eta^\prime)\phi_{k}^*(\eta) \Theta(\eta-\eta^\prime) - \phi_{k}(\eta)\phi_{k}(\eta^\prime)\right]\, .
    \label{eq:BBProp_FS}
\end{equation}

As it is by now well-appreciated \cite{CosmoReview}, the graph time integrals in \eqref{eq:time_int_rep} have a remarkably simple structure -- after time-integration we get a rational function with poles in the locus where the sums of energies of sub-graphs add up to zero, which appear combined into what are called \textit{tubings} of the graph. For example, consider the $4$-point diagram:

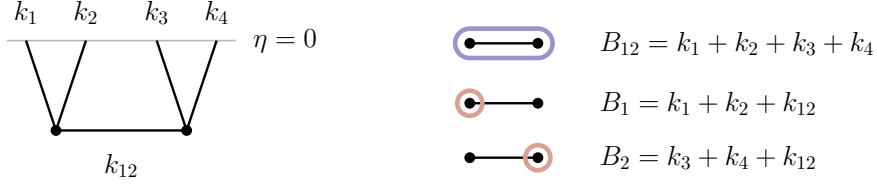
\begin{figure}[H]
\centering

\resizebox{0.75\textwidth}{!}{%
\begin{tikzpicture}[
    line/.style={line width=1.25pt},
    vertex/.style={circle, fill=black, inner sep=2pt},
    boundary/.style={gray!55, line width=0.9pt},
    yellowtube/.style={draw=Blue!40!white, line width=2.4pt},
    redcircle/.style={draw=Maroon!40!white, line width=2.4pt},
    every node/.style={font=\Large}
]

% =====================================================
% Parameters
% =====================================================

\def\boundaryY{1.65}
\def\tubeR{0.27}

% =====================================================
% Left: four-point exchange diagram
% =====================================================

\begin{scope}[shift={(0,0)}]

    \coordinate (X1) at (0,0);
    \coordinate (X2) at (2.4,0);

    \coordinate (E1) at (-0.55,\boundaryY);
    \coordinate (E2) at (0.55,\boundaryY);
    \coordinate (E3) at (1.85,\boundaryY);
    \coordinate (E4) at (2.95,\boundaryY);

    % eta = 0 boundary, only slightly past the diagram
    \draw[boundary] (-0.9,\boundaryY) -- (3.3,\boundaryY)
        node[right=5pt, black] {$\eta=0$};

    \node[above=5pt] at (E1) {$k_1$};
    \node[above=5pt] at (E2) {$k_2$};
    \node[above=5pt] at (E3) {$k_3$};
    \node[above=5pt] at (E4) {$k_4$};

    \draw[line] (E1) -- (X1);
    \draw[line] (E2) -- (X1);
    \draw[line] (E3) -- (X2);
    \draw[line] (E4) -- (X2);

    \draw[line] (X1) -- node[below=8pt] {$k_{12}$} (X2);

    \node[vertex] at (X1) {};
    \node[vertex] at (X2) {};

%    \node[below=7pt] at (X1) {$X_1$};
%    \node[below=7pt] at (X2) {$X_2$};

\end{scope}

% =====================================================
% Right: cut structure
% =====================================================

\begin{scope}[shift={(7.6,0.15)}]

    % -------------------------
    % Total-energy cut
    % -------------------------

    \coordinate (A1) at (0,1.45);
    \coordinate (A2) at (1.25,1.45);

    % yellow tube/capsule
    \draw[yellowtube]
        ($(A1)+(0,\tubeR)$)
        -- ($(A2)+(0,\tubeR)$)
        arc[start angle=90,end angle=-90,radius=\tubeR]
        -- ($(A1)+(0,-\tubeR)$)
        arc[start angle=270,end angle=90,radius=\tubeR];

    \draw[line] (A1) -- (A2);
    \node[vertex] at (A1) {};
    \node[vertex] at (A2) {};

    \node[right=1.0cm, black] at (A2)
        {$B_{12} = k_1 + k_2 + k_3 + k_4$};

    % -------------------------
    % Left partial-energy cut
    % -------------------------

    \coordinate (B1) at (0,0.35);
    \coordinate (B2) at (1.25,0.35);

    \draw[line] (B1) -- (B2);
    \node[vertex] at (B1) {};
    \node[vertex] at (B2) {};

    \draw[redcircle] (B1) circle[radius=0.23];

    \node[right=1.0cm, black] at (B2)
        {$B_{1} = k_1 + k_2 + k_{12}$};

    % -------------------------
    % Right partial-energy cut
    % -------------------------

    \coordinate (C1) at (0,-0.65);
    \coordinate (C2) at (1.25,-0.65);

    \draw[line] (C1) -- (C2);
    \node[vertex] at (C1) {};
    \node[vertex] at (C2) {};

    \draw[redcircle] (C2) circle[radius=0.23];

    \node[right=1.0cm, black] at (C2)
        {$B_{2} = k_3+k_4+k_{12}$};

\end{scope}

\end{tikzpicture}
}
\end{figure}
\noindent where on the right we draw the diagram with the external legs deleted, and identify the different subgraphs via \textit{tubes}, to which we associate variables $B_{\tau}$ corresponding to the sum of the energies entering the subgraph. The contribution to $\psi_4$ from this graph is then
\begin{equation}
    \Psi_4 \supset \frac{1}{(k_1+k_2+k_3+k_4)(k_1+k_2+ k_{12})(k_3+k_4+ k_{34})}\, \,  \equiv \, \, \, \begin{gathered}
\begin{tikzpicture}[
    line/.style={line width=1.25pt},
    vertex/.style={circle, fill=black, inner sep=1.6pt},
    boundary/.style={gray!55, line width=0.9pt},
    yellowtube/.style={draw=Blue!40!white, line width=2.2pt},
    redcircle/.style={draw=Maroon!40!white, line width=2.2pt},
    scale=0.95
]
\def\tubeR{0.27}

    % -------------------------
    % Total-energy cut
    % -------------------------

    \coordinate (A1) at (0,0);
    \coordinate (A2) at (1.25,0);

    % yellow tube/capsule
    \draw[yellowtube]
        ($(A1)+(0,\tubeR)$)
        -- ($(A2)+(0,\tubeR)$)
        arc[start angle=90,end angle=-90,radius=\tubeR]
        -- ($(A1)+(0,-\tubeR)$)
        arc[start angle=270,end angle=90,radius=\tubeR];

    \draw[line] (A1) -- (A2);
    \node[vertex] at (A1) {};
    \node[vertex] at (A2) {};

    \draw[redcircle] (A1) circle[radius=0.17];

    \draw[redcircle] (A2) circle[radius=0.17];

\end{tikzpicture}        
    \end{gathered} \,\,,
\end{equation}
where each pole is then associated with a tube and all the full fraction corresponds to a tubing of the graph -- $i.e.$ a maximal collection of non-crossing subgraphs/tubes. While the $4$-point exchange graph has a single tubing, in general, a given graph has many tubings which contribute to the wavefunction coefficient with fractions analogous to the one above. As a result, the wavefunction for a graph $G$ can be written as:
\begin{equation}
    \Psi \vert_G = \sum_{\mathbf{T} \text{ of G}}  \, \,\prod_{\tau\in\mathbf{T}} \frac{1}{B_{\tau}}\, ,
\end{equation}
where the sum is over all tubings, $\mathbf{T}$, of the graph, and the singularities, $B_{\tau}$, are the sum of energies entering each tube $\tau$ in $\mathbf{T}$.

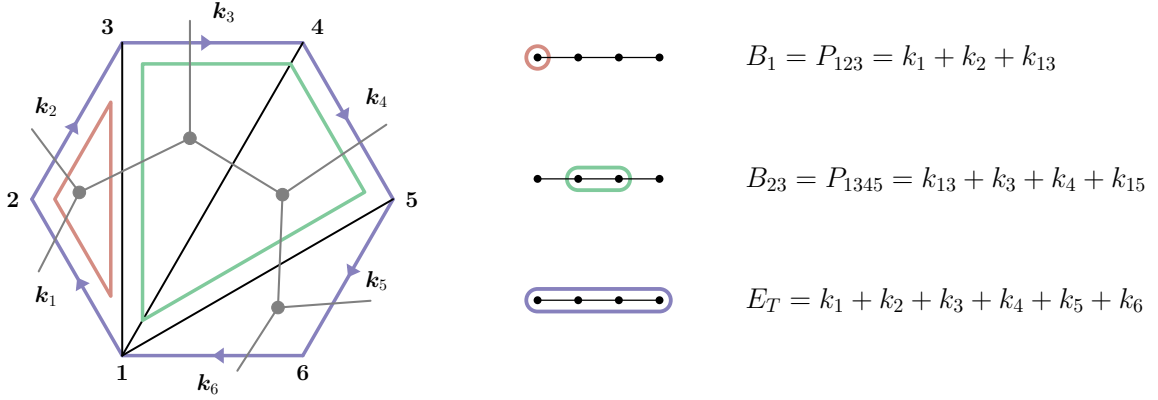
\begin{figure}[t]
\centering
\resizebox{0.98\textwidth}{!}{%
\begin{tikzpicture}[
    line cap=round,
    line join=round,
    outer/.style={
        draw=Blue!50!white,
        line width=2.2pt,
        postaction={decorate},
        decoration={
            markings,
            mark=at position 0.5 with {
                \arrow{Triangle[length=2.5mm,width=3.4mm]}
            }
        }
    },
    blackdiag/.style={draw=black,line width=1.25pt},
    redshape/.style={draw=Maroon!50!white,line width=2.2pt},
    greenshape/.style={draw=Green!50!white,line width=2.2pt},
    mom/.style={draw=gray,line width=1.3pt},
    pnode/.style={circle,fill=gray,draw=gray,inner sep=2.8pt},
    vlab/.style={font=\bfseries\Large},
    plab/.style={font=\Large},
    chainline/.style={draw=black,line width=0.8pt},
    equation/.style={font=\Large,anchor=west},
    redring/.style={draw=Maroon!50!white,line width=2.4pt},
    greenring/.style={draw=Green!50!white,line width=2.4pt},
    bluering/.style={draw=Blue!50!white,line width=2.4pt}
]

\def\s{4.0}
\pgfmathsetmacro{\h}{0.8660254*\s}
\def\RedScale{0.62}
\def\GreenScale{0.82}

\coordinate (V1) at ({-0.5*\s},{-\h});
\coordinate (V2) at ({-1.0*\s},{0});
\coordinate (V3) at ({-0.5*\s},{\h});
\coordinate (V4) at ({ 0.5*\s},{\h});
\coordinate (V5) at ({ 1.0*\s},{0});
\coordinate (V6) at ({ 0.5*\s},{-\h});

\draw[outer] (V1)--(V2);
\draw[outer] (V2)--(V3);
\draw[outer] (V3)--(V4);
\draw[outer] (V4)--(V5);
\draw[outer] (V5)--(V6);
\draw[outer] (V6)--(V1);

\draw[blackdiag] (V1)--(V3);
\draw[blackdiag] (V1)--(V4);
\draw[blackdiag] (V1)--(V5);

\path let
    \p1=(V1), \p2=(V2), \p3=(V3)
in coordinate (CR) at
    ({(\x1+\x2+\x3)/3},{(\y1+\y2+\y3)/3});

\coordinate (R1) at ($(CR)!\RedScale!(V1)$);
\coordinate (R2) at ($(CR)!\RedScale!(V2)$);
\coordinate (R3) at ($(CR)!\RedScale!(V3)$);
\draw[redshape] (R1)--(R2)--(R3)--cycle;

\path let
    \p1=(V1), \p2=(V3), \p3=(V4), \p4=(V5)
in coordinate (CG) at
    ({(\x1+\x2+\x3+\x4)/4},
     {(\y1+\y2+\y3+\y4)/4});

\coordinate (G1) at ($(CG)!\GreenScale!(V1)$);
\coordinate (G3) at ($(CG)!\GreenScale!(V3)$);
\coordinate (G4) at ($(CG)!\GreenScale!(V4)$);
\coordinate (G5) at ($(CG)!\GreenScale!(V5)$);
\draw[greenshape] (G1)--(G3)--(G4)--(G5)--cycle;

\coordinate (N12) at (-2.95, 0.15);
\coordinate (N3)  at (-0.50, 1.35);
\coordinate (N45) at ( 1.55, 0.10);
\coordinate (N6)  at ( 1.45,-2.40);

\draw[mom] (N12)--(N3)--(N45)--(N6);
\draw[mom] (N12)--(-3.85,-1.60);
\draw[mom] (N12)--(-4.00, 1.55);
\draw[mom] (N3) --(-0.50, 3.95);
\draw[mom] (N45)--( 3.85, 1.65);
\draw[mom] (N6) --( 3.50,-2.25);
\draw[mom] (N6) --( 0.55,-3.80);

\node[pnode] at (N12) {};
\node[pnode] at (N3)  {};
\node[pnode] at (N45) {};
\node[pnode] at (N6)  {};

\node[vlab,below=2pt]       at (V1) {1};
\node[vlab,left=4pt]        at (V2) {2};
\node[vlab,above left=2pt]  at (V3) {3};
\node[vlab,above right=2pt] at (V4) {4};
\node[vlab,right=4pt]       at (V5) {5};
\node[vlab,below=2pt]       at (V6) {6};

\node[plab,left]  at (-3.3,-2.15) {$\vec{k}_1$};
\node[plab,left]  at (-3.3, 2.05) {$\vec{k}_2$};
\node[plab,above] at ( 0.25,3.80) {$\vec{k}_3$};
\node[plab,right] at ( 3.25,2.25) {$\vec{k}_4$};
\node[plab,right] at ( 3.25,-1.80) {$\vec{k}_5$};
\node[plab,below] at (-0.10,-3.75) {$\vec{k}_6$};

\begin{scope}[
    shift={(7.2,0.2)},
    scale=1.25,
    every node/.style={transform shape}
]

\coordinate (A1) at (0.00,2.35);
\coordinate (A2) at (0.72,2.35);
\coordinate (A3) at (1.44,2.35);
\coordinate (A4) at (2.16,2.35);

\draw[redring] (A1) circle[radius=0.20];
\draw[chainline] (A1)--(A4);
\fill (A1) circle[radius=2.0pt];
\fill (A2) circle[radius=2.0pt];
\fill (A3) circle[radius=2.0pt];
\fill (A4) circle[radius=2.0pt];

\coordinate (B1) at (0.00,0.20);
\coordinate (B2) at (0.72,0.20);
\coordinate (B3) at (1.44,0.20);
\coordinate (B4) at (2.16,0.20);

\draw[greenring]
    ($(B2)+(0,0.20)$)
    --
    ($(B3)+(0,0.20)$)
    arc[start angle=90,end angle=-90,radius=0.20]
    --
    ($(B2)+(0,-0.20)$)
    arc[start angle=270,end angle=90,radius=0.20]
    -- cycle;

\draw[chainline] (B1)--(B4);
\fill (B1) circle[radius=2.0pt];
\fill (B2) circle[radius=2.0pt];
\fill (B3) circle[radius=2.0pt];
\fill (B4) circle[radius=2.0pt];

\coordinate (C1) at (0.00,-1.95);
\coordinate (C2) at (0.72,-1.95);
\coordinate (C3) at (1.44,-1.95);
\coordinate (C4) at (2.16,-1.95);

\draw[bluering]
    ($(C1)+(0,0.20)$)
    --
    ($(C4)+(0,0.20)$)
    arc[start angle=90,end angle=-90,radius=0.20]
    --
    ($(C1)+(0,-0.20)$)
    arc[start angle=270,end angle=90,radius=0.20]
    -- cycle;

\draw[chainline] (C1)--(C4);
\fill (C1) circle[radius=2.0pt];
\fill (C2) circle[radius=2.0pt];
\fill (C3) circle[radius=2.0pt];
\fill (C4) circle[radius=2.0pt];

\node[equation] at (3.55,2.35)
    {$B_1=P_{123}=k_1+k_2+k_{13}$};

\node[equation] at (3.55,0.20)
    {$B_{23}=P_{1345}=k_{13}+k_3+k_4+k_{15}$};

\node[equation] at (3.55,-1.95)
    {$E_T=k_1+k_2+k_3+k_4+k_5+k_6$};

\end{scope}
\end{tikzpicture}%
}
\caption{Triangulation of the hexagon corresponding to a six-point process for the path-graph topology, together with the associated chain subgraphs and perimeter variables.}
\label{fig:Polygon6Main}
\end{figure}

To obtain the full wavefunction coefficient one must further sum over all possible graphs -- at tree- and  loop-level, multiplied by the respective powers of $\lambda_3$. This step is in particular important when obtaining the NLSM wavefunction by $\delta$-shifting the Tr$(\phi^3)$ one, since, just like for the amplitude, the shift of individual diagrams contains spurious terms which cancel in the full sum. It is therefore useful to re-express the tubings in a way which also keeps track of the different underlying graphs. This can be trivially achieved by thinking of the different diagrams of $\Psi_n$ as triangulations of the $n$-gon\footnote{this is at tree-level, at higher loop-order, cubic graphs are dual to triangulations of higher genus surfaces.}, and the respective tubings are given by \textit{Russian dolls} of the $n$-gon. This language was first used in \cite{Cosmohedron, Figueiredo:2025daa} to find a geometry for the wavefunction and for the correlator of Tr$(\phi^3)$. The $n$-gon is obtained by drawing the spatial momenta, $\vec{k}_i$, tip-to-toe following the underlying color-ordering, producing a closed polygon due to spatial momentum conservation (see Fig.~\ref{fig:Polygon6Main}). 

Given a triangulation of the $n$-gon, we can obtain the dual graph by assigning a vertex to each triangle as shown in Fig.~\ref{fig:Polygon6Main}; the different tubes of the graph are in one-to-one correspondence with the different subpolygons of the $n$-gon -- for example the tube enclosing the full graph corresponds to the full $n$-gon (marked in purple), the tube enclosing the leftmost vertex is given by the red triangle, and the tube enclosing the two middle vertices by the green quadrilateral subpolygon in Fig.~\ref{fig:Polygon6Main}. Quite nicely, the variables $B_\tau$ associated to each tube are simply given by the perimeter, $P_\tau$, of the respective subpolygons in the momentum $n$-gon. The tubings of graphs are then given by maximal collections of non-crossing subpolygons -- called Russian dolls of the $n$-gon. 

As such, the tree-level contribution to the wavefunction coefficient, $\Psi_n$, can be written as:
\begin{equation}
    \Psi_n = \sum_{\mathcal{T} \text{of n-gon}} \, \, \sum_{\mathbf{R}_{\mathcal{T}}}\prod_{\tau\in\mathbf{R}_{\mathcal{T}}}\frac{1}{P_{\tau}}\, ,
\end{equation}
where the first sum is over all triangulations, $\mathcal{T}$, the second sum is over all Russian dolls compatible with $\mathcal{T}$ -- this is which contain the set of triangles specified by $\mathcal{T}$ -- and for each Russian doll we get a product over the perimeters of the subpolygons present in $\mathbf{R}_\mathcal{T}$. 

\subsection{De Sitter wavefunction}
\label{sec:DSSpacePsi}

When considering the same Tr$(\phi^3)$ theory in a de Sitter background, the action is instead
\begin{equation}
S[\phi]
=
\int d\eta \, d^3 x \, \sqrt{-g}
\left[
-\frac{1}{2}\operatorname{Tr}\!\left(\partial_\mu \phi \, \partial^\mu \phi\right)
-\frac{1}{2}\left(m^2+\xi R\right)\operatorname{Tr}\!\left(\phi^2\right)
-\frac{\lambda_3}{3!}\operatorname{Tr}\!\left(\phi^3\right)
\right]\, ,
\end{equation}
where the metric is just the de Sitter metric:
\begin{equation}
    ds^2= \frac{-d\eta^2 + d\vec{x}^2}{H^2\eta^2}\, , \quad \text{ and } \sqrt{-g} = \frac{1}{H^4 \eta^4}
\end{equation}
with $H$ the de Sitter scale. The Ricci scalar, $R=12H^2$, so we can redefine our mass to be $\mu^2(m,\xi)\equiv m^2+ \xi\,12 H^2$. The special case of $\mu^2=2H^2$ -- where the scalars are conformally coupled -- has been extensively studied over the past few years \cite{CosmoReview} as it can be related to the massless flat-space theory described in Sec.~\ref{sec:FlatSpacePsi} where the cubic coupling is time dependent, $\lambda_3 \equiv \lambda_3(\eta)$. In this manuscript, we will be interested in more general $\mu^2$ and in particular in the case of minimally coupled massless scalars, $\mu^2=0$. 

For general $\mu^2$, the free classical equations of motion in momentum space read
\begin{equation}
\partial_\eta^2\phi_{k}^{(\mu)}
-
\frac{2}{\eta}\partial_\eta\phi_{k}^{(\mu)}
+
\left[
k^2
+
\frac{\mu^2}{H^2\eta^2}
\right]\phi_{k}^{(\mu)}
=
0\, ,
\end{equation}
and the solution satisfying the Bunch-Davies condition in the past, giving the bulk-to-boundary de Sitter propagator is 
\begin{equation}
\phi^{(\mu)}_{k}(\eta)
=
\frac{\sqrt{\pi}}{2}
H
e^{-\frac{i\pi}{2}\left(\nu+\frac12\right)}
(-\eta)^{3/2}
H_{\nu}^{(2)}(-k\eta)\, ,
\quad \text{ with } \, 
\nu
=
\sqrt{
\frac{9}{4}
-
\frac{\mu^2}{H^2}
}\, .
\label{eq:MassiveBulkBdy}
\end{equation}
where $H^{(2)}_\nu$ is a Hankel function of the second kind. For a massless scalar ($\mu^2=0$), this simplifies significantly into:
\begin{equation}
\phi^{(0)}_{k}(\eta)
=
\frac{H}{\sqrt{2k^3}}
\left(
1-ik \eta
\right)
e^{i k \eta} \quad \Rightarrow \quad \phi^{\text{dS}}_{k}(\eta) = \left(
1-ik \eta
\right)
e^{i k \eta},
\label{eq:MasslessBulkBdy}
\end{equation}
where $\phi^{\text{dS}}_{k}$ simply corresponds to the boundary-normalized bulk-to-boundary propagator which satisfies $\phi^{\text{dS}}_{k}(\eta=0)=1$, just like \eqref{eq:FS_BulktoBdy} in flat-space. 
As for the bulk-to-bulk propagator it is given by Green's function that obeys the following differential equation:
\begin{equation}
   \left( \partial_{\eta}^2 - \frac{2}{\eta}\partial_{\eta} +\left[k^2+\frac{\mu^2}{H^2\eta^2}\right]\right)G^{(\mu)}_{BB}(\eta,\eta^{\prime};k) = -i H^2 \eta^2 \delta(\eta-\eta^{\prime})\, ,
\end{equation}
which in the massless case yields
\begin{equation}
   G^{(0)}_{BB}(\eta,\eta^{\prime};k) = \frac{H^2}{2k^3}\left( \phi^{\text{dS}}_{k}(\eta) \phi^{\text{dS}}_{-k}(\eta^\prime)\Theta(\eta^\prime-\eta)+\phi^{\text{dS}}_{k}(\eta^\prime) \phi^{\text{dS}}_{-k}(\eta)\Theta(\eta-\eta^{\prime}) - \phi^{\text{dS}}_{k}(\eta) \phi^{\text{dS}}_{k}(\eta^\prime) \right)
\end{equation}

Therefore, for a graph, $G$, with massless external legs, contributing to $\Psi^{\text{dS}}_n$, the de Sitter time-integral associated with it follows exactly the same structure of \eqref{eq:time_int_rep}, except with a different measure:
\begin{equation}
    \Psi \vert_G = \int_{-\infty(1-i \epsilon)}^{0}\prod_{v\in \mathcal{V}(G)} \frac{d\eta_v}{\eta_v^4} \; \prod_{i\in L(v)}\phi^{dS}_{k_i}(\eta_v)\prod_{e\in \mathcal{E}(G)} G^{(\mu)}_{BB}(\eta_{v_e},\eta_{v_e^{\prime}}; k_e) \; .
    \label{eq:time_int_rep_dS}
\end{equation}
The factor of $1/\eta^4$ in the measure is responsible for the familiar fact that massless fields in de Sitter lead to physical divergences near $\eta \sim 0$, which appear as logarithmic functions of the type $(\log[-k\eta])^L$, see \cite{Ford:1984hs,Gorbenko:2019rza,Burgess:2015ajz,Rajaraman:2010xd,Beneke:2012kn,Mirbabayi:2019qtx,Baumgart:2020oby,Cespedes:2023aal, Benincasa:2024lxe, Benincasa:2024rfw, Beneke:2023wmt, Cohen:2020php}. These divergences signal the breakdown of perturbation theory at sufficiently late times, and one needs to re-sum them and compute their contributions to the correlator.

As we will see in section \ref{sec:PionWf_DS}, in the case of (massless) NLSM in de Sitter, there are no physical late time divergences at tree-level, they only occur for loop diagrams coming from the loop integration \cite{Kitamoto:2010et,Armstrong:2022csc}\footnote{In general, tree-level wavefunction coefficients for scale-invariant, parity-even theories involving interactions with at least two derivatives in even spacetime dimensions (the NLSM included) are rational functions \cite{Goodhew:2022ayb}.}.  However tree-diagrams do contain power-divergences, which are not physical as they are purely imaginary and cancel in the correlator. As a result, we can simply remove them with a suitable analytic regulator, or an appropriate subtraction scheme to remove boundary divergences (see for example  \cite{Skenderis:2002wp}). Moreover, the finite part of the tree-level NLSM wavefunction with a suitable analytic regulator is real (see \cite{Goodhew:2024eup}); this will be particularly useful to compute correlators as we we show in  Sec. \ref{sec:Correlators}.

\section{Pion wavefunction in flat-space} \label{sec:NLSM_WF_FS}

As stated in the introduction, NLSM amplitudes can be obtained from Tr$(\phi^3)$ amplitudes via the $\delta$-shift. As explained in \cite{Arkani-Hamed:2024nhp}, the reason why this kinematic shift yields the correct answer, is simply because the $\delta$-shifted Tr$(\phi^3)$ amplitudes are themselves amplitudes in a theory with a spontaneously broken $U(N) \to U(k)\times U(N-k)$ symmetry. Concretely, starting with a colored $N\times N$ scalar, $\Phi_I^J$, with the $U(N)$-invariant potential: 
\begin{equation}
    V(\Phi) = \frac{g}{3} \text{Tr}(\Phi^3) +\frac{m^2}{2}\text{Tr}(\Phi^2) - c\text{Tr}(\Phi), \quad  \quad\Rightarrow \quad \quad \begin{gathered}       \begin{tikzpicture}[scale=0.9, redlab/.style={
        text=red!70!black,
        font=\normalsize
    }]
\begin{axis}[
    axis lines=none,
    xmin=-9, xmax=5,
    ymin=-20, ymax=40,
    domain=-9:5,
    samples=200,
    xtick=\empty,
    ytick=\empty,
    width=5cm,
    height=4cm
]
\addplot[Blue, line width=1pt] {x^3/3 + 5*x^2/2 - 2*x};
\addplot[only marks, mark=*, mark size=1.5pt,color=Maroon] coordinates {
    ({(-5-sqrt(33))/2}, {(185 + 33*sqrt(33))/12})
    ({(-5+sqrt(33))/2}, {(185 - 33*sqrt(33))/12})
};

\draw[->, line width=0.6pt] (axis cs:-9,2) -- (axis cs:4.8,2);
\draw[->,line width=0.6pt] (axis cs:-3,-80) -- (axis cs:-3,40);

\node[redlab] at (0.4,6.35) {$r_+$};
\node[redlab] at (-4.8,21.35) {$r_-$};
\end{axis}
\end{tikzpicture}
    \end{gathered} 
\end{equation}
by choosing the VeV, $v = \text{Diag}(r_+,\cdots,r_+,r_-,\cdots,r_-)$, with the first $k$ diagonals at $r_+$, and the remaining $(N-k)$, at $r_-$; the $U(N)$ is spontaneously broken to $U(k) \times U(N-k)$. In the broken theory, the massive fields are then the upper $k\times k$ block of $\Phi$, $\phi$, and the lower $(N-k)\times (N-k)$ block, $\tilde{\phi}$, whose masses satisfy, $m_\phi^2 = - m_{\tilde{\phi}}^2 \equiv \delta^2$; and the Goldstones are the off-diagonal blocks, $\chi$ and $\chi^\dagger$. Besides the Tr$(\phi^3)$ and Tr$(\tilde{\phi}^3)$ interactions, the $\chi/\chi^\dagger$ fields interact with the $\phi/\tilde{\phi}$ via cubic interactions: Tr$(\chi \chi^\dagger \phi)$ and Tr$(\chi^\dagger  \chi \tilde{\phi})$. The claim is then that the scattering amplitude for ($\chi \chi^\dagger \cdots \chi \chi^\dagger$) is simply the $\delta$-shifted Tr$(\phi^3)$ amplitude. As a result, the large $\delta$ limit -- corresponding to integrating out the massive scalars -- at leading order, yields pure pion amplitude.

From this picture, it is then clear that one should be able to extract the pion wavefunction by studying the $\chi \chi^\dagger$ wavefunction in this theory, and taking the $\delta \to \infty$ limit, on the masses of the internal scalars. In this section we will start by doing this for the case on the flat-space wavefunction, as a warm up to De Sitter.

%So we start with the de Sitter wavefunction for massless states ($\chi \chi^\dagger \cdots \chi \chi^\dagger$) but where we allow for the exchange of massive in the bulk as dictated by the cubic interactions Tr$(\chi \chi^\dagger \phi)$ and Tr$(\chi^\dagger  \chi \tilde{\phi})$. These are exactly such that the propagators $k_{i,j}$ with $i,j$ both even/odd are massive $\phi$ or $\tilde{\phi}$ propagators, while $k_{e,o}$ are massless -- exactly the picture we had in flat space. 

\subsection{The \texorpdfstring{$\delta$}{delta}-shift for spatial momenta} \label{sec:SqrtShift_FS}

In section \ref{sec:FlatSpacePsi}, we described how to obtain the flatspace wavefunction for massless (colored) scalars with cubic interactions. In order to apply the $\delta$-shift, we simply need to generalize this construction for the case in which there are massive fields propagating in the bulk, corresponding to $\phi$ and $\tilde{\phi}$. The massive bulk-to-bulk propagator satisfies
\begin{equation}
    \left[ \partial_{\eta_{v_e}}^2 + k_e^2 +m^2 \right] G_{BB}^{m^2}(\eta_{v_e},\eta_{v^\prime_e}; k_e) = -i\delta( \eta_{v_e}-\eta_{v^\prime_e})
\end{equation}
where $\vec{k}_e$ is the spatial momentum of the propagator, and $m$ its mass. This is precisely the equation defining the massless propagator, eqn.~\eqref{eq:GreensFunc_FS}, where we redefine the massless momenta, $k_e^2 \to k_e^2 + m^2$. Therefore, the  wavefunction coefficients in the spontaneously broken theory, $\Psi_{2n}\left[\chi(k_1) \chi^\dagger (k_2) \cdots \chi(k_{2n-1})\chi^\dagger (k_{2n}) \right]$, can be obtained from the massless flat-space wavefunction by performing the appropriate kinematic redefinition: 
\begin{equation}
    k_{i,j} \to \sqrt{k_{i,j}^2 +\delta_{i,j}^2}, \quad \text{with } \quad \delta_{i,j}^2 = \begin{cases}
        +\delta^2, \quad \text{if $(i,j)$ are both even}, \\
        -\delta^2, \quad \text{if $(i,j)$ are both odd}, \\
        0, \quad \quad \text{otherwise,}
    \end{cases}  
\label{eq:DeltaShift_k}
\end{equation}
which correctly turns the massless-bulk propagators into massive $\phi/\tilde{\phi}$ propagators. Finally, to extract the NLSM wavefunction we simply need to take the $\delta \to \infty$ limit. 

\paragraph{4-point example} Let us see this in action for the simplest example at $n=4$. In this case the tree-level massless flat-space wavefunction is simply: 
\begin{equation}
    \Psi_4 = \frac{1}{P_{1234}} \times \left[ \frac{1}{(k_1+k_2+k_{1,3})(k_3+k_4+k_{1,3})} + \frac{1}{(k_2+k_3+k_{2,4})(k_1+k_4+k_{2,4})} \right],
\end{equation}
where $P_{1234} = k_1+k_2+k_3+k_4$. Applying the $\delta$-shift, $k_{1,3} \to \sqrt{k_{1,3}^2 -\delta^2}$ and $k_{2,4} \to \sqrt{k_{2,4}^2 +\delta^2}$, and expanding at large $\delta$, we find:
\begin{equation}
    \Psi_4^{\delta} \to -\frac{1+i}{\delta^3} + \frac{1}{\delta^4} \left[ P_{1234} + \frac{ P_{123} +P_{124} +P_{134} +P_{234}}{2} - \frac{P_{123}P_{134}+P_{124}P_{234}}{P_{1234}}\right] + \mathcal{O}(\delta^{-5}),
    \label{eq:Pion4pt_FS}
\end{equation}
where again $P_{\tau}$ denotes the perimeter of the subpolygon with vertices in the list $\tau$. At leading order in $\delta$, we find a pure constant term which is not part of the NLSM wavefunction as it does not have the correct units. The presence of the $\delta^{-3}$ term in the expansion of the wavefunction is to be contrasted with the case of the amplitude -- where we see explicitly the wrong-unit leading order terms fully canceling out. However, this discrepancy is not surprising as the wavefunction is an off-shell object where pure-contact contributions are not meaningful since they can, in general, be removed via field-redefinitions. 

As for the $\delta^{-4}$ piece, it is precisely the $4$-point pion wavefunction\footnote{One can check explicitly it agrees with the result in \cite{Bittermann:2022nfh}, computed via the pion feynman rules, up to a factor of $4 P_{1234}/3$ -- which can again can be removed via field redefinitions.}! In \eqref{eq:Pion4pt_FS}, we wrote it in terms of the perimeter variables to make manifest that on the total energy pole, $P_{1234}=0$, the wavefunction turns into the amplitude since 
\begin{equation}
    P_{123}P_{134} \xrightarrow{P_{1234}=0} X_{1,3} \equiv s \quad \quad P_{124}P_{234} \xrightarrow{P_{1234}=0} X_{2,4} \equiv t\, .
\end{equation}

\paragraph{6-point example} At $6$-points the tree-level Tr$(\phi^3)$ wavefunction has already $72$ terms, corresponding to the $72$ different Russian-dolls of the hexagon, and therefore we will not write it here explicitly. But applying the $\delta$-shift, which at $6$-points is simply: 
\begin{equation}
    k_{1,3} \to \sqrt{k_{1,3}^2 -\delta^2}\, , \quad  k_{2,4} \to \sqrt{k_{2,4}^2 +\delta^2}\, , \quad \text{ [and cyclic $(i,j) \to (i+2,j+2)$]}\,,
\end{equation}
expanding at large $\delta$, we find again spurious terms at orders $\delta^{-7}$, and at order $\delta^{-8}$ the $6$-point pion wavefunction. Once again, writing in terms of the perimeter variables we find a simple way of writing the wavefunction which makes manifest in the total energy pole, $E_t \equiv P_{123456}=0$ it turns  into the amplitude: 
\begin{equation}
\begin{aligned}
    \Psi_6 &= \frac{1}{E_t} \times \left[ \frac{(L_{13} R_{13} + L_{24}R_{24} +E_t Q^L_{14} ) (L_{15} R_{15} + L_{46}R_{46} +E_t Q^R_{14} )}{L_{14}R_{14}} + \, \text{(cyclic)}  \right. \\
    &\quad \quad \quad \quad \quad  \quad \quad \left. - (L_{13}R_{13} + L_{24}R_{24} +L_{35}R_{35} + L_{46}R_{46} + L_{15}R_{15} + L_{26}R_{26}    ) \right] \\
    & \quad + \text{(pure contact, no $E_t$ pole)}\\[0.5em]
    \text{with } & \quad Q_{14}^L =2L_{14} -L_{13}-L_{24} \, , \quad Q_{14}^R =2R_{14} -R_{13}-R_{24}\,;
\end{aligned}
\end{equation}
where for simplicity of notation we use $L_{ij}/R_{ij}$ to denote the perimeter to the left/right of chord $(ij)$; so for example $L_{13}=P_{123}$ and $R_{13}=P_{13456}$. Once again these variables are nice because when $E_t \to0$, we have $L_{ij} R_{ij} \to X_{i,j}$; so from the expression above it is clear the $E_t$ pole of $\Psi_6$ is precisely the $6$-point pion amplitude. 

It would be interesting to understand whether the numerator of the pion wavefunction can always be written as the sum over the left-right products that manifestly become the flat-space amplitude when taking the $E_T=0$ residue plus a deformation, here written as $E_T Q^{L/R}$. We found that at least up to 10 points it is possible, however the complexity of the $Q^{L/R}$ simply seems to keep increasing, and no general pattern was found.

\subsection{Pion wavefunction from frequency representation}
\label{sec:Freq_Rep_FS}

One interesting feature of the $\delta$-shift for the wavefunction when compared to the amplitude is the presence of the square-root in \eqref{eq:DeltaShift_k} -- which is a consequence of the wavefunction depending directly on the magnitudes of the spatial momenta. Turns out there is a different representation for the wavefunction which allows for an ``amplitude-like'' formulation graph-by-graph, where the dependence on the $k_{i,j}$ variables appears through it's mod-square, and therefore the $\delta$-shift resembles that of the amplitude -- the \textit{frequency representation} \cite{Bittermann:2022nfh,Werth:2024mjg,Raju:2011mp,Albayrak:2020isk,Mei:2024abu,Mei:2024sqz}.

While the frequency representation is practically less useful to directly compute the NLSM wavefunction in flat space -- when compared with the direct $\delta$-shift method described in the previous section -- it plays a crucial role in defining universal soft-limits for the wavefunction, as well as understanding how the $\delta$-shift construction generalizes to de Sitter in section \ref{sec:PionWf_DS}. 

The frequency representation for the time integrals in  the wavefunction hinges on the fact that the different terms in the propagator bulk-to-bulk can be written dispersively as
\begin{equation}
\begin{aligned}
   & \frac{e^{i k (\eta-\eta^{\prime})} \Theta(\eta^\prime-\eta) + e^{i k(\eta^\prime-\eta)} \Theta(\eta-\eta^\prime)}{2\,k}  = - \int_{-\infty}^{+\infty} \frac{d \omega}{2 \pi i } \,  \frac{e^{-i \omega (\eta -\eta^\prime)}}{\omega
    ^2 - k^2 + i \epsilon}\\
    &  \frac{e^{i k (\eta+\eta^{\prime})} }{2\,k} = -\int_{-\infty}^{+\infty} \frac{d \omega}{2 \pi i } \,  \frac{e^{-i \omega (\eta +\eta^\prime)}}{\omega
    ^2 - k^2 + i \epsilon}
\end{aligned}
\end{equation}
where each integral is evaluated as a sum over the relevant residues provided $\eta, \eta^\prime <0$; putting these together one finds
\begin{equation}
\begin{aligned}
    G_{BB}(\eta,\eta^\prime; k) &= \frac{1}{2}\int_{-\infty}^{\infty}\frac{d\omega}{2 \pi i} \,\frac{(\phi_{\omega}(\eta)-\phi_{-\omega}(\eta))(\phi_{\omega}(\eta^\prime)-\phi_{-\omega}(\eta^\prime))}{\omega^2 - k^2 + i \epsilon}\, \\
    &=\frac{1}{2}\int_{-\infty}^{\infty}\frac{d\omega}{2 \pi i} \,\frac{(e^{i \omega \eta}-e^{-i \omega \eta})(e^{i \omega \eta^\prime}-e^{-i \omega \eta^\prime})}{\omega^2 - k^2 + i \epsilon}\, .
\end{aligned}
    \label{eq:freq_rep_BB}
\end{equation}
with $\phi_\omega(\eta)$ being the bulk-to-boundary propagator from \eqref{eq:FS_BulktoBdy}. The fact that the numerator factors into a piece depending on $\eta$ and another on  $\eta^\prime$ means that once we plug this back into \eqref{eq:time_int_rep}, the time integrals all factorize and can be trivially integrated. After integration, we are left with the denominators $(\omega^2 -k^2)$ from \eqref{eq:freq_rep_BB}, as well as new ``vertex-factors'' coming from the $\eta$-integrals -- these look like discontinuities of the $3$-point wavefunction coefficient over the different bulk-to-bulk edges attached to the respective vertex:
\begin{equation}
\begin{aligned}
    & \left.\begin{gathered}
\begin{tikzpicture}[
    line cap=round,
    line width=1pt,
    offshell/.style={
        draw=Maroon,
        line width=1.9pt
    },
    onshell/.style={
        draw=black,
        line width=0.9pt
    },
]
    % External legs
    \draw[onshell] (-0.7, 0.6) -- (0,0);
    \draw[onshell] (-0.7,-0.6) -- (0,0);
    \draw[offshell] (0,0) -- (0.9,0);

    % Labels
    \node at (-0.98, 0.55) {$k$};
    \node at (0.1, -0.35) {$\eta$};
    \node at (-0.98,-0.55) {$k^\prime$};
    \node at ( 1.3, 0) {$\omega_0$};

    % Black vertex
    \filldraw[black] (0,0) circle (3.2pt);
\end{tikzpicture}
    \end{gathered} \, \, \right]\,\,  \begin{aligned}  C^{(1)}(k,k^\prime; \omega_0)& = \int_{-\infty}^0 d \eta \, e^{i(k+k^\prime)\eta} (e^{i\omega_0\eta}-e^{-i\omega_0\eta}) \\
    &=\frac{-2 i \omega_0}{\omega_0^2-(k+k^\prime)^2}\end{aligned} \\[0.8em]
&\left. \begin{gathered}
\begin{tikzpicture}[
    line cap=round,
    line width=1pt,
    offshell/.style={
        draw=Maroon,
        line width=1.9pt
    },
    onshell/.style={
        draw=black,
        line width=0.9pt
    },
]
    % External legs
    \draw[onshell] (-0.7, 0.6) -- (0,0);
    \draw[offshell] (-0.7,-0.6) -- (0,0);
    \draw[offshell] (0,0) -- (0.9,0);

    % Labels
    \node at (-0.98, 0.55) {$k$};
     \node at (0.1, -0.35) {$\eta$};
    \node at (-0.98,-0.65) {$\omega_1$};
    \node at ( 1.3, 0) {$\omega_0$};

    % Black vertex
    \filldraw[black] (0,0) circle (3.2pt);
\end{tikzpicture}
    \end{gathered} \, \, \right]\, \, \begin{aligned} C^{(2)}(k;\omega_0,\omega_1) &=\int_{-\infty}^0 d \eta \, e^{ik\eta} \prod_{j=0,1}(e^{i\omega_j\eta}-e^{-i\omega_j\eta}) \\
    &=\frac{-8 i k \omega_0 \omega_1}{\left((\omega_0+\omega_1)^2-k^2 + i \epsilon \right)\left((\omega_0-\omega_1)^2-k^2+i\epsilon \right)} \end{aligned} 
\\[0.8em]
&\left. \begin{gathered}
\begin{tikzpicture}[
    line cap=round,
    line width=1pt,
    offshell/.style={
        draw=Maroon,
        line width=1.9pt
    },
    onshell/.style={
        draw=black,
        line width=0.9pt
    },
]
    % External legs
    \draw[offshell] (-0.7, 0.6) -- (0,0);
    \draw[offshell] (-0.7,-0.6) -- (0,0);
    \draw[offshell] (0,0) -- (0.9,0);

    % Labels
    \node at (-0.98, 0.55) {$\omega_2$};
     \node at (0.1, -0.35) {$\eta$};
    \node at (-0.98,-0.65) {$\omega_1$};
    \node at ( 1.3, 0) {$\omega_0$};

    % Black vertex
    \filldraw[black] (0,0) circle (3.2pt);
\end{tikzpicture}  
    \end{gathered}\, \, \right] \, \,  \begin{aligned} C^{(3)}(\omega_0,\omega_1,\omega_2) &=\int_{-\infty}^0 d \eta \,  \prod_{j=0,1,2}(e^{i\omega_j\eta}-e^{-i\omega_j\eta}) \\
    &=\sum_{\sigma_{j}=\pm 1} \sigma_{0} \sigma_{1} \sigma_2 \times \frac{-i}{(\sigma_0 \omega_0 + \sigma_1 \omega_1 +\sigma_2 \omega_2) - i \epsilon} \end{aligned} 
\end{aligned}
\label{eq:3ptFactors}
\end{equation}
where the red lines denote the internal propagators and the black ones those corresponding to bulk-to-boundary legs. Hence, given a graph contributing to $\Psi_n$, corresponding to some triangulation $\mathcal{T}_G$ with propagators $(i,j) \in \mathcal{T}$, the respective frequency representation is simply
\begin{equation}
    \Psi_n \vert_{\mathcal{T}_G} = \frac{1}{2^{n-3}}\int_{\mathbb{R}} \prod_{(i,j) \in \mathcal{T}_G} \frac{d \omega_{i,j}}{2 \pi i (w_{i,j}^2 - k_{i,j}^2)} \times \prod_{t \in \mathcal{T}_G} C^{(t)}(\{k_i\},\{\omega_{i,j}\}) ,
    \label{eq:FreqSpace_FlatSpace}
\end{equation}
where the second product is over all the triangles (vertices), $t$, in the triangulation (graph), $\mathcal{T}$; and $C^{(t)}(\{k_i\},\{\omega_{i,j}\})$ the respective $3$-point factors from \eqref{eq:3ptFactors}. The frequency integrals are ultimately meant to turn into a sum over residues once we close the contour in the upper/lower-half plane. 

Interestingly, in this representation we find that the dependence on $k_{i,j}$ enters via $k_{i,j}^2$, since the $C^{(t)}$ only depend on the external, $k_i$, and the frequency variables, $\omega_{i,j}$. Hence, the $\delta$-shift in the frequency representation is a simply linear shift, $k_{i,j}^2 \to k_{i,j}^2+\delta_{i,j}^2$; but as we will discuss now, expanding in the $\delta \to \infty$ limit to obtain the pion wavefunction is a bit more subtle. 

Let's consider the simplest case for $n=4$. From the ``s-channel'' diagram we get the following contribution: 
\begin{equation}
\begin{aligned}
\Psi_{4} \bigg \vert_{\begin{gathered}
    \begin{tikzpicture}[scale=0.8,
    line cap=round,
    line width=1pt,
    offshell/.style={
        draw=Maroon,
        line width=1.9pt
    },
    onshell/.style={
        draw=black,
        line width=0.9pt
    },
]
    % External legs
    \draw[onshell] (-0.5, 0.6) -- (0,0);
    \draw[onshell] (-0.1,0.6) -- (0,0);
    \draw[offshell] (0,0) -- (0.9,0);
    \draw[onshell] (0.9, 0) -- (1,0.6);
    \draw[onshell] (0.9,0) -- (1.4,0.6);

    % Black vertex
    \filldraw[black] (0,0) circle (2.5pt);
    \filldraw[black] (0.9,0) circle (2.5pt);
\end{tikzpicture}
\end{gathered}} &= \frac{1}{2}\int_{\mathbb{R}}\frac{d\omega_{1,3}}{2 \pi i} C^{(1)}(k_1,k_2;\omega_{1,3})\times  \frac{1}{\omega_{1,3}^2-k_{1,3}^2}  \times C^{(1)}(k_3,k_4;\omega_{1,3}) \\
&=-2 \int_{\mathbb{R}} \frac{d\omega_{1,3}}{2 \pi i}  \frac{\omega_{1,3}^2}{(\omega_{13}^2 -(k_1+k_2)^2) (\omega_{13}^2 -k_{13}^2)(\omega_{13}^2 -(k_3+k_4)^2)} \, .
\end{aligned}
\label{eq:4ptFreqRep}
\end{equation}
and so the $\delta$-shift simply takes $k_{1,3}^2 \to k_{1,3}^2 -\delta^2$. However, as we expand the integrand for large $\delta$, we get higher powers of $\omega_{1,3}$ in the numerator which yield divergent integrals, already at order $1/\delta^4$ -- precisely the one giving the pion wavefunction. Another naive problem is that from the expansion of the integrand we only obtain even powers of $1/\delta$, which we know is not the full answer (recall \eqref{eq:Pion4pt_FS}).
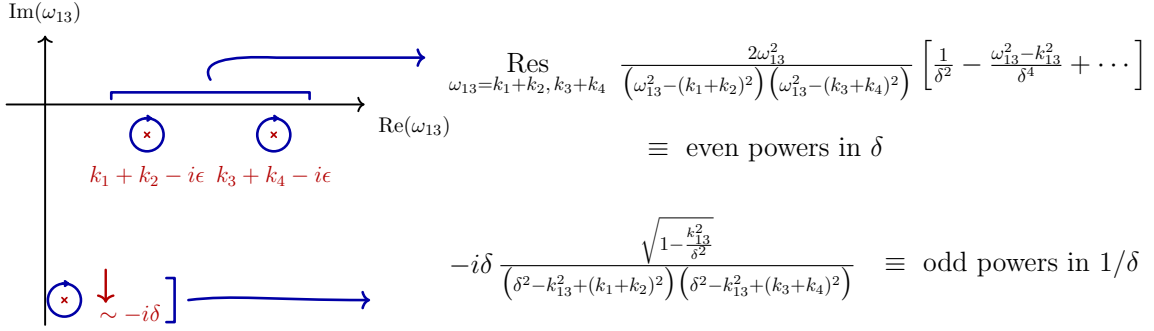
\begin{figure}[t]
\centering
\resizebox{0.98\textwidth}{!}{%
\begin{tikzpicture}[
    line cap=round,
    line join=round,
    axis/.style={draw=black,line width=1.0pt,->},
    blue/.style={draw=blue!65!black,line width=1.35pt},
    bluearrow/.style={draw=blue!65!black,line width=1.35pt,->},
    redarrow/.style={draw=red!70!black,line width=1.25pt,->}
]

% The real axis is raised by 1.00; the vertical endpoints are unchanged.
\draw[axis] (-0.70,1.00) -- (5.80,1.00)
    node[anchor=north west,xshift=3pt,yshift=-2pt]
    {$\mathrm{Re}(\omega_{13})$};
\draw[axis] (0,-3.00) -- (0,2.25)
    node[anchor=south,yshift=2pt] {$\mathrm{Im}(\omega_{13})$};

% The two poles and their bracket move upward with the real axis.
\coordinate (P1) at (1.85,0.45);
\coordinate (P2) at (4.15,0.45);
\foreach \P in {P1,P2} {
    \draw[blue] (\P) circle[radius=0.30];
    \draw[blue] ($(\P)+(-0.035,0.30)$) --
                ($(\P)+( 0.065,0.30)$);
    \draw[blue] ($(\P)+(0.065,0.30)$) --
                ($(\P)+(0.010,0.342)$);
    \draw[blue] ($(\P)+(0.065,0.30)$) --
                ($(\P)+(0.010,0.258)$);
    \draw[red!70!black,line width=0.9pt]
        ($(\P)+(-0.045,-0.045)$) -- ($(\P)+(0.045,0.045)$);
    \draw[red!70!black,line width=0.9pt]
        ($(\P)+(-0.045,0.045)$) -- ($(\P)+(0.045,-0.045)$);
}
\node[text=red!70!black,font=\large,anchor=north]
    at ($(P1)+(0,-0.43)$) {$k_1+k_2-i\epsilon$};
\node[text=red!70!black,font=\large,anchor=north]
    at ($(P2)+(0,-0.43)$) {$k_3+k_4-i\epsilon$};

\draw[blue]
    (1.20,1.10) -- (1.20,1.22) --
    (4.80,1.22) -- (4.80,1.10);
\draw[bluearrow]
    (3.00,1.45)
    .. controls (2.80,1.78) and (3.30,1.89) ..
    (4.20,1.85) -- (6.90,1.85);

% Lower pole and contour retain their original coordinates.
\coordinate (Pd) at (0.35,-2.55);
\draw[blue] (Pd) circle[radius=0.30];
\draw[blue] ($(Pd)+(-0.035,0.30)$) -- ($(Pd)+(0.065,0.30)$);
\draw[blue] ($(Pd)+(0.065,0.30)$) -- ($(Pd)+(0.010,0.342)$);
\draw[blue] ($(Pd)+(0.065,0.30)$) -- ($(Pd)+(0.010,0.258)$);
\draw[red!70!black,line width=0.9pt]
    ($(Pd)+(-0.045,-0.045)$) -- ($(Pd)+(0.045,0.045)$);
\draw[red!70!black,line width=0.9pt]
    ($(Pd)+(-0.045,0.045)$) -- ($(Pd)+(0.045,-0.045)$);
\draw[redarrow] (1.10,-2) -- (1.10,-2.60);
\node[text=red!70!black,font=\large,anchor=west]
    at (0.85,-2.80) {$\sim -i\delta$};

\draw[blue]
    (2.20,-2.15) -- (2.35,-2.15) --
    (2.35,-2.95) -- (2.20,-2.95);
\draw[bluearrow]
    (2.60,-2.55)
    .. controls (3.40,-2.45) and (4.10,-2.55) ..
    (5.90,-2.55);

% Expressions remain at precisely the original positions.
\node[font=\Large,anchor=north west] at (7.20,2.3) {%
    $\mathop{\mathrm{Res}}\limits_{\omega_{13}=k_1+k_2,\,k_3+k_4}
    \;
    \frac{2\omega_{13}^{2}}
         {\bigl(\omega_{13}^{2}-(k_1+k_2)^2\bigr)
          \bigl(\omega_{13}^{2}-(k_3+k_4)^2\bigr)}
    \left[
        \frac{1}{\delta^2}
        -\frac{\omega_{13}^{2}-k_{13}^{2}}{\delta^4}
        +\cdots
    \right]$%
};
\node[font=\Large,anchor=west] at (10.80,0.20)
    {$\equiv\ \mathrm{even\ powers\ in}\ \delta$};

\node[font=\Large,anchor=west] at (7.20,-1.80) {%
    $-i\delta\,
    \frac{\sqrt{1-\frac{k_{13}^{2}}{\delta^2}}}
         {\bigl(\delta^2-k_{13}^{2}+(k_1+k_2)^2\bigr)
          \bigl(\delta^2-k_{13}^{2}+(k_3+k_4)^2\bigr)}$%
};
\node[font=\Large,anchor=west] at (15.10,-1.9)
    {$\equiv\ \mathrm{odd\ powers\ in}\ 1/\delta$};

\end{tikzpicture}%
}
\caption{Closing the contour in the LHP leads to a sum over three residues from $\Psi\vert_{(1,3)}$. One of  these moves to $-i\infty$, and only contributes with odd powers in the $1/\delta$-expansion. The finite ones resemble the even powers in $1/\delta$ from which we extract the NLSM wavefunction.}
\label{fig:ResidueSum}
\end{figure}

To correctly recover the result from \eqref{eq:Pion4pt_FS}, one should instead recall that the frequency integral \eqref{eq:4ptFreqRep} is really a residue sum, so if we close the contour in the lower-half plane we must sum over the residues $w_{1,3}= \{k_1+k_2 - i \epsilon, k_3+k_4 - i \epsilon,-\sqrt{k_{1,3}^2-\delta^2 - i \epsilon}\} $. For the first two residues we can indeed expand the integrand at large $\delta$, since in this case the contour integral is localized around finite values of $\omega$; however, the same is not true for the later. The results from the different residues are shown in figure \ref{fig:ResidueSum}. From the residue at $\omega_{1,3} \sim -i \delta$, we precisely obtain the odd-power contributions in $1/\delta$ that are missing from the naive integrand expansion; the remaining two residues give the even powers in $1/\delta$, in particular the piece in $1/\delta^4$ relevant to the pion wavefunction. Note that from these residues there is also a term in $1/\delta^2$ which is absent in \eqref{eq:Pion4pt_FS}, and which indeed cancels when we add the contribution from the ``t-channel'' diagram.

Having understood how we can correctly recover the $\delta$-expansion from the frequency representation, we learn that, at least in this simple $n=4$ example, to extract the pion wavefunction we can directly expand the integrand at large $\delta $, extract the piece with the correct power in $1/\delta$ and sum over the subset of residues that survive the expansion. This strategy works trivially at $n=4$ since the $\delta$-shifted residues do not contribute to the relevant even power in $1/\delta$ giving the pion wavefunction. Yet, as we increase $n$, instead of a single $\omega$ residue, we must take consecutive residues in all $\omega_{i,j}$'s for the propagators of the diagram. As a result, it can indeed happen that residues where multiple $\omega_{i,j} \sim -i \delta$ give a contribution with the correct power of $1/\delta$ -- which we miss if we blindly expand the integrand and sum over the remaining residues. We will now show that all of these contributions are nonetheless  local-boundary terms which can be removed via field redefinitions. 

Since at higher $n$ keeping track of the residues like we did at $n=4$ becomes cumbersome, we will argue this by simply looking at an individual bulk-to-bulk propagator. Without loss of generality, let's pick a propagator $k_{i,j}$ where $(i,j)$ are both odd from a given diagram. From the frequency representation we get an integral of the form: 
\begin{equation}
    \begin{gathered}
        \begin{tikzpicture}[scale=0.9,
    line cap=round,
    line width=1pt,
    offshell/.style={
        draw=Maroon,
        line width=1.9pt
    },
    onshell/.style={
        draw=gray,
        line width=0.9pt
    },
]
    % External legs
    \draw[onshell] (-0.8, 0.6) -- (0,0);
    \draw[onshell] (-0.4,0.6) -- (0,0);
    \draw[offshell] (0,0) -- (1,0);
    \draw[onshell] (1, 0) -- (1.4,0.6);
    \draw[onshell] (1,0) -- (1.8,0.6);
    \node at (0.5, 0.4) {$k_{i,j}$};
    \node at (-0.3, -0.2) {$\eta_{L}$};
    \node at (1.4, -0.2) {$\eta_{R}$};

    % Black vertex
    \filldraw[black] (0,0) circle (2.5pt);
    \filldraw[black] (1,0) circle (2.5pt);
\end{tikzpicture}
    \end{gathered} \quad = \,  \frac{1}{2} \int_{\mathbb{R}} \frac{d \omega_{ij}}{2\pi i} F_L(\omega_{ij}) \times  \frac{1}{(\omega_{ij}-k_\delta)(\omega_{ij}+k_{\delta})} \times F_R(\omega_{ij}) 
\label{eq:DeltExpmth}
\end{equation}
where $k_\delta = \sqrt{k_{ij}^2-\delta^2-i\epsilon}$ and $F_L/F_R$ stand for the functions coming from the left/right parts of the graph, depending on the respective kinematics as well as $\delta$, which take the form: 
\begin{equation}
    F_{L}(\omega) = \int_{-\infty}^0 d \eta_{L} f_{L}(\eta_{L}) (e^{i \omega \eta_{L}} - e^{-i \omega \eta_{L}}),
    \label{eq:defFL}
\end{equation}
and similar for $F_R(\omega)$. Now let's say we choose to close the contour in the lower half plane, then we have: 
\begin{equation}
    \int_\mathbb{R}  \frac{d \omega_{ij}}{2\pi i} I_m(\omega_{i,j}) = -\sum_{\omega_A \text{ in LHP}} \text{Res}_{\omega_{i,j}=\omega_A}(I_m) - \text{Res}_{\omega_{i,j}=-k_\delta} (I_m) ,
\end{equation}
where $\omega_A$ stand for the remaining simple poles which happen at some finite location. From the form of $F_L/F_R$ we have that the pole in $-k_\delta \to -i \delta$ as $\delta \to \infty$, we get: 
\begin{equation}
    - \text{Res}_{\omega_{i,j}=-k_\delta} (I_m) = \frac{F_L (-k_\delta) F_R(-k_\delta)}{2 k_\delta }, \quad \text{ with } -k_\delta \sim - i \delta,
    \label{eq:Deltaresidue}
\end{equation}
so that the contribution from this residue is fully determined by the assymptotic expansion of $F_L/F_R$ at large $\omega$. If $\omega$ is real, then under the suited $i \epsilon$ regularization we can use \eqref{eq:defFL} to derive an asymptotic expansion for $F_L$ using integration by parts: 
\begin{equation}
    \int_{-\infty}^0 d \eta_{L} f_{L}(\eta_{L}) e^{i \omega \eta_{L}} = \frac{f_L(0)}{i \omega} - \frac{f^{\prime}_L(0)}{(i\omega)^2} + \mathcal{O}(1/\omega^3)
\end{equation}
and so subtracting the contribution from $f_L(\eta_L)e^{-i \omega \eta_L}$, all the odd derivatives cancel and we are left with 
\begin{equation}
    F_L(\omega) = -2i \left( \frac{f_L(0)}{\omega} - \frac{f_L^{(2)}(0)}{w^3} + \frac{f_L^{(4)}(0)}{w^5} + \cdots \right), 
    \label{eq:FL_exp}
\end{equation}
and similar for $F_R(\omega)$. Now since we know that $F_L/F_R$ are meromorphic functions of $\omega$, the asymptotic expansion at large $\omega$ must be unique. So plugging this back into \eqref{eq:Deltaresidue} we learn that the contributions to the $\delta$ expansion coming from the shifted residues are totally boundary local -- totally localized at $\eta_{L/R}=0$. 

In addition, we see from here the result found at $n=4$: the product of the asymptotic expansions of $F_L/F_R$ gives only even powers in $1/\delta$ and the $1/k_\delta$ from \eqref{eq:Deltaresidue} means we only get odd powers from the shifted pole; at higher points however $f_L/f_R$ are themselves functions of $\delta$, so $f_L \equiv f_{L,\delta}$ which at large $\delta$ turn into 
\begin{equation}
    f_{L,\delta}(\eta_L) \sim f^{(0)}_{L} \vert_{\delta=\infty}(\eta_L) + \frac{1}{\delta} \times \frac{d \, f_{L}}{d \, (1/\delta)}  \bigg \vert_{\delta= \infty}  + \frac{1}{2\delta^2} \times \frac{d^2 \, f_{L}}{d \,  (1/\delta)^2}  \bigg \vert_{\delta= \infty} +\cdots,  \quad \text{(similar for $f_R$)}
\end{equation}
which means that we can indeed get even powers of $1/\delta$ in \eqref{eq:Deltaresidue} some of which do contribute to the pion wavefunction. However, as we just showed, these are completely boundary local terms which we can therefore ignore. 

Note that this sum over finite residues in frequency space, is providing a new prescription to compute the pion wavefunction -- $i.e.$ a different off-shell action -- which is not the same as the flat-out $\delta$-shift (so not the $\chi,\chi^\dagger$ action described early) -- as it differs by boundary terms. But remarkably, it is this prescription which reveals a simple, universal behavior under the soft limit. 
\subsection{Universal soft limits}\label{sec:Soft_Limits_FS}

Given the prescription described in the previous section to compute the pion wavefunction, let's now show how in this representation something remarkably simple happens in the soft limit: at leading order it turns into the sum of two mixed pion-scalar wavefunctions. To argue this we start by dividing the different diagrams contributing to the respective wavefunction into the different terms in the standard Berends-Giele recursion. Since in the frequency representation the $\omega$-integrand has precisely the form of an amplitude, we can write the $n$-point wavefunction as follows:
\begin{align}
\Psi_n
&= \sum_{j=3}^{n}
\begin{tikzpicture}[baseline=(current bounding box.center),line width=1pt,scale=10]
    \def\R{0.101}
    \coordinate (J)   at (87:\R);
    \coordinate (Two) at (220:\R);
    \coordinate (One) at (261:\R);
    \fill[black!50,opacity=0.2]
        (J) -- (Two)
        arc[start angle=220,end angle=87,radius=\R cm]
        -- cycle;
    \fill[black!50,opacity=0.2]
        (J) -- (One)
        arc[start angle=261,end angle=447,radius=\R cm]
        -- cycle;
    \draw (0,0) circle[radius=\R cm];
    \draw[red!50!black] (J) -- (Two);
    \draw[red!50!black] (J) -- (One);
    \fill (J) circle[radius=0.1pt];
    \fill[red!50!black] (Two) circle[radius=0.15pt];
    \fill[red!50!black] (One) circle[radius=0.15pt];
    \node[scale=0.8,yshift=7pt] at (J) {$j$};
    \node[scale=0.8,xshift=-10pt] at (Two) {$2$};
    \node[scale=0.8,xshift=-7pt,yshift=-5pt] at (One) {$1$};
    \node[scale=0.8] at (-0.070,0.025) {$\Psi_L$};
    \node[scale=0.8] at (0.050,0.010) {$\Psi_R$};
    \node[scale=0.65] at (-0.033,-0.030) {$C^{(t)}$};
\end{tikzpicture} \\
&=\sum_{j=3}^{n}\int \frac{d \omega_{2,j}}{2\pi i}\int \frac{d \omega_{1,j}}{2\pi i}
\Psi_L(2,3,\cdots,\hat{j})
\frac{C^{(t)}(k_1,\omega_{1,j},\omega_{2,j})}
{(\omega_{2,j}^{2}-k_{2,j}^2)(\omega_{1,j}^{2}-k_{1,j}^2)}
\Psi_R(\hat{1},j,\cdots,n).
\label{eq:BerendsGiele_Rec}
\end{align}
where the hat denotes a particle which is off-shell -- in the sense of being regarded as an internal propagator when considering the $C$-factors. This form of the recursion is ``anchored'' in the external leg with momenta $k_1$ and is therefore appropriate to study the limit in which $k_1 \to 0$\footnote{Note that as usual it is important to take the soft-limit under the support of (spatial) momentum conservation, which we automatically ensure by dealing with the variables of the momentum polygon which remains closed throughout the limit}. 

Now to derive the behavior under the soft-limit we will analyze the different terms in the sum and show that they always turn into shifted lower-point wavefunctions. Let's start with the case of $j=3$, applying the $\delta$ expansion to the $k_{1,3}$ propagator, we have
\begin{equation}
    \begin{gathered}
    \begin{tikzpicture}[line width=1,scale=10]
    \node[regular polygon, regular polygon sides=22,  minimum size=2cm] (p) at (0,0) {};
     \draw (0,0) circle [radius=0.101cm];

    \node[scale=0.8,xshift=-10,yshift=0] at (p.corner 7) {$2$};
    \node[scale=0.8,xshift=-7,yshift=-5] at (p.corner 10) {$1$};    
    \node[scale=0.8,xshift=-7,yshift=7] at (p.corner 4) {$3$};
   % \node[scale=0.8,xshift=7,yshift=7] at (p.corner 20) {$j_2$};
      \draw[Maroon] (p.corner 4) -- (p.corner 10);

    \foreach \i in {7}
    {\draw[fill] (p.corner \i) circle [radius=0.1pt];}
    \foreach \i in {4}
    {\draw[fill] (p.corner \i) circle [radius=0.1pt];}
    \draw[fill] (p.corner 10) circle [radius=0.1pt];

     %\fill[Blue,opacity=0.2]  (p.corner 8) -- (p.corner 7) -- (p.corner 6) -- (p.corner 5) -- (p.corner 4) -- (p.corner 3) -- (p.corner 2) -- (p.corner 1) -- (p.corner 22)-- cycle;

     \fill[Gray,opacity=0.2]   (p.corner 10) -- (p.corner 11) -- (p.corner 12) -- (p.corner 13) -- (p.corner 14) -- (p.corner 15) -- (p.corner 16) -- (p.corner 17) -- (p.corner 18)-- (p.corner 19)-- (p.corner 20) -- (p.corner 21) -- (p.corner 22) -- (p.corner 1) -- (p.corner 2) -- (p.corner 3) -- (p.corner 4) -- cycle;

     %\node[scale=0.8,xshift=18,yshift=30] at (p.corner 8) {$\mathcal{S}_L$};
     \node[scale=0.8,xshift=25,yshift=30] at (p.corner 10) {$\Psi_R$};
   
    \fill[Maroon] (p.corner 7) circle [radius=0.15pt];
     \fill[Maroon] (p.corner 10) circle [radius=0.15pt];

    \end{tikzpicture}     
\end{gathered} = \sum_{m=0}^\infty\frac{(-1)^m}{2 \, \delta^{2m+2}} \int \frac{d \omega_{1,3}}{2\pi i} \frac{-2 i \omega_{1,3}}{(\omega_{1,3}^2-(k_1+k_2)^2)}(\omega_{1,3}^2-k_{1,3}^2)^m \times \Psi_R \, ,
\label{eq:13Soft}
\end{equation}
where $\Psi_R$ also depends on $\omega_{1,3}$. To make this dependence explicit, we need to distinguish between the triangulations where the adjacent triangle to $(1,2,3)$ is of type $C^{(2)}$ and those of type $C^{(3)}$. In the first case, say for triangle $(1,3,n)$, we have an integral of the form:
\begin{equation}
    \begin{gathered}
    \begin{tikzpicture}[line width=1,scale=10]
    \node[regular polygon, regular polygon sides=22,  minimum size=2cm] (p) at (0,0) {};
     \draw (0,0) circle [radius=0.101cm];

    \node[scale=0.8,xshift=-10,yshift=0] at (p.corner 7) {$2$};
    \node[scale=0.8,xshift=-7,yshift=-5] at (p.corner 10) {$1$};   
    \node[scale=0.8,xshift=3,yshift=-7] at (p.corner 12) {$n$};
    \node[scale=0.8,xshift=-7,yshift=7] at (p.corner 4) {$3$};
   % \node[scale=0.8,xshift=7,yshift=7] at (p.corner 20) {$j_2$};

    \foreach \i in {7}
    {\draw[fill] (p.corner \i) circle [radius=0.1pt];}
    \foreach \i in {4}
    {\draw[fill] (p.corner \i) circle [radius=0.1pt];}
    \draw[fill] (p.corner 10) circle [radius=0.1pt];

     %\fill[Blue,opacity=0.2]  (p.corner 8) -- (p.corner 7) -- (p.corner 6) -- (p.corner 5) -- (p.corner 4) -- (p.corner 3) -- (p.corner 2) -- (p.corner 1) -- (p.corner 22)-- cycle;

     \fill[Gray,opacity=0.2]   (p.corner 10) -- (p.corner 11) -- (p.corner 12) -- (p.corner 13) -- (p.corner 14) -- (p.corner 15) -- (p.corner 16) -- (p.corner 17) -- (p.corner 18)-- (p.corner 19)-- (p.corner 20) -- (p.corner 21) -- (p.corner 22) -- (p.corner 1) -- (p.corner 2) -- (p.corner 3) -- (p.corner 4) -- cycle;

     %\node[scale=0.8,xshift=18,yshift=30] at (p.corner 8) {$\mathcal{S}_L$};
     \node[scale=0.8,xshift=30,yshift=30] at (p.corner 10) {$\tilde{\Psi}_R$};
      \draw[Maroon] (p.corner 4) -- (p.corner 10);
    \draw[Gray] (p.corner 4) -- (p.corner 12);
    \fill[Gray] (p.corner 12) circle [radius=0.15pt];
    \fill[Maroon] (p.corner 10) circle [radius=0.15pt];
     \fill[Maroon] (p.corner 7) circle [radius=0.15pt];
     \fill[Black] (p.corner 4) circle [radius=0.15pt];

    \end{tikzpicture}     
\end{gathered} = \sum_{m=0}^\infty \frac{(-1)^m}{2 \, \delta^{2m+2}} \int \frac{d \omega_{1,3}}{2\pi i} \frac{-2 i \omega_{1,3}(\omega_{1,3}^2-k_{1,3}^2)^m }{(\omega_{1,3}^2-(k_1+k_2)^2)}C^{(2)}(k_n;\omega_{1,3} ,\omega_{3,n}) \times \tilde{\Psi}_R \, ,
\end{equation}
where $\tilde{\Psi}_R$ contains the $(\omega_{3,n}^2-k_{3,n}^2)$ denominator; we have omitted the integration over $\omega_{3,n}$ for simplicity. Taking the soft limit, $k_1 \to 0$, also implies $k_{1,3} \to k_2$, and thus the integral simplifies to: 
\begin{equation}
    \sum_{m=0}^\infty \frac{(-1)^m}{2 \, \delta^{2m+2}} \int \frac{d \omega_{1,3}}{2\pi i} \frac{-16  \omega_{1,3}^2\omega_{3,n} k_n (\omega_{1,3}^2-k_2^2)^{m-1} }{((\omega_{1,3}+\omega_{3,n})^2-k_n^2)((\omega_{1,3}-\omega_{3,n})^2-k_n^2)} \times \tilde{\Psi}_R \, .
\end{equation}
Recall that for the pion wavefunction we are interested in the piece at order $\delta^{-(2n-4)}$; but since $\tilde{\Psi}_R$ depends on $\delta$, in general, different $m$ terms play a role. For $m=0$, to compute the $\omega_{1,3}$ integral we sum over the residues $\omega_{1,3} = \{k_2,\omega_{3,n}+k_n,-\omega_{3,n}+k_n\}$, and find: 
\begin{equation}
    \underline{m=0:}\quad \frac{i}{\delta^2} \int \frac{d \omega_{3,n}}{2\pi i} \frac{-2i \omega_{3,n}}{(\omega_{3,n}^2-(k_n+k_2)^2)} \times \tilde{\Psi}_R  \equiv \frac{i}{\delta^2} \times  \Psi^R_{n-1}(2,3,\cdots,n), 
\end{equation}
which we can recognize precisely as the $(n-1)$-point $\delta$-shifted wavefunction! Since this is an odd-point wavefunction, the $\delta$-shift does not yield a pure pion wavefunction, but instead a mixed wavefunction of pions and colored scalars -- as described in \cite{Arkani-Hamed:2024nhp}. The mixed wavefunction occurs at order $\delta^{-(2n-6)}$ which together with the $1/\delta^2$ pre-factor precisely gives the order corresponding to the pion wavefunction. The factor of $(i)$ ensures the final result is real, since inside $\Psi_n-1$ we have an odd number of $(i)$ from the different $C^{(t)}$ functions. 

Finally, for the $m \geq 1$ terms, we only have to sum over the two latter residues, $\omega_{1,3}=\{\omega_{3,n}+k_n,-\omega_{3,n}+k_n\}$, and we find
\begin{equation}
\begin{aligned}
    \underline{m\geq1:}\quad \frac{(-1)^m}{\delta^{2m+2}} \int \frac{d \omega_{3,n}}{2\pi i} &\left[(\omega_{3,n}-k_n)( (\omega_{3,n}-k_n)^2 - k_2^2)^{m-1} \right.\\
    & \quad \left.+ 
 (k_n + \omega_{3,n}) ((\omega_{3,n}+k_n)^2 - k_2^2)^{m-1})\right] \times \tilde{\Psi}_R , 
\end{aligned}
\end{equation}
now interestingly the factor multiplying $\tilde{\Psi}_R$ is precisely the expansion of the of the three-point factor $C^{(1)}(k_2,k_n;\omega_{3,n})$, obtained after collapsing $2 \leftrightarrow 1$, assuming $k_2$ is $\delta$-shifted; this is: 
\begin{equation}
\begin{aligned}
&C^{(1)}\left(\sqrt{k_2^2-\delta^2},k_n;\omega_{3,n}\right) \xrightarrow{\delta \to \infty} \quad -i \sum_{j=1}^{\infty} \frac{a_{2j}}{\delta^{2j}}  + (\text{odd powers in }1/\delta) \\
& \text{with } a_{2j} = (-1)^j \left[(\omega_{3,n}-k_n)( (\omega_{3,n}-k_n)^2 - k_2^2)^{j-1} +
 (k_n + \omega_{3,n}) ((\omega_{3,n}+k_n)^2 - k_2^2)^{j-1})\right] 
\end{aligned}
\end{equation}
but since the odd powers do not contribute to the pion wavefunction, we can interpret the remaining terms in the series, $m \geq 1$, as: 
\begin{equation}
    \underline{m\geq1:} \quad \frac{-i}{\delta^2} \times \Psi^R_{n-1}(\hat{2},3,\cdots,n),
    \label{eq:PsiR}
\end{equation}
where $\Psi^R_{n-1}$ is the $\delta$-shifted wavefunction associated to the $(n-1)$-gon we get by collapsing $1$ and $2$, but where crucially the hatted state, $\hat{2}$, is $\delta$-shifted, $k_2 \to \sqrt{k_2^2-\delta^2}$. Note that the sign of the $\delta$-shift for $k_2$ is the one of the propagator $k_{1,3}$ which in the soft limit is precisely identified with $k_2$. 

Now, analogous to   what happens for the amplitude \cite{Zeros,Arkani-Hamed:2024nhp,circles}, the $\delta$-shift of an odd-point wavefunction  precisely gives a mixed wavefunction for pions and scalars, NLSM$\oplus \phi^3$. In this particular case note that the chords ending in $2$ are shifted as if $2$ was \textit{odd} -- since in the underlying $n$-point problem it comes from boundary point $1$. As a result we have that 
\begin{equation}
    \Psi^R_{n-1}(\hat{2},3,4,5,\cdots,n-1,n) \equiv \Psi^{\text{NLSM}\oplus \phi^3}_{n-1}(\hat{2}^{\phi},3^{\phi},4^\pi,5^\pi,\cdots,(n-1)^\pi,n^\phi) \vert_{k_2 \to \sqrt{k_2^2-\delta^2}}
\end{equation}
this is the wavefunction in the mixed theory where $(2,3,n)$ are $\phi$'s and the remaining states are $\pi$'s; and where the external momenta, $k_2$ has been also $\delta$-shifted.

If instead the triangle adjacent to $(1,2,3)$ is of type $C^{(3)}$, in the soft limit, we get the following integral:
\begin{equation}
    \begin{gathered}
    \begin{tikzpicture}[line width=1,scale=10]
    \node[regular polygon, regular polygon sides=22,  minimum size=2cm] (p) at (0,0) {};
     \draw (0,0) circle [radius=0.101cm];

    \node[scale=0.8,xshift=-10,yshift=0] at (p.corner 7) {$2$};
    \node[scale=0.8,xshift=-7,yshift=-5] at (p.corner 10) {$1$};   
    \node[scale=0.8,xshift=8,yshift=5] at (p.corner 17) {$p$};
    \node[scale=0.8,xshift=-7,yshift=7] at (p.corner 4) {$3$};
   % \node[scale=0.8,xshift=7,yshift=7] at (p.corner 20) {$j_2$};

    \foreach \i in {7}
    {\draw[fill] (p.corner \i) circle [radius=0.1pt];}
    \foreach \i in {4}
    {\draw[fill] (p.corner \i) circle [radius=0.1pt];}
    \draw[fill] (p.corner 10) circle [radius=0.1pt];

     %\fill[Blue,opacity=0.2]  (p.corner 8) -- (p.corner 7) -- (p.corner 6) -- (p.corner 5) -- (p.corner 4) -- (p.corner 3) -- (p.corner 2) -- (p.corner 1) -- (p.corner 22)-- cycle;

     \fill[Gray,opacity=0.2]   (p.corner 10) -- (p.corner 11) -- (p.corner 12) -- (p.corner 13) -- (p.corner 14) -- (p.corner 15) -- (p.corner 16) -- (p.corner 17) -- (p.corner 18)-- (p.corner 19)-- (p.corner 20) -- (p.corner 21) -- (p.corner 22) -- (p.corner 1) -- (p.corner 2) -- (p.corner 3) -- (p.corner 4) -- cycle;

     %\node[scale=0.8,xshift=18,yshift=30] at (p.corner 8) {$\mathcal{S}_L$};
     \node[scale=0.7,xshift=35,yshift=60] at (p.corner 10) {$\tilde{\Psi}_{(1)}$};
     \node[scale=0.7,xshift=40,yshift=5] at (p.corner 10) {$\tilde{\Psi}_{(2)}$};
     \node[scale=0.7,xshift=27,yshift=33] at (p.corner 10) {$C^{(3)}$};
      \draw[Maroon] (p.corner 4) -- (p.corner 10);
    \draw[Gray] (p.corner 4) -- (p.corner 18);
    \draw[Gray] (p.corner 10) -- (p.corner 18);
    \fill[Gray] (p.corner 18) circle [radius=0.15pt];
    \fill[Maroon] (p.corner 10) circle [radius=0.15pt];
     \fill[Maroon] (p.corner 7) circle [radius=0.15pt];
     \fill[Black] (p.corner 4) circle [radius=0.15pt];

    \end{tikzpicture}     
\end{gathered} = \sum_{m=0}^\infty \frac{(-1)^m}{2 \, \delta^{2m+2}} \int \frac{d \omega_{1,3}}{2\pi i} \left[-2 i \omega_{1,3}(\omega_{1,3}^2-k_{1,3}^2)^{m-1} \right] C^{(3)} \times \tilde{\Psi}_{(1)}\tilde{\Psi}_{(2)}
\end{equation}
where $C^{(3)} \equiv C^{(3)}(\omega_{1,3},\omega_{1,p} ,\omega_{3,p})$. Now given the form of $C^{(3)}$ from eqn.$\,$\eqref{eq:3ptFactors}, we have for all $m\geq 1$ the sum over the finite residues in $\omega_{1,3}$ coming from $C^{(3)}$ vanishes, and so we are only left with the contribution from $m=0$, which gives
\begin{equation}
     \underline{m=0:}\quad \frac{i}{\delta^2} \int  \frac{ d \omega_{1,p} d \omega_{3,p}}{(2\pi i)^2}\underbracket[0.4pt]{ \frac{-8i k_2 \omega_{1,p} \omega_{3,p}}{((\omega_{1,p}+\omega_{3,p})^2-k_2^2)((\omega_{1,p}-\omega_{3,p})^2-k_2^2)} }_{C^{(2)}(k_2;\omega_{1,p},\omega_{3,p})}\times  \tilde{\Psi}_{(1)} \tilde{\Psi}_{(2)}  \equiv \frac{i}{\delta^2} \times  \Psi^R_{n-1}, 
\end{equation}
which is precisely the result found previously: the $(n-1)$-point $\delta$-shifted wavefunction. Note that from the triangulations with a $C^{(2)}$-type triangle adjacent to $(1,2,3)$, we also obtained a piece corresponding to $\Psi_{n-1}^R$ but with $k_2$ $\delta$-shifted; and this contribution seems to be missing here since the $m\geq 1$ terms in the expansion vanished. As it turns out this is simply reflecting the fact that the $\delta$-shift of $k_2$ inside $C^{(2)}(k_2;\omega_{1,p},\omega_{3,p})$ only leads to odd-powers in $1/\delta$ and therefore gives no contribution to the relevant order in $1/\delta$: 
\begin{equation}
    C^{(2)}\left[\sqrt{k_2^2-\delta^2};\omega_{1,p},\omega_{3,p}\right] \xrightarrow[\delta \gg 1]{} \delta \times \frac{8\omega_{1,p} \omega_{3,p} \sqrt{1 - k_2^2/\delta^2}}{((\omega_{1,p}+\omega_{3,p})^2-k_2^2+\delta^2)((\omega_{1,p}-\omega_{3,p})^2-k_2^2+\delta^2)} \,.
\end{equation}

Therefore, we have that for all the triangulations containing triangle $(1,2,3)$, in the limit where $k_1 \to 0$, we obtain the sum over two terms:
\begin{equation}
    \begin{gathered}
    \begin{tikzpicture}[line width=1,scale=10]
    \node[regular polygon, regular polygon sides=22,  minimum size=2cm] (p) at (0,0) {};
     \draw (0,0) circle [radius=0.101cm];
    \node[scale=0.8,xshift=-10,yshift=0] at (p.corner 7) {$2$};
    \node[scale=0.8,xshift=-7,yshift=-5] at (p.corner 10) {$1$};   
    \node[scale=0.8,xshift=3,yshift=-7] at (p.corner 12) {$n$};
    \node[scale=0.8,xshift=-7,yshift=7] at (p.corner 4) {$3$};
   % \node[scale=0.8,xshift=7,yshift=7] at (p.corner 20) {$j_2$};

    \foreach \i in {7}
    {\draw[fill] (p.corner \i) circle [radius=0.1pt];}
    \foreach \i in {4}
    {\draw[fill] (p.corner \i) circle [radius=0.1pt];}
    \draw[fill] (p.corner 10) circle [radius=0.1pt];

     %\fill[Blue,opacity=0.2]  (p.corner 8) -- (p.corner 7) -- (p.corner 6) -- (p.corner 5) -- (p.corner 4) -- (p.corner 3) -- (p.corner 2) -- (p.corner 1) -- (p.corner 22)-- cycle;

     \fill[Gray,opacity=0.2]   (p.corner 10) -- (p.corner 11) -- (p.corner 12) -- (p.corner 13) -- (p.corner 14) -- (p.corner 15) -- (p.corner 16) -- (p.corner 17) -- (p.corner 18)-- (p.corner 19)-- (p.corner 20) -- (p.corner 21) -- (p.corner 22) -- (p.corner 1) -- (p.corner 2) -- (p.corner 3) -- (p.corner 4) -- cycle;

     %\node[scale=0.8,xshift=18,yshift=30] at (p.corner 8) {$\mathcal{S}_L$};
     \node[scale=0.8,xshift=30,yshift=30] at (p.corner 10) {$\Psi_R$};
      \draw[Maroon] (p.corner 4) -- (p.corner 10);
    %\draw[Gray] (p.corner 4) -- (p.corner 12);
    \fill[Black] (p.corner 12) circle [radius=0.15pt];
    \fill[Maroon] (p.corner 10) circle [radius=0.15pt];
     \fill[Maroon] (p.corner 7) circle [radius=0.15pt];
     \fill[Black] (p.corner 4) circle [radius=0.15pt];
\draw[-{Stealth[length=1.5mm,width=0.9mm]}]
  (220.09:0.115cm)
  arc[
    start angle=220.09,
    end angle=185,
    radius=0.115cm
  ];

    \end{tikzpicture}     
\end{gathered} \quad \xrightarrow[k_1 \to 0]{}\quad   \frac{i}{\delta^2} \times \left[\begin{gathered}
    \begin{tikzpicture}[line width=1,scale=10]
    \node[regular polygon, regular polygon sides=22,  minimum size=2cm] (p) at (0,0) {};
     \draw (0,0) circle [radius=0.101cm];
   % \draw[Blue!80!White,dashed,line width=2]
  %({90-180/22+3*360/22}:0.101cm)
  %arc[
  %  start angle={90-180/22+3*360/22},
  %  end angle={90-180/22+6*360/22},
   % radius=0.101cm
  %];

    \node[scale=0.8,xshift=-9,yshift=6] at (p.corner 7) {$2$};
    \node[scale=0.8,xshift=-7,yshift=-5] at (p.corner 7) {$1_{\text{\tiny{R}}}$};   
    \node[scale=0.8,xshift=-6,yshift=-5] at (p.corner 10) {$n$};
    \node[scale=0.8,xshift=-7,yshift=7] at (p.corner 4) {$3$};
   % \node[scale=0.8,xshift=7,yshift=7] at (p.corner 20) {$j_2$};
\draw[Maroon]
  ({90-180/22+3*360/22}:0.101cm)
  arc[
    start angle={90-180/22+3*360/22},
    end angle={90-180/22+6*360/22},
    radius=0.101cm
  ];

    \foreach \i in {7}
    {\draw[Maroon,fill] (p.corner \i) circle [radius=0.1pt];}
    \foreach \i in {4}
    {\draw[fill] (p.corner \i) circle [radius=0.1pt];}
    \draw[fill] (p.corner 10) circle [radius=0.1pt];

     \fill[Gray,opacity=0.2]   (p.corner 10) -- (p.corner 11) -- (p.corner 12) -- (p.corner 13) -- (p.corner 14) -- (p.corner 15) -- (p.corner 16) -- (p.corner 17) -- (p.corner 18)-- (p.corner 19)-- (p.corner 20) -- (p.corner 21) -- (p.corner 22) -- (p.corner 1) -- (p.corner 2) -- (p.corner 3) -- (p.corner 4) -- (p.corner 5) -- (p.corner 6) -- (p.corner 7) -- (p.corner 8) -- (p.corner 9)-- cycle;

     %\node[scale=0.8,xshift=18,yshift=30] at (p.corner 8) {$\mathcal{S}_L$};
     \node[scale=0.8,xshift=25,yshift=30] at (p.corner 10) {$\Psi^R_{n-1}$};
      %\draw[Gray] (p.corner 4) -- (p.corner 10);

    \fill[Black] (p.corner 10) circle [radius=0.15pt];
     \fill[Black] (p.corner 4) circle [radius=0.15pt];

    \end{tikzpicture}     
\end{gathered} \quad - \quad  \begin{gathered}
    \begin{tikzpicture}[line width=1,scale=10]
    \node[regular polygon, regular polygon sides=22,  minimum size=2cm] (p) at (0,0) {};
     \draw (0,0) circle [radius=0.101cm];
    \draw[Blue!80!White,dashed,line width=2]
  ({90-180/22+3*360/22}:0.101cm)
  arc[
    start angle={90-180/22+3*360/22},
    end angle={90-180/22+6*360/22},
    radius=0.101cm
  ];

    \node[scale=0.8,xshift=-9,yshift=6] at (p.corner 7) {$2$};
    \node[scale=0.8,xshift=-7,yshift=-5] at (p.corner 7) {$1_{\text{\tiny{R}}}$};   
    \node[scale=0.8,xshift=-6,yshift=-5] at (p.corner 10) {$n$};
    \node[scale=0.8,xshift=-7,yshift=7] at (p.corner 4) {$3$};
   % \node[scale=0.8,xshift=7,yshift=7] at (p.corner 20) {$j_2$};

    \foreach \i in {7}
    {\draw[Maroon,fill] (p.corner \i) circle [radius=0.1pt];}
    \foreach \i in {4}
    {\draw[fill] (p.corner \i) circle [radius=0.1pt];}
    \draw[fill] (p.corner 10) circle [radius=0.1pt];

     \fill[Gray,opacity=0.2]   (p.corner 10) -- (p.corner 11) -- (p.corner 12) -- (p.corner 13) -- (p.corner 14) -- (p.corner 15) -- (p.corner 16) -- (p.corner 17) -- (p.corner 18)-- (p.corner 19)-- (p.corner 20) -- (p.corner 21) -- (p.corner 22) -- (p.corner 1) -- (p.corner 2) -- (p.corner 3) -- (p.corner 4) -- (p.corner 5) -- (p.corner 6) -- (p.corner 7) -- (p.corner 8) -- (p.corner 9)-- cycle;

     %\node[scale=0.8,xshift=18,yshift=30] at (p.corner 8) {$\mathcal{S}_L$};
     \node[scale=0.8,xshift=25,yshift=30] at (p.corner 10) {$\Psi^R_{n-1}$};
      %\draw[Gray] (p.corner 4) -- (p.corner 10);

    \fill[Black] (p.corner 10) circle [radius=0.15pt];
     \fill[Black] (p.corner 4) circle [radius=0.15pt];

    \end{tikzpicture}     
\end{gathered} \right]
\label{eq:Softj3}
\end{equation}
where the first term is simply the $\delta$-shifted lower-point wavefunction, coming from the $m=0$ term in the expansion, and the second term the $m\geq1$, where now also $k_2$ is $\delta$-shifted, denoted by the dashed blue line. {From the complete right-hand side of
\eqref{eq:Softj3}, including the prefactor $i/\delta^2$,
we extract the coefficient of $\delta^{-(2n-4)}$.
Equivalently, the expression inside the brackets must be
expanded to order $\delta^{-(2n-6)}$.} In both cases, we added a subscript, $1_R$, to remind ourselves that the $\delta$-shift in $\Psi^R_{n-1}$ is such that the parity of the red vertex is determined by $1$ and not $2$ -- so that it coincides with the original $\delta$-shift before taking the soft limit. 

Before proceeding to the more general triangle to $(1,2,j)$, let us look at the other extreme where $j=n$. This case is totally analogous to the case of $j=3$, with the only difference that now $k_{2,n}^2$ is shifted by $+\delta^2$, and so instead of \eqref{eq:13Soft} we get
\begin{equation}
    \begin{gathered}
    \begin{tikzpicture}[line width=1,scale=10]
    \node[regular polygon, regular polygon sides=22,  minimum size=2cm] (p) at (0,0) {};
     \draw (0,0) circle [radius=0.101cm];

    \node[scale=0.8,xshift=-10,yshift=0] at (p.corner 7) {$2$};
    \node[scale=0.8,xshift=-7,yshift=-5] at (p.corner 10) {$1$};    
    \node[scale=0.8,xshift=-3,yshift=-7] at (p.corner 13) {$n$};
   % \node[scale=0.8,xshift=7,yshift=7] at (p.corner 20) {$j_2$};
      \draw[Maroon] (p.corner 7) -- (p.corner 13);

    \foreach \i in {13}
    {\draw[fill] (p.corner \i) circle [radius=0.1pt];}
    \draw[fill] (p.corner 10) circle [radius=0.1pt];

     %\fill[Blue,opacity=0.2]  (p.corner 8) -- (p.corner 7) -- (p.corner 6) -- (p.corner 5) -- (p.corner 4) -- (p.corner 3) -- (p.corner 2) -- (p.corner 1) -- (p.corner 22)-- cycle;

     \fill[Gray,opacity=0.2]   (p.corner 13) -- (p.corner 14) -- (p.corner 15) -- (p.corner 16) -- (p.corner 17) -- (p.corner 18)-- (p.corner 19)-- (p.corner 20) -- (p.corner 21) -- (p.corner 22) -- (p.corner 1) -- (p.corner 2) -- (p.corner 3) -- (p.corner 4) --(p.corner 5) -- (p.corner 6) -- (p.corner 7) --cycle;

     %\node[scale=0.8,xshift=18,yshift=30] at (p.corner 8) {$\mathcal{S}_L$};
     \node[scale=0.8,xshift=25,yshift=30] at (p.corner 10) {$\Psi_L$};
   
    \fill[Maroon] (p.corner 7) circle [radius=0.15pt];
     \fill[Maroon] (p.corner 10) circle [radius=0.15pt];

    \end{tikzpicture}     
\end{gathered} = \sum_{m=0}^\infty\frac{(-1)}{2 \, \delta^{2m+2}} \int \frac{d \omega_{2,n}}{2\pi i} \frac{-2 i \omega_{2,n}}{(\omega_{2,n}^2-(k_1+k_n)^2)}(\omega_{2,n}^2-k_{2,n}^2)^m \times \Psi_L \, ,
\label{eq:2nSoft}
\end{equation}
with the main difference compared to \eqref{eq:13Soft} being the factor of $(-1)$ outfront, instead of $(-1)^m$. As a result we find that in the soft limit we instead get
\begin{equation}
    \begin{gathered}
    \begin{tikzpicture}[line width=1,scale=10]
    \node[regular polygon, regular polygon sides=22,  minimum size=2cm] (p) at (0,0) {};
     \draw (0,0) circle [radius=0.101cm];
    
    \node[scale=0.8,xshift=-10,yshift=0] at (p.corner 7) {$2$};
    \node[scale=0.8,xshift=-7,yshift=-5] at (p.corner 10) {$1$};    
    \node[scale=0.8,xshift=-3,yshift=-7] at (p.corner 13) {$n$};
   % \node[scale=0.8,xshift=7,yshift=7] at (p.corner 20) {$j_2$};

   \draw[Maroon] (p.corner 7) -- (p.corner 13);

    \foreach \i in {13}
    {\draw[fill] (p.corner \i) circle [radius=0.1pt];}
    \draw[fill] (p.corner 10) circle [radius=0.1pt];

     %\fill[Blue,opacity=0.2]  (p.corner 8) -- (p.corner 7) -- (p.corner 6) -- (p.corner 5) -- (p.corner 4) -- (p.corner 3) -- (p.corner 2) -- (p.corner 1) -- (p.corner 22)-- cycle;

     \fill[Gray,opacity=0.2]   (p.corner 13) -- (p.corner 14) -- (p.corner 15) -- (p.corner 16) -- (p.corner 17) -- (p.corner 18)-- (p.corner 19)-- (p.corner 20) -- (p.corner 21) -- (p.corner 22) -- (p.corner 1) -- (p.corner 2) -- (p.corner 3) -- (p.corner 4) --(p.corner 5) -- (p.corner 6) -- (p.corner 7) --cycle;

     %\node[scale=0.8,xshift=18,yshift=30] at (p.corner 8) {$\mathcal{S}_L$};
     \node[scale=0.8,xshift=25,yshift=30] at (p.corner 10) {$\Psi_L$};
      %\draw[Maroon] (p.corner 4) -- (p.corner 10);
    %\draw[Gray] (p.corner 4) -- (p.corner 12);

    \fill[Maroon] (p.corner 10) circle [radius=0.15pt];
     \fill[Maroon] (p.corner 7) circle [radius=0.15pt];
\draw[-{Stealth[length=1.5mm,width=0.9mm]}]
  (220.09:0.115cm)
  arc[
    start angle=220.09,
    end angle=185,
    radius=0.115cm
  ];

    \end{tikzpicture}     
\end{gathered} \quad \xrightarrow[k_1 \to 0]{}\quad   \frac{-i}{\delta^2} \times \left[ \begin{gathered}
    \begin{tikzpicture}[line width=1,scale=10]
    \node[regular polygon, regular polygon sides=22,  minimum size=2cm] (p) at (0,0) {};
     \draw (0,0) circle [radius=0.101cm];
   % \draw[Blue!80!White,dashed,line width=2]
  %({90-180/22+3*360/22}:0.101cm)
  %arc[
  %  start angle={90-180/22+3*360/22},
  %  end angle={90-180/22+6*360/22},
   % radius=0.101cm
  %];

    \node[scale=0.8,xshift=-7,yshift=6] at (p.corner 7) {$2_{\text{\tiny{L}}}$};
    \node[scale=0.8,xshift=-8,yshift=-5] at (p.corner 7) {$1$};   
    \node[scale=0.8,xshift=-6,yshift=-5] at (p.corner 10) {$n$};
    \node[scale=0.8,xshift=-7,yshift=7] at (p.corner 4) {$3$};
   % \node[scale=0.8,xshift=7,yshift=7] at (p.corner 20) {$j_2$};
\draw[Maroon]
  (180:0.101cm)
  arc[
    start angle=180,
    end angle=229.09,
    radius=0.101cm
  ];
    
    \foreach \i in {7}
    {\draw[Maroon,fill] (p.corner \i) circle [radius=0.1pt];}
    \foreach \i in {4}
    {\draw[fill] (p.corner \i) circle [radius=0.1pt];}
    \draw[fill] (p.corner 10) circle [radius=0.1pt];

     \fill[Gray,opacity=0.2]   (p.corner 10) -- (p.corner 11) -- (p.corner 12) -- (p.corner 13) -- (p.corner 14) -- (p.corner 15) -- (p.corner 16) -- (p.corner 17) -- (p.corner 18)-- (p.corner 19)-- (p.corner 20) -- (p.corner 21) -- (p.corner 22) -- (p.corner 1) -- (p.corner 2) -- (p.corner 3) -- (p.corner 4) -- (p.corner 5) -- (p.corner 6) -- (p.corner 7) -- (p.corner 8) -- (p.corner 9)-- cycle;

     %\node[scale=0.8,xshift=18,yshift=30] at (p.corner 8) {$\mathcal{S}_L$};
     \node[scale=0.8,xshift=25,yshift=30] at (p.corner 10) {$\Psi^L_{n-1}$};
      %\draw[Gray] (p.corner 4) -- (p.corner 10);

    \fill[Black] (p.corner 10) circle [radius=0.15pt];
     \fill[Black] (p.corner 4) circle [radius=0.15pt];

    \end{tikzpicture}     
\end{gathered} \quad - \quad  \begin{gathered}
    \begin{tikzpicture}[line width=1,scale=10]
    \node[regular polygon, regular polygon sides=22,  minimum size=2cm] (p) at (0,0) {};
     \draw (0,0) circle [radius=0.101cm];

  \draw[Blue!80!White,dashed,line width=2]
  (180:0.101cm)
  arc[
    start angle=180,
    end angle=229.09,
    radius=0.101cm
  ];

    \node[scale=0.8,xshift=-7,yshift=6] at (p.corner 7) {$2_{\text{\tiny{L}}}$};
    \node[scale=0.8,xshift=-8,yshift=-5] at (p.corner 7) {$1$};   
    \node[scale=0.8,xshift=-6,yshift=-5] at (p.corner 10) {$n$};
    \node[scale=0.8,xshift=-7,yshift=7] at (p.corner 4) {$3$};
   % \node[scale=0.8,xshift=7,yshift=7] at (p.corner 20) {$j_2$};

    \foreach \i in {7}
    {\draw[Maroon,fill] (p.corner \i) circle [radius=0.1pt];}
    \foreach \i in {4}
    {\draw[fill] (p.corner \i) circle [radius=0.1pt];}
    \draw[fill] (p.corner 10) circle [radius=0.1pt];

     \fill[Gray,opacity=0.2]   (p.corner 10) -- (p.corner 11) -- (p.corner 12) -- (p.corner 13) -- (p.corner 14) -- (p.corner 15) -- (p.corner 16) -- (p.corner 17) -- (p.corner 18)-- (p.corner 19)-- (p.corner 20) -- (p.corner 21) -- (p.corner 22) -- (p.corner 1) -- (p.corner 2) -- (p.corner 3) -- (p.corner 4) -- (p.corner 5) -- (p.corner 6) -- (p.corner 7) -- (p.corner 8) -- (p.corner 9)-- cycle;

     %\node[scale=0.8,xshift=18,yshift=30] at (p.corner 8) {$\mathcal{S}_L$};
     \node[scale=0.8,xshift=25,yshift=30] at (p.corner 10) {$\Psi^L_{n-1}$};
      %\draw[Gray] (p.corner 4) -- (p.corner 10);

    \fill[Black] (p.corner 10) circle [radius=0.15pt];
     \fill[Black] (p.corner 4) circle [radius=0.15pt];

    \end{tikzpicture}     
\end{gathered} \right], 
\label{eq:Softjn}
\end{equation}
where now in the $(n-1)$-point wavefunction, $\Psi^L_{n-1}$, the parity of the red vertex is even -- as denoted by $2_\text{L}$; in addition in the second term we have that $k_n$ is $\delta$-shifted, $k_n \to \sqrt{k_n^2+\delta^2}$, where again the sign of the shift comes from the chord $k_{2,n}$ which is identified with $k_n$ in the soft limit. Precisely as for $\Psi^R_{n-1}$ in \eqref{eq:PsiR}, we can also recognize $\Psi^L_{n-1}$ as a wavefunction in the mixed theory, explicitly:
\begin{equation}
    \Psi^L_{n-1} \equiv \Psi^{\text{NLSM}\oplus \phi^3}_{n-1}(2^{\phi},3^{\pi},4^\pi,\cdots,(n-2)^\pi,(n-1)^\phi,n^\phi) \, . 
\end{equation}

Now for the remaining terms in the recursion \eqref{eq:BerendsGiele_Rec} corresponding to $j \neq 3,n$, we get an integral of the form
\begin{equation}
    \begin{gathered}
    \begin{tikzpicture}[line width=1,scale=10]
    \node[regular polygon, regular polygon sides=22,  minimum size=2cm] (p) at (0,0) {};
     \draw (0,0) circle [radius=0.101cm];

    \node[scale=0.8,xshift=-10,yshift=0] at (p.corner 7) {$2$};
    \node[scale=0.8,xshift=-7,yshift=-5] at (p.corner 10) {$1$};    
    \node[scale=0.8,xshift=7,yshift=7] at (p.corner 19) {$j$};
   % \node[scale=0.8,xshift=7,yshift=7] at (p.corner 20) {$j_2$};

    \foreach \i in {7}
    {\draw[fill] (p.corner \i) circle [radius=0.1pt];}
    \foreach \i in {19}
    {\draw[fill] (p.corner \i) circle [radius=0.1pt];}
    \draw[fill] (p.corner 10) circle [radius=0.1pt];

     %\fill[Blue,opacity=0.2]  (p.corner 8) -- (p.corner 7) -- (p.corner 6) -- (p.corner 5) -- (p.corner 4) -- (p.corner 3) -- (p.corner 2) -- (p.corner 1) -- (p.corner 22)-- cycle;

     \fill[Gray,opacity=0.2]   (p.corner 10) --  (p.corner 19)--(p.corner 18) -- (p.corner 17) -- (p.corner 16) -- (p.corner 15) -- (p.corner 14) -- (p.corner 13) -- (p.corner 12)-- (p.corner 11)-- cycle;
    
    \fill[Gray,opacity=0.2]   (p.corner 7) --  (p.corner 6)--(p.corner 5) -- (p.corner 4) -- (p.corner 3) -- (p.corner 2) -- (p.corner 1) -- (p.corner 21) -- (p.corner 20)-- (p.corner 19)-- cycle;

     %\node[scale=0.8,xshift=18,yshift=30] at (p.corner 8) {$\mathcal{S}_L$};
     \node[scale=0.8,xshift=38,yshift=10] at (p.corner 10) {$\Psi_R$};
     \node[scale=0.8,xshift=20,yshift=48] at (p.corner 10) {$\Psi_L$};
      \draw[Maroon] (p.corner 7) -- (p.corner 19);
      \draw[Maroon] (p.corner 10) -- (p.corner 19);
    \fill[Maroon] (p.corner 10) circle [radius=0.15pt];
     \fill[Maroon] (p.corner 7) circle [radius=0.15pt];
     \fill[Black] (p.corner 19) circle [radius=0.15pt];

    \end{tikzpicture}     
\end{gathered} = \sum_{m=0}^\infty\frac{(-1)}{4 \, \delta^{2m+2}} \int \frac{d \omega_{2,j} d \omega_{1,j}}{(2\pi i)^2} \Psi_L \times \frac{(\omega_{2,j}^2-k_{2,j}^2)^m C^{(2)}(k_1;\omega_{1,j},\omega_{2,j})}{(\omega_{1,j}^2-k_{1,j}^2)}  \times \Psi_R \, ,
\end{equation}
where here we have assumed, without loss of generality, that $j$ is even so that only $k_{2,j}$ is $\delta$-shifted. Now the numerator factor of $k_1$ inside $C^{(2)}$ means that in the soft limit, the only contribution comes from the poles in $C^{(2)}$ which collide when we send $k_1 \to 0$; these are at $\omega_{1,j} = \pm \omega_{2,j}$. Reproducing the analysis done above for these residues, both in the case where $j$ is even/odd, we find the following behavior in the soft limit:
\begin{equation}
\begin{gathered}
    \begin{tikzpicture}[line width=1,scale=10]
    \node[regular polygon, regular polygon sides=22,  minimum size=2cm] (p) at (0,0) {};
     \draw (0,0) circle [radius=0.101cm];

    \node[scale=0.8,xshift=-10,yshift=0] at (p.corner 7) {$2$};
    \node[scale=0.8,xshift=-7,yshift=-5] at (p.corner 10) {$1$};    
    \node[scale=0.8,xshift=7,yshift=7] at (p.corner 19) {$j$};
   % \node[scale=0.8,xshift=7,yshift=7] at (p.corner 20) {$j_2$};

    \foreach \i in {7}
    {\draw[fill] (p.corner \i) circle [radius=0.1pt];}
    \foreach \i in {19}
    {\draw[fill] (p.corner \i) circle [radius=0.1pt];}
    \draw[fill] (p.corner 10) circle [radius=0.1pt];

     %\fill[Blue,opacity=0.2]  (p.corner 8) -- (p.corner 7) -- (p.corner 6) -- (p.corner 5) -- (p.corner 4) -- (p.corner 3) -- (p.corner 2) -- (p.corner 1) -- (p.corner 22)-- cycle;

     \fill[Gray,opacity=0.2]   (p.corner 10) --  (p.corner 19)--(p.corner 18) -- (p.corner 17) -- (p.corner 16) -- (p.corner 15) -- (p.corner 14) -- (p.corner 13) -- (p.corner 12)-- (p.corner 11)-- cycle;
    
    \fill[Gray,opacity=0.2]   (p.corner 7) --  (p.corner 6)--(p.corner 5) -- (p.corner 4) -- (p.corner 3) -- (p.corner 2) -- (p.corner 1) -- (p.corner 21) -- (p.corner 20)-- (p.corner 19)-- cycle;

     %\node[scale=0.8,xshift=18,yshift=30] at (p.corner 8) {$\mathcal{S}_L$};
     \node[scale=0.8,xshift=38,yshift=10] at (p.corner 10) {$\Psi_R$};
     \node[scale=0.8,xshift=20,yshift=48] at (p.corner 10) {$\Psi_L$};
      \draw[Maroon] (p.corner 7) -- (p.corner 19);
      \draw[Maroon] (p.corner 10) -- (p.corner 19);
    \fill[Maroon] (p.corner 10) circle [radius=0.15pt];
     \fill[Maroon] (p.corner 7) circle [radius=0.15pt];
     \fill[Black] (p.corner 19) circle [radius=0.15pt];
     \draw[-{Stealth[length=1.5mm,width=0.9mm]}]
  (220.09:0.115cm)
  arc[
    start angle=220.09,
    end angle=185,
    radius=0.115cm
  ];

    \end{tikzpicture}     
\end{gathered} \quad \xrightarrow[k_1 \to 0]{}\quad (-1)^{j+1}\frac{i}{\delta^2} \times \left[ \begin{gathered}
    \begin{tikzpicture}[line width=1,scale=10]
    \node[regular polygon, regular polygon sides=22,  minimum size=2cm] (p) at (0,0) {};
     \draw (0,0) circle [radius=0.101cm];

    \node[scale=0.8,xshift=-10,yshift=5] at (p.corner 7) {$2_{\text{\tiny{L}}}$};
    \node[scale=0.8,xshift=-10,yshift=-5] at (p.corner 7) {$1_{\text{\tiny{R}}}$};    
    \node[scale=0.8,xshift=7,yshift=7] at (p.corner 19) {$j$};
   % \node[scale=0.8,xshift=7,yshift=7] at (p.corner 20) {$j_2$};

    \foreach \i in {7}
    {\draw[fill] (p.corner \i) circle [radius=0.1pt];}
    \foreach \i in {19}
    {\draw[fill] (p.corner \i) circle [radius=0.1pt];}
    %\draw[fill] (p.corner 10) circle [radius=0.1pt];

     %\fill[Blue,opacity=0.2]  (p.corner 8) -- (p.corner 7) -- (p.corner 6) -- (p.corner 5) -- (p.corner 4) -- (p.corner 3) -- (p.corner 2) -- (p.corner 1) -- (p.corner 22)-- cycle;

     \fill[Gray,opacity=0.2]   (p.corner 10) -- (p.corner 11) -- (p.corner 12) -- (p.corner 13) -- (p.corner 14) -- (p.corner 15) -- (p.corner 16) -- (p.corner 17) -- (p.corner 18)-- (p.corner 19)-- (p.corner 20) -- (p.corner 21) -- (p.corner 22) -- (p.corner 1) -- (p.corner 2) -- (p.corner 3) -- (p.corner 4) -- (p.corner 5) -- (p.corner 6) -- (p.corner 7) -- (p.corner 8) -- (p.corner 9)-- cycle;

     %\node[scale=0.8,xshift=18,yshift=30] at (p.corner 8) {$\mathcal{S}_L$};
     \node[scale=0.8,xshift=25,yshift=15] at (p.corner 10) {$\Psi_R$};
     \node[scale=0.8,xshift=20,yshift=48] at (p.corner 10) {$\Psi_L$};
      \draw[Maroon] (p.corner 7) -- (p.corner 19);
      %\draw[Maroon] (p.corner 10) -- (p.corner 19);
    %\fill[Maroon] (p.corner 10) circle [radius=0.15pt];
     \fill[Maroon] (p.corner 7) circle [radius=0.15pt];
     %\fill[Black] (p.corner 19) circle [radius=0.15pt];

    \end{tikzpicture}     
\end{gathered} \quad - \quad \begin{gathered}
    \begin{tikzpicture}[line width=1,scale=10]
    \node[regular polygon, regular polygon sides=22,  minimum size=2cm] (p) at (0,0) {};
     \draw (0,0) circle [radius=0.101cm];

    \node[scale=0.8,xshift=-10,yshift=5] at (p.corner 7) {$2_{\text{\tiny{L}}}$};
    \node[scale=0.8,xshift=-10,yshift=-5] at (p.corner 7) {$1_{\text{\tiny{R}}}$};    
    \node[scale=0.8,xshift=7,yshift=7] at (p.corner 19) {$j$};
   % \node[scale=0.8,xshift=7,yshift=7] at (p.corner 20) {$j_2$};

    \foreach \i in {7}
    {\draw[fill] (p.corner \i) circle [radius=0.1pt];}
    \foreach \i in {19}
    {\draw[fill] (p.corner \i) circle [radius=0.1pt];}
    %\draw[fill] (p.corner 10) circle [radius=0.1pt];

     %\fill[Blue,opacity=0.2]  (p.corner 8) -- (p.corner 7) -- (p.corner 6) -- (p.corner 5) -- (p.corner 4) -- (p.corner 3) -- (p.corner 2) -- (p.corner 1) -- (p.corner 22)-- cycle;

     \fill[Gray,opacity=0.2]   (p.corner 10) -- (p.corner 11) -- (p.corner 12) -- (p.corner 13) -- (p.corner 14) -- (p.corner 15) -- (p.corner 16) -- (p.corner 17) -- (p.corner 18)-- (p.corner 19)-- (p.corner 20) -- (p.corner 21) -- (p.corner 22) -- (p.corner 1) -- (p.corner 2) -- (p.corner 3) -- (p.corner 4) -- (p.corner 5) -- (p.corner 6) -- (p.corner 7) -- (p.corner 8) -- (p.corner 9)-- cycle;

     %\node[scale=0.8,xshift=18,yshift=30] at (p.corner 8) {$\mathcal{S}_L$};
     \node[scale=0.8,xshift=25,yshift=15] at (p.corner 10) {$\Psi_R$};
     \node[scale=0.8,xshift=20,yshift=48] at (p.corner 10) {$\Psi_L$};
      \draw[Maroon] (p.corner 7) -- (p.corner 19);
      \draw[Blue!80!White,dashed,line width=2] (p.corner 7) -- (p.corner 19);
      %\draw[Maroon] (p.corner 10) -- (p.corner 19);
    %\fill[Maroon] (p.corner 10) circle [radius=0.15pt];
     \fill[Maroon] (p.corner 7) circle [radius=0.15pt];
     %\fill[Black] (p.corner 19) circle [radius=0.15pt];

    \end{tikzpicture}     
\end{gathered} \right]
\label{eq:softLim2}
\end{equation}
which is precisely the analogue of what we found for the $j=3,n$ terms! In the soft limit, the $j$-th contribution gives the part of $\Psi_{n-1} (2,3,\cdots, n)$ containing propagator $(2,j)$; where in the first term this propagator is unshifted, and in the second it is shifted with $+\delta^2$ if $j$ is even, and $-\delta^2$ if $j$ is odd. Again subscripts on $1$ and $2$, reflect the fact that the red vertex has even parity for $\Psi_L$ but odd for $\Psi_R$, so that the $\delta$-shift in the lower-point wavefunctions matches that of the original  $\Psi_n$. To be explicit the integral associated with the first term in the r.h.s. of \eqref{eq:softLim2} is 
\begin{equation}
    \begin{gathered}
    \begin{tikzpicture}[line width=1,scale=10]
    \node[regular polygon, regular polygon sides=22,  minimum size=2cm] (p) at (0,0) {};
     \draw (0,0) circle [radius=0.101cm];

    \node[scale=0.8,xshift=-10,yshift=5] at (p.corner 7) {$2_{\text{\tiny{L}}}$};
    \node[scale=0.8,xshift=-10,yshift=-5] at (p.corner 7) {$1_{\text{\tiny{R}}}$};    
    \node[scale=0.8,xshift=7,yshift=7] at (p.corner 19) {$j$};
   % \node[scale=0.8,xshift=7,yshift=7] at (p.corner 20) {$j_2$};

    \foreach \i in {7}
    {\draw[fill] (p.corner \i) circle [radius=0.1pt];}
    \foreach \i in {19}
    {\draw[fill] (p.corner \i) circle [radius=0.1pt];}
    %\draw[fill] (p.corner 10) circle [radius=0.1pt];

     %\fill[Blue,opacity=0.2]  (p.corner 8) -- (p.corner 7) -- (p.corner 6) -- (p.corner 5) -- (p.corner 4) -- (p.corner 3) -- (p.corner 2) -- (p.corner 1) -- (p.corner 22)-- cycle;

     \fill[Gray,opacity=0.2]   (p.corner 10) -- (p.corner 11) -- (p.corner 12) -- (p.corner 13) -- (p.corner 14) -- (p.corner 15) -- (p.corner 16) -- (p.corner 17) -- (p.corner 18)-- (p.corner 19)-- (p.corner 20) -- (p.corner 21) -- (p.corner 22) -- (p.corner 1) -- (p.corner 2) -- (p.corner 3) -- (p.corner 4) -- (p.corner 5) -- (p.corner 6) -- (p.corner 7) -- (p.corner 8) -- (p.corner 9)-- cycle;

     %\node[scale=0.8,xshift=18,yshift=30] at (p.corner 8) {$\mathcal{S}_L$};
     \node[scale=0.8,xshift=25,yshift=28] at (p.corner 10) {$\omega$};
     \node[scale=0.8,xshift=25,yshift=10] at (p.corner 10) {$\Psi_R$};
     \node[scale=0.8,xshift=20,yshift=48] at (p.corner 10) {$\Psi_L$};
      \draw[Maroon] (p.corner 7) -- (p.corner 19);
      %\draw[Maroon] (p.corner 10) -- (p.corner 19);
    %\fill[Maroon] (p.corner 10) circle [radius=0.15pt];
     \fill[Maroon] (p.corner 7) circle [radius=0.15pt];
     %\fill[Black] (p.corner 19) circle [radius=0.15pt];

    \end{tikzpicture}     
\end{gathered} \equiv \int \frac{d \omega}{2\pi i}  \Psi_L(2^{\text{even}},3,\cdots,\underline{j}) \times \frac{1}{2(\omega^2 - k_{2,j}^2)} \times \Psi_R(\underline{2}^{\text{odd}},j,\cdots,n),
\end{equation}
where  the underlines denote the internal state; and the superscripts  refer to the parity of the collapsed vertex in the respective lower-point wavefunctions. The second term on the r.h.s. of \eqref{eq:softLim2} is given by exactly the same integral but where now $k_{2,j}^2 \to k_{2,j}^2 \pm \delta^2$, depending on whether $j$ is even or odd.

Putting these together, we have that under the soft limit, $k_1 \to 0$, the $n$-point wavefunction turns into the sum of $\delta$-shifted lower-point wavefunctions coming from each term in \eqref{eq:BerendsGiele_Rec}: 
\begin{equation}
    \Psi_n \quad \xrightarrow{k_1 \to 0} \quad \frac{i}{\delta^2} \sum_{j=3}^n (-1)^{j+1} \left[ \begin{gathered}
    \begin{tikzpicture}[line width=1,scale=10]
    \node[regular polygon, regular polygon sides=22,  minimum size=2cm] (p) at (0,0) {};
     \draw (0,0) circle [radius=0.101cm];

    \node[scale=0.8,xshift=-10,yshift=5] at (p.corner 7) {$2_{\text{\tiny{L}}}$};
    \node[scale=0.8,xshift=-10,yshift=-5] at (p.corner 7) {$1_{\text{\tiny{R}}}$};    
    \node[scale=0.8,xshift=7,yshift=7] at (p.corner 19) {$j$};
   % \node[scale=0.8,xshift=7,yshift=7] at (p.corner 20) {$j_2$};

    \foreach \i in {7}
    {\draw[fill] (p.corner \i) circle [radius=0.1pt];}
    \foreach \i in {19}
    {\draw[fill] (p.corner \i) circle [radius=0.1pt];}
    %\draw[fill] (p.corner 10) circle [radius=0.1pt];

     %\fill[Blue,opacity=0.2]  (p.corner 8) -- (p.corner 7) -- (p.corner 6) -- (p.corner 5) -- (p.corner 4) -- (p.corner 3) -- (p.corner 2) -- (p.corner 1) -- (p.corner 22)-- cycle;

     \fill[Gray,opacity=0.2]   (p.corner 10) -- (p.corner 11) -- (p.corner 12) -- (p.corner 13) -- (p.corner 14) -- (p.corner 15) -- (p.corner 16) -- (p.corner 17) -- (p.corner 18)-- (p.corner 19)-- (p.corner 20) -- (p.corner 21) -- (p.corner 22) -- (p.corner 1) -- (p.corner 2) -- (p.corner 3) -- (p.corner 4) -- (p.corner 5) -- (p.corner 6) -- (p.corner 7) -- (p.corner 8) -- (p.corner 9)-- cycle;

     %\node[scale=0.8,xshift=18,yshift=30] at (p.corner 8) {$\mathcal{S}_L$};
     \node[scale=0.8,xshift=25,yshift=15] at (p.corner 10) {$\Psi_R$};
     \node[scale=0.8,xshift=20,yshift=48] at (p.corner 10) {$\Psi_L$};
      \draw[Maroon] (p.corner 7) -- (p.corner 19);
      %\draw[Maroon] (p.corner 10) -- (p.corner 19);
    %\fill[Maroon] (p.corner 10) circle [radius=0.15pt];
     \fill[Maroon] (p.corner 7) circle [radius=0.15pt];
     %\fill[Black] (p.corner 19) circle [radius=0.15pt];

    \end{tikzpicture}     
\end{gathered} \quad - \quad \begin{gathered}
    \begin{tikzpicture}[line width=1,scale=10]
    \node[regular polygon, regular polygon sides=22,  minimum size=2cm] (p) at (0,0) {};
     \draw (0,0) circle [radius=0.101cm];

    \node[scale=0.8,xshift=-10,yshift=5] at (p.corner 7) {$2_{\text{\tiny{L}}}$};
    \node[scale=0.8,xshift=-10,yshift=-5] at (p.corner 7) {$1_{\text{\tiny{R}}}$};    
    \node[scale=0.8,xshift=7,yshift=7] at (p.corner 19) {$j$};
   % \node[scale=0.8,xshift=7,yshift=7] at (p.corner 20) {$j_2$};

    \foreach \i in {7}
    {\draw[fill] (p.corner \i) circle [radius=0.1pt];}
    \foreach \i in {19}
    {\draw[fill] (p.corner \i) circle [radius=0.1pt];}
    %\draw[fill] (p.corner 10) circle [radius=0.1pt];

     %\fill[Blue,opacity=0.2]  (p.corner 8) -- (p.corner 7) -- (p.corner 6) -- (p.corner 5) -- (p.corner 4) -- (p.corner 3) -- (p.corner 2) -- (p.corner 1) -- (p.corner 22)-- cycle;

     \fill[Gray,opacity=0.2]   (p.corner 10) -- (p.corner 11) -- (p.corner 12) -- (p.corner 13) -- (p.corner 14) -- (p.corner 15) -- (p.corner 16) -- (p.corner 17) -- (p.corner 18)-- (p.corner 19)-- (p.corner 20) -- (p.corner 21) -- (p.corner 22) -- (p.corner 1) -- (p.corner 2) -- (p.corner 3) -- (p.corner 4) -- (p.corner 5) -- (p.corner 6) -- (p.corner 7) -- (p.corner 8) -- (p.corner 9)-- cycle;

     %\node[scale=0.8,xshift=18,yshift=30] at (p.corner 8) {$\mathcal{S}_L$};
     \node[scale=0.8,xshift=25,yshift=15] at (p.corner 10) {$\Psi_R$};
     \node[scale=0.8,xshift=20,yshift=48] at (p.corner 10) {$\Psi_L$};
      \draw[Maroon] (p.corner 7) -- (p.corner 19);
      \draw[Blue!80!White,dashed,line width=2] (p.corner 7) -- (p.corner 19);
      %\draw[Maroon] (p.corner 10) -- (p.corner 19);
    %\fill[Maroon] (p.corner 10) circle [radius=0.15pt];
     \fill[Maroon] (p.corner 7) circle [radius=0.15pt];
     %\fill[Black] (p.corner 19) circle [radius=0.15pt];

    \end{tikzpicture}     
\end{gathered} \right] \, . 
\label{eq:FullSoftLimit}
\end{equation}

Now, quite remarkably, most terms in this sum actually \textit{cancel} each other! So we only get contributions from the limiting cases $j=3,n$, where $k_2$ and $k_n$ are shifted, respectively -- this is shifted terms in \eqref{eq:Softj3}-\eqref{eq:Softjn}. In the total energy pole, the wavefunction turns into the amplitude and so the result must vanish in the soft limit. Indeed out of all the terms in \eqref{eq:FullSoftLimit}, the only ones that don't have the total energy pole are precisely those where one of the external momenta is $\delta$-shifted. So indeed we knew in the total energy pole all the remaining terms should cancel each other, but it's remarkable that this still holds \textit{away} from the total energy pole. 

To prove the cancellation of all the intermediate terms in the sum, we group all triangulations of the $(n-1)$-gon into subsets listed by the collection of chords touching the red vertex. For example, any triangulation with no chords ending on the red vertex, must have chord $(3,n)$ -- and therefore come from the $j=\{3,n\}$ terms in the sum, but these terms come precisely with opposite sign so they cancel out:
\begin{equation}
     \frac{i}{\delta^2} \times \begin{gathered}
    \begin{tikzpicture}[line width=1,scale=10]
    \node[regular polygon, regular polygon sides=22,  minimum size=2cm] (p) at (0,0) {};
     \draw (0,0) circle [radius=0.101cm];
   % \draw[Blue!80!White,dashed,line width=2]
  %({90-180/22+3*360/22}:0.101cm)
  %arc[
  %  start angle={90-180/22+3*360/22},
  %  end angle={90-180/22+6*360/22},
   % radius=0.101cm
  %];

    \node[scale=0.8,xshift=-9,yshift=6] at (p.corner 7) {$2$};
    \node[scale=0.8,xshift=-7,yshift=-5] at (p.corner 7) {$1_{\text{\tiny{R}}}$};   
    \node[scale=0.8,xshift=-6,yshift=-5] at (p.corner 10) {$n$};
    \node[scale=0.8,xshift=-7,yshift=7] at (p.corner 4) {$3$};
   % \node[scale=0.8,xshift=7,yshift=7] at (p.corner 20) {$j_2$};
\draw[Maroon]
  ({90-180/22+3*360/22}:0.101cm)
  arc[
    start angle={90-180/22+3*360/22},
    end angle={90-180/22+6*360/22},
    radius=0.101cm
  ];

    \foreach \i in {7}
    {\draw[Maroon,fill] (p.corner \i) circle [radius=0.1pt];}
    \foreach \i in {4}
    {\draw[fill] (p.corner \i) circle [radius=0.1pt];}
    \draw[fill] (p.corner 10) circle [radius=0.1pt];

     \fill[Gray,opacity=0.2]   (p.corner 10) -- (p.corner 11) -- (p.corner 12) -- (p.corner 13) -- (p.corner 14) -- (p.corner 15) -- (p.corner 16) -- (p.corner 17) -- (p.corner 18)-- (p.corner 19)-- (p.corner 20) -- (p.corner 21) -- (p.corner 22) -- (p.corner 1) -- (p.corner 2) -- (p.corner 3) -- (p.corner 4) -- cycle;

     %\node[scale=0.8,xshift=18,yshift=30] at (p.corner 8) {$\mathcal{S}_L$};
     \node[scale=0.8,xshift=25,yshift=30] at (p.corner 10) {$\Psi_{n-2}$};
      \draw[Gray] (p.corner 4) -- (p.corner 10);

    \fill[Black] (p.corner 10) circle [radius=0.15pt];
     \fill[Black] (p.corner 4) circle [radius=0.15pt];

    \end{tikzpicture}     
\end{gathered}\quad - \quad \frac{i}{\delta^2} \times \begin{gathered}
    \begin{tikzpicture}[line width=1,scale=10]
    \node[regular polygon, regular polygon sides=22,  minimum size=2cm] (p) at (0,0) {};
     \draw (0,0) circle [radius=0.101cm];
   % \draw[Blue!80!White,dashed,line width=2]
  %({90-180/22+3*360/22}:0.101cm)
  %arc[
  %  start angle={90-180/22+3*360/22},
  %  end angle={90-180/22+6*360/22},
   % radius=0.101cm
  %];

    \node[scale=0.8,xshift=-7,yshift=6] at (p.corner 7) {$2_{\text{\tiny{L}}}$};
    \node[scale=0.8,xshift=-8,yshift=-5] at (p.corner 7) {$1$};   
    \node[scale=0.8,xshift=-6,yshift=-5] at (p.corner 10) {$n$};
    \node[scale=0.8,xshift=-7,yshift=7] at (p.corner 4) {$3$};
   % \node[scale=0.8,xshift=7,yshift=7] at (p.corner 20) {$j_2$};
\draw[Maroon]
  (180:0.101cm)
  arc[
    start angle=180,
    end angle=229.09,
    radius=0.101cm
  ];
    
    \foreach \i in {7}
    {\draw[Maroon,fill] (p.corner \i) circle [radius=0.1pt];}
    \foreach \i in {4}
    {\draw[fill] (p.corner \i) circle [radius=0.1pt];}
    \draw[fill] (p.corner 10) circle [radius=0.1pt];

     \fill[Gray,opacity=0.2]   (p.corner 10) -- (p.corner 11) -- (p.corner 12) -- (p.corner 13) -- (p.corner 14) -- (p.corner 15) -- (p.corner 16) -- (p.corner 17) -- (p.corner 18)-- (p.corner 19)-- (p.corner 20) -- (p.corner 21) -- (p.corner 22) -- (p.corner 1) -- (p.corner 2) -- (p.corner 3) -- (p.corner 4) -- cycle;

     %\node[scale=0.8,xshift=18,yshift=30] at (p.corner 8) {$\mathcal{S}_L$};
     \node[scale=0.8,xshift=25,yshift=30] at (p.corner 10) {$\Psi_{n-2}$};
    \draw[Gray] (p.corner 4) -- (p.corner 10);

    \fill[Black] (p.corner 10) circle [radius=0.15pt];
     \fill[Black] (p.corner 4) circle [radius=0.15pt];

    \end{tikzpicture}     
\end{gathered} \quad \equiv 0 \, !
\end{equation}

Now let's say we consider triangulations where there are $l$ chords touching the red vertex, which end on vertices $\{v_1,v_2,\cdots,v_l\}$ -- which will now receive contributions from multiple terms in \eqref{eq:FullSoftLimit}. Starting from $j=3$ term we get a contribution where the chords are shifted assuming the parity of the red vertex is odd. Now let's say $v_1$ is even, so it will not be shifted in the $j=3$ contribution; but we also get a contribution with precisely the same shift assignment coming from the unshifted term in $j=v_1$, but these precisely cancel:
\begin{equation} 
\frac{i}{\delta^2}\times 
    \begin{gathered}
    \begin{tikzpicture}[line width=1,scale=10]
    \node[regular polygon, regular polygon sides=22,  minimum size=2cm] (p) at (0,0) {};
     \draw (0,0) circle [radius=0.101cm];
   % \draw[Blue!80!White,dashed,line width=2]
  %({90-180/22+3*360/22}:0.101cm)
  %arc[
  %  start angle={90-180/22+3*360/22},
  %  end angle={90-180/22+6*360/22},
   % radius=0.101cm
  %];

    \node[scale=0.8,xshift=-8,yshift=6] at (p.corner 7) {$2$};
    \node[scale=0.8,xshift=-8,yshift=-5] at (p.corner 7) {$1_{\text{\tiny{R}}}$};   
    \node[scale=0.8,xshift=-6,yshift=-5] at (p.corner 10) {$n$};
    \node[scale=0.8,xshift=-7,yshift=7] at (p.corner 4) {$3$};
   % \node[scale=0.8,xshift=7,yshift=7] at (p.corner 20) {$j_2$};
\draw[Maroon]
  ({90-180/22+3*360/22}:0.101cm)
  arc[
    start angle={90-180/22+3*360/22},
    end angle={90-180/22+6*360/22},
    radius=0.101cm
  ];
    
    \foreach \i in {7}
    {\draw[Maroon,fill] (p.corner \i) circle [radius=0.1pt];}
    \foreach \i in {4}
    {\draw[fill] (p.corner \i) circle [radius=0.1pt];}
    \draw[fill] (p.corner 10) circle [radius=0.1pt];

     \fill[Gray,opacity=0.2]   (p.corner 10) -- (p.corner 11) -- (p.corner 12) -- (p.corner 13) -- (p.corner 14) -- (p.corner 15) -- (p.corner 16) -- (p.corner 17) -- (p.corner 18)-- (p.corner 19)-- (p.corner 20) -- (p.corner 21) -- (p.corner 22) -- (p.corner 1) -- (p.corner 2) -- (p.corner 3) -- (p.corner 4) -- (p.corner 5) -- (p.corner 6) -- (p.corner 7) -- (p.corner 8) -- (p.corner 9)-- cycle;

     %\node[scale=0.8,xshift=18,yshift=30] at (p.corner 8) {$\mathcal{S}_L$};
    \draw[Gray] (p.corner 7) -- (p.corner 1);
    \node[scale=0.8,xshift=5,yshift=5] at (p.corner 1) {$v_1$};
    \draw[Gray] (p.corner 7) -- (p.corner 20);
    \node[scale=0.8,xshift=5,yshift=5] at (p.corner 20) {$v_2$};
    \draw[Gray] (p.corner 7) -- (p.corner 16);
    \draw[Gray] (p.corner 7) -- (p.corner 14);
    \node[scale=0.8,xshift=5,yshift=-5] at (p.corner 14) {$v_l$};

    \fill[Black] (p.corner 10) circle [radius=0.15pt];
     \fill[Black] (p.corner 4) circle [radius=0.15pt];
     \fill[Maroon] (p.corner 7) circle [radius=0.15pt];

    \end{tikzpicture}     
\end{gathered} + \frac{(-1)^{v_1+1}i}{\delta^2} \times 
    \begin{gathered}
    \begin{tikzpicture}[line width=1,scale=10]
    \node[regular polygon, regular polygon sides=22,  minimum size=2cm] (p) at (0,0) {};
     \draw (0,0) circle [radius=0.101cm];
   % \draw[Blue!80!White,dashed,line width=2]
  %({90-180/22+3*360/22}:0.101cm)
  %arc[
  %  start angle={90-180/22+3*360/22},
  %  end angle={90-180/22+6*360/22},
   % radius=0.101cm
  %];

    \node[scale=0.8,xshift=-8,yshift=6] at (p.corner 7) {$2_{\text{\tiny{L}}}$};
    \node[scale=0.8,xshift=-8,yshift=-5] at (p.corner 7) {$1_{\text{\tiny{R}}}$};   
    \node[scale=0.8,xshift=-6,yshift=-5] at (p.corner 10) {$n$};
    \node[scale=0.8,xshift=-7,yshift=7] at (p.corner 4) {$3$};
   % \node[scale=0.8,xshift=7,yshift=7] at (p.corner 20) {$j_2$};
    
    \foreach \i in {7}
    {\draw[Maroon,fill] (p.corner \i) circle [radius=0.1pt];}
    \foreach \i in {4}
    {\draw[fill] (p.corner \i) circle [radius=0.1pt];}
    \draw[fill] (p.corner 10) circle [radius=0.1pt];

     \fill[Gray,opacity=0.2]   (p.corner 10) -- (p.corner 11) -- (p.corner 12) -- (p.corner 13) -- (p.corner 14) -- (p.corner 15) -- (p.corner 16) -- (p.corner 17) -- (p.corner 18)-- (p.corner 19)-- (p.corner 20) -- (p.corner 21) -- (p.corner 22) -- (p.corner 1) -- (p.corner 2) -- (p.corner 3) -- (p.corner 4) -- (p.corner 5) -- (p.corner 6) -- (p.corner 7) -- (p.corner 8) -- (p.corner 9)-- cycle;

     %\node[scale=0.8,xshift=18,yshift=30] at (p.corner 8) {$\mathcal{S}_L$};
    \draw[Maroon] (p.corner 7) -- (p.corner 1);
    \node[scale=0.8,xshift=5,yshift=5] at (p.corner 1) {$v_1$};
    \draw[Gray] (p.corner 7) -- (p.corner 20);
    \node[scale=0.8,xshift=5,yshift=5] at (p.corner 20) {$v_2$};
    \draw[Gray] (p.corner 7) -- (p.corner 16);
    \draw[Gray] (p.corner 7) -- (p.corner 14);
    \node[scale=0.8,xshift=5,yshift=-5] at (p.corner 14) {$v_l$};

    \fill[Black] (p.corner 10) circle [radius=0.15pt];
     \fill[Black] (p.corner 4) circle [radius=0.15pt];
     \fill[Maroon] (p.corner 7) circle [radius=0.15pt];

    \end{tikzpicture}     
\end{gathered} \equiv 0 \, .
\end{equation}

Note that here for the shifted chords to agree it is important that there are no chords connecting the red vertex to a vertex between $[3,v_1]$, as it would be shifted differently in both terms -- of course this is guaranteed by the fact that we are looking at triangulations with \textit{only} chords $\{v_1,v_2,\cdots,v_l\}$. Now the other piece coming from $j=v_1$, where $(2,v_1)$ is shifted with $+\delta^2$, will cancel with the contribution from $j=v_2$ where $(2,v_2)$ unshifted/shifted depending on whether $v_2$ is odd/even, respectively. This cancellation continues telescopically until eventually the $j=v_l$ cancels with the unshifted part of $j=n$. 

Putting these together, we have that under the soft limit, $k_1 \to 0$, the $n$-point wavefunction turns into the sum of $\delta$-shifted lower-point wavefunctions coming from each term in \eqref{eq:BerendsGiele_Rec}: 
\begin{equation}
\begin{aligned}
    \Psi_n  \to & -\frac{i}{\delta^2} \left[ \Psi^{\text{NLSM}\oplus \phi^3}_{n-1}(2^{\phi},3^{\phi},4^\pi,\cdots,(n-1)^\pi,n^\phi) \vert_{\sqrt{k_2^2-\delta^2}} \right. \\
    & \left. \quad \quad \quad \quad \quad \quad \quad -  \Psi^{\text{NLSM}\oplus \phi^3}_{n-1}(2^{\phi},3^{\pi},\cdots,(n-2)^\pi,(n-1)^\phi,n^\phi) \vert_{\sqrt{k_n^2+\delta^2}} \right] \\
    &\equiv  - \frac{i}{\delta^2} \times   
    \begin{gathered}
    \begin{tikzpicture}[line width=1,scale=10]
    \node[regular polygon, regular polygon sides=22,  minimum size=2cm] (p) at (0,0) {};
     \draw (0,0) circle [radius=0.101cm];
    \draw[Blue!80!White,dashed,line width=2]
  ({90-180/22+3*360/22}:0.101cm)
  arc[
    start angle={90-180/22+3*360/22},
    end angle={90-180/22+6*360/22},
    radius=0.101cm
  ];
     
    \node[scale=0.8,xshift=-9,yshift=6] at (p.corner 7) {$2$};
    \node[scale=0.8,xshift=-7,yshift=-5] at (p.corner 7) {$1_{\text{\tiny{R}}}$};   
    \node[scale=0.8,xshift=-6,yshift=-5] at (p.corner 10) {$n$};
    \node[scale=0.8,xshift=-7,yshift=7] at (p.corner 4) {$3$};

    \foreach \i in {7}
    {\draw[Maroon,fill] (p.corner \i) circle [radius=0.1pt];}
    \foreach \i in {4}
    {\draw[fill] (p.corner \i) circle [radius=0.1pt];}
    \draw[fill] (p.corner 10) circle [radius=0.1pt];

     \fill[Gray,opacity=0.2]   (p.corner 10) -- (p.corner 11) -- (p.corner 12) -- (p.corner 13) -- (p.corner 14) -- (p.corner 15) -- (p.corner 16) -- (p.corner 17) -- (p.corner 18)-- (p.corner 19)-- (p.corner 20) -- (p.corner 21) -- (p.corner 22) -- (p.corner 1) -- (p.corner 2) -- (p.corner 3) -- (p.corner 4) -- (p.corner 5) -- (p.corner 6) -- (p.corner 7) -- (p.corner 8) -- (p.corner 9)-- cycle;

     %\node[scale=0.8,xshift=18,yshift=30] at (p.corner 8) {$\mathcal{S}_L$};
     \node[scale=0.8,xshift=25,yshift=30] at (p.corner 10) {$\Psi^R_{n-1}$};
      %\draw[Gray] (p.corner 4) -- (p.corner 10);

    \fill[Black] (p.corner 10) circle [radius=0.15pt];
     \fill[Black] (p.corner 4) circle [radius=0.15pt];

    \end{tikzpicture}    
\end{gathered}
\quad   +    \quad \frac{i}{\delta^2} \begin{gathered}
    \begin{tikzpicture}[line width=1,scale=10]
    \node[regular polygon, regular polygon sides=22,  minimum size=2cm] (p) at (0,0) {};
     \draw (0,0) circle [radius=0.101cm];

  \draw[Blue!80!White,dashed,line width=2]
  (180:0.101cm)
  arc[
    start angle=180,
    end angle=229.09,
    radius=0.101cm
  ];

    \node[scale=0.8,xshift=-7,yshift=6] at (p.corner 7) {$2_{\text{\tiny{L}}}$};
    \node[scale=0.8,xshift=-8,yshift=-5] at (p.corner 7) {$1$};   
    \node[scale=0.8,xshift=-6,yshift=-5] at (p.corner 10) {$n$};
    \node[scale=0.8,xshift=-7,yshift=7] at (p.corner 4) {$3$};
   % \node[scale=0.8,xshift=7,yshift=7] at (p.corner 20) {$j_2$};

    \foreach \i in {7}
    {\draw[Maroon,fill] (p.corner \i) circle [radius=0.1pt];}
    \foreach \i in {4}
    {\draw[fill] (p.corner \i) circle [radius=0.1pt];}
    \draw[fill] (p.corner 10) circle [radius=0.1pt];

     \fill[Gray,opacity=0.2]   (p.corner 10) -- (p.corner 11) -- (p.corner 12) -- (p.corner 13) -- (p.corner 14) -- (p.corner 15) -- (p.corner 16) -- (p.corner 17) -- (p.corner 18)-- (p.corner 19)-- (p.corner 20) -- (p.corner 21) -- (p.corner 22) -- (p.corner 1) -- (p.corner 2) -- (p.corner 3) -- (p.corner 4) -- (p.corner 5) -- (p.corner 6) -- (p.corner 7) -- (p.corner 8) -- (p.corner 9)-- cycle;

     %\node[scale=0.8,xshift=18,yshift=30] at (p.corner 8) {$\mathcal{S}_L$};
     \node[scale=0.8,xshift=25,yshift=30] at (p.corner 10) {$\Psi^L_{n-1}$};
      %\draw[Gray] (p.corner 4) -- (p.corner 10);

    \fill[Black] (p.corner 10) circle [radius=0.15pt];
     \fill[Black] (p.corner 4) circle [radius=0.15pt];

    \end{tikzpicture}  \, ,    
\end{gathered} 
\end{aligned}
\label{eq:FullSoftLimit_final}
\end{equation}
which is precisely the statement in \eqref{eq:SoftStatement}.

Our derivation of eq.~\eqref{eq:FullSoftLimit_final} made no assumption on the order of $1/\delta$ we were extracting, therefore it obviously holds for all orders and not only the $1/\delta^{2n-4}$ coefficient. One can then wonder whether it holds for the full square root shift, $k_{ij}\to\sqrt{k_{ij}^2\pm\delta^2}$ -- from which we can obtain the wavefunction without going to the frequency representation, as we did in section \ref{sec:SqrtShift_FS}. However, it is not obvious it should simply because when we expand the frequency integrands we discard the $\omega\sim \pm i\delta$ residues. Quite interestingly, by checking explicitly, we do find that at least up to eight-points the standard square-root shift also obeys eq.~\eqref{eq:FullSoftLimit_final}, even at finite $\delta$. In particular, this means that eq.~\eqref{eq:FullSoftLimit_final} is independent of the boundary field redefinitions we make when subtracting the $\omega\sim \pm i\delta$ poles, which is surprising as we expect soft theorems to depend on boundary field re-definitions.

\subsection{Simple recursion for pion wavefunction}
\label{sec:RecursionFlat}

To end this section, we now show how the frequency representation lets us write a simple recursion for the pion wavefunction. 
At $n$ points, where $n$ is even, take all possible collections, $\mathcal{C}$, of compatible chords, $k_{i,j}$, where $(i,j)$ have different parity. Then the $n$-point pion wavefunction can be written as 
\begin{equation}
    \Psi_n = \sum_{\mathcal{C}} \int_{-\infty}^{+\infty} \prod_{(i,j) \in \mathcal{C}} \frac{d \omega_{i,j}}{(2\pi i) [2(w_{i,j}^2-k_{i,j}^2)]} \ \prod_{P_{e} \subset \mathcal{C}} \hat{\Psi}_{P_e} + \hat{\Psi}_{n}
    \label{eq:RecursionPion1}
\end{equation}
where the second product is over the even subpolygons, $P_e$, that we obtain by dividing the original $n$-gon into the subpolygons specified by the collection of chords in $\mathcal{C}$. $\hat{\Psi}_{P_e}$ corresponds to the ``pure contact'' part of the NLSM wavefunction associated with the subpolygon $P_e$ with a number of off-shell legs corresponding to the edges of $P_e$ associated with chords $k_{i,j}$ in the collection:
\begin{equation}
    \begin{gathered}
    \begin{tikzpicture}[line width=1,scale=10]
    \node[regular polygon, regular polygon sides=22,  minimum size=2cm] (p) at (0,0) {};
     \draw (0,0) circle [radius=0.101cm];

     \fill[Gray,opacity=0.2]   (p.corner 10) -- (p.corner 11) -- (p.corner 12) -- (p.corner 13) -- (p.corner 14) -- (p.corner 15) -- (p.corner 16) -- (p.corner 17) -- (p.corner 18)-- (p.corner 19)-- (p.corner 20) -- (p.corner 21) -- (p.corner 22) -- (p.corner 1) -- (p.corner 2) -- (p.corner 3) -- (p.corner 4) -- (p.corner 5) -- (p.corner 6) -- (p.corner 7) -- (p.corner 8) -- (p.corner 9)-- cycle;
     \node[scale=0.8,xshift=33,yshift=43] at (p.corner 10) {$\hat{\Psi}_{n_2}$};
     \node[scale=0.8,xshift=40,yshift=10] at (p.corner 10) {$\hat{\Psi}_{n_3}$};
     \node[scale=0.8,xshift=3,yshift=38] at (p.corner 10) {$\hat{\Psi}_{n_1}$};
      \draw[Maroon] (p.corner 1) -- (p.corner 10);

      \draw[Maroon] (p.corner 20) -- (p.corner 11);
     \fill[Maroon] (p.corner 20) circle [radius=0.15pt];
     \fill[Maroon] (p.corner 11) circle [radius=0.15pt];
     \fill[Maroon] (p.corner 10) circle [radius=0.15pt];
     \fill[Maroon] (p.corner 1) circle [radius=0.15pt];
    \node[scale=0.8,xshift=-5,yshift=-5] at (p.corner 10) {$i_1$};     \node[scale=0.8,xshift=5,yshift=8] at (p.corner 1) {$j_1$};

     \node[scale=0.8,xshift=3,yshift=-8] at (p.corner 11) {$i_2$};     \node[scale=0.8,xshift=5,yshift=8] at (p.corner 20) {$j_2$};

    \end{tikzpicture}     
\end{gathered} \quad \begin{aligned}
    &\text{ with } \hat{\Psi}_{n_1}(k_{\text{ext}};\omega_{i_1,j_1}) \sim \frac{1}{\delta^{2n_1 -4}}, \\
    &\quad \text{[similar for $\hat{\Psi}_{n_2},\hat{\Psi}_{n_3}$]}
\end{aligned} \quad \text{s.t.}\quad  \sum_{i=1}^3 (2n_i-4) = (2n-4)
\label{eq:Even-gons}
\end{equation}
where with ``pure contact'' we refer the piece which in the underlying $\delta$-shifted theory comes from the contribution of pure odd-odd or even-even triangulations of $P_e$ -- this is where all chords are $\delta$-shifted. Similarly, $\hat{\Psi}_n$ is then simply the ``pure-contact'' piece for the full $n$-gon. 

So from this recursion, just like for the amplitude, in order to build the $n$-point pion wavefunction, we only need to know the contact ``vertices'' of all lower point wavefunctions as well as the $n$-point vertex, $\hat{\Psi}_n$.

As explained above, at each $n_i$ we can compute $\hat{\Psi}_{n_i}$ by looking only at the triangulations of the $n_i$-gon where all chords are shifted and extracting the piece with the correct units, $\delta^{-(2n_i-4)}$. Starting from this object, we can turn some of the bulk-to-boundary propagators for the external legs into bulk-to-bulk ones, as need to enter in \eqref{eq:RecursionPion1}, by applying a discontinuity, for example  
\begin{equation}
    \hat{\Psi}_{n_1}(k_{\text{ext}};\omega_{i_1,j_1}) \equiv \hat{\Psi}_{n_1}(k_{\text{ext}};k=\omega_{i,j}) - \hat{\Psi}_{n_1}(k_{\text{ext}};k=-\omega_{i,j})\, .
\end{equation}

Now to prove this recursion, all we need to do is reorganize the cubic diagrams entering in the $\delta$-shift picture into the different terms in \eqref{eq:RecursionPion1}. This is precisely the division introduced in \cite{circles}: given a triangulation of the $n$-gon, list all chord $(i,j)$ in the triangulation that have different parity; this will form one of the collections $\mathcal{C}$ in \eqref{eq:RecursionPion1}, and this diagram will precisely contribute to that term in the expansion. 

Now the only non-trivial aspect of the formula in \eqref{eq:RecursionPion1} is that given a diagram which on top of the chords in $\mathcal{C}$ also defines some triangulation of the respective $P_e$, when we do the $\delta$-expansion of this term, we only ask for the total power in $\delta$ to be $\delta^{-(2n-4)}$. As a result from each subpolygon, we might get powers of $1/\delta$ which differ from the $1/\delta^{2n_e-4}$ power in \eqref{eq:Even-gons}. 

We now show that all different powers which do not correspond to those in \eqref{eq:Even-gons}, precisely cancel out when we add up all the diagrams. This turns out to be very simple to see, but let's just look at the simplest example at $6$-points. Take the triangulation containing chords $T =\{(1,3),(1,4),(1,5)\}$ to which we would associate the frequency integral
\begin{equation}
\begin{aligned}
    \Psi \vert_T = \frac{1}{2^3} &\int_{\mathbb{R}} \prod_{j=3}^5\frac{ d \omega_{1,j}}{(2\pi i)} \frac{C^{(1)}(k_1,k_2,\omega_{13})C^{(2)}(k_3,\omega_{13},\omega_{14}) }{\omega_{1,3}^2-k_{1,3}^2+\delta^2}  \times \frac{1}{\omega_{1,4}^2-k_{1,4}^2} \\
    & \quad \quad \quad \quad \quad \times \frac{C^{(2)}(k_4,\omega_{14},\omega_{15}) C^{(1)}(k_5,k_6,\omega_{15})}{\omega_{15}^2-k_{1,5}^2+\delta^2} ,
\end{aligned}
\end{equation}
from which we want to extract the piece in $1/\delta^8$. As highlighted in \eqref{eq:Even-gons}, one piece comes from extracting the $1/\delta^4$ parts of the two $4$-point factors -- the first one involving $\omega_{13}$ and the other $\omega_{15}$. But clearly we will also have a contribution where we keep the $1/\delta^2$ part in $\omega_{13}$; and $1/\delta^6$ from the $\omega_{15}$ (and vice-versa). 

Now this piece cancels for a simple reason: since the $1/\delta^2$ piece corresponds to the term in the expansion which is independent of $k_{1,3}$, it cancels with the diagram corresponding to the other triangulation of the square $(1,2,3,4)$, where we instead have $\omega_{2,4}$, which is shifted instead with $+\delta^2$: 
\begin{equation}
    \begin{gathered}
    \begin{tikzpicture}[line width=1,scale=10]
    \node[regular polygon, regular polygon sides=6,  minimum size=2cm] (p) at (0,0) {};

     \fill[Gray,opacity=0.2]   (p.corner 1) -- (p.corner 2) -- (p.corner 3) -- (p.corner 4) -- (p.corner 5) -- (p.corner 6)-- cycle;
     
     \draw[Black,opacity=0.8]    (p.corner 1) -- (p.corner 2) -- (p.corner 3) -- (p.corner 4) -- (p.corner 5) -- (p.corner 6)-- cycle;
     
     \node[scale=0.7,xshift=14,yshift=-20] at (p.corner 2) {$\delta^{-2}$};
    
      \draw[Maroon] (p.corner 1) -- (p.corner 4);

       \draw[Blue] (p.corner 2) -- (p.corner 4);
       \draw[Blue] (p.corner 4) -- (p.corner 6);
       \node[scale=0.7,xshift=-20,yshift=0] at (p.corner 6) {$\delta^{-6}$};

     \fill[Maroon] (p.corner 4) circle [radius=0.15pt];
     \fill[Maroon] (p.corner 1) circle [radius=0.15pt];
    \node[scale=0.8,xshift=-5,yshift=-5] at (p.corner 4) {$1$};
    \node[scale=0.8,xshift=-7,yshift=0] at (p.corner 3) {$2$};
    \node[scale=0.8,xshift=-7,yshift=0] at (p.corner 2) {$3$};
    \node[scale=0.8,xshift=7,yshift=0] at (p.corner 1) {$4$};

    \node[scale=0.8,xshift=7,yshift=0] at (p.corner 6) {$5$};

    \node[scale=0.8,xshift=7,yshift=0] at (p.corner 5) {$6$};

    \end{tikzpicture}     
\end{gathered} \quad +\quad   \begin{gathered}
    \begin{tikzpicture}[line width=1,scale=10]
    \node[regular polygon, regular polygon sides=6,  minimum size=2cm] (p) at (0,0) {};

     \fill[Gray,opacity=0.2]   (p.corner 1) -- (p.corner 2) -- (p.corner 3) -- (p.corner 4) -- (p.corner 5) -- (p.corner 6)-- cycle;
     
     \draw[Black,opacity=0.8]    (p.corner 1) -- (p.corner 2) -- (p.corner 3) -- (p.corner 4) -- (p.corner 5) -- (p.corner 6)-- cycle;
     
     \node[scale=0.7,xshift=12,yshift=-28] at (p.corner 2) {$\delta^{-2}$};
    
      \draw[Maroon] (p.corner 1) -- (p.corner 4);

       \draw[Blue] (p.corner 3) -- (p.corner 1);
       \draw[Blue] (p.corner 4) -- (p.corner 6);
       \node[scale=0.7,xshift=-20,yshift=0] at (p.corner 6) {$\delta^{-6}$};

     \fill[Maroon] (p.corner 4) circle [radius=0.15pt];
     \fill[Maroon] (p.corner 1) circle [radius=0.15pt];
    \node[scale=0.8,xshift=-5,yshift=-5] at (p.corner 4) {$1$};
    \node[scale=0.8,xshift=-7,yshift=0] at (p.corner 3) {$2$};
    \node[scale=0.8,xshift=-7,yshift=0] at (p.corner 2) {$3$};
    \node[scale=0.8,xshift=7,yshift=0] at (p.corner 1) {$4$};

    \node[scale=0.8,xshift=7,yshift=0] at (p.corner 6) {$5$};

    \node[scale=0.8,xshift=7,yshift=0] at (p.corner 5) {$6$};

    \end{tikzpicture}     
\end{gathered} \quad =0 \, . 
\end{equation}

Now this cancellation mechanism generalizes trivially to higher points. The one key point is the following: from any given $n_e$-gon (triangulated with pure even/odd chords) the lowest power in $1/\delta$ we can get is where we collect the $1/\delta^2$ piece from all propagators in the triangulation of the $n_e$-gon. This gives a power of $(1/\delta^{2n_e-6})$, which is only $1/\delta^2$ away from $(1/\delta^{2n_e-4})$. Therefore, given a collection of even-gons tiling the $n$-gon, to move away from the powers of $1/\delta$ given in \eqref{eq:Even-gons}, we need to go to one lower order in $1/\delta^2$ to one of the even-gons and one higher in another. But now in the one where we went to a lower order, we precisely get $(1/\delta^{2n_e-6})$, which corresponds to the piece where the dependence from all the $k_{e,e}$/$k_{o,o}$ entering the particular triangulation drop -- and therefore once we add over all possible even/odd triangulations of the even-gon the result simply cancels out! Therefore, at the end of the day, the answer is simply given by the pieces of the $\delta$ expansion of each lower-even-gon which precisely reproduce the pion wavefunction of the respective even-gon, which is precisely the statement in \eqref{eq:RecursionPion1}.

\section{Pion wavefunction in de Sitter}\label{sec:PionWf_DS}

Now let us move to de Sitter space. Just like in flat-space, to compute the NLSM de Sitter wavefunction we need to perform a large mass expansion of the cubic scalar theory, corresponding to the $\delta \to \infty$ limit of the $\delta$-shift. However, in de Sitter, the mode functions are Hankel functions whose mass dependence lies on the Hankel parameter (see eq.~\eqref{eq:MassiveBulkBdy}), so it's unclear how the answer simplifies in that limit. What we show here is that, quite remarkably, precisely for the orders in $1/\delta$ needed to extract the pion wavefunction, the $\delta$-expansion of the massive modes in de Sitter can be reorganized into the frequency integrals almost identical to those we found in flat space. 

\subsection{\texorpdfstring{$\delta$}{delta}-shift, recursion and frequency representation in de Sitter}

The easiest way to proceed is by expanding the Green's equation for the bulk-to-bulk propagators, since only the internal modes become very massive under the $\delta$-shift. Recall that the Green's equation for a massive mode in de Sitter is:
\begin{equation}
   \left( \partial_{\eta}^2 - \frac{2}{\eta}\partial_{\eta} +\left[k^2+M(\eta)\right]\right)G_{BB}^{(M)}(\eta,\eta^{\prime};k) = -i H^2 \eta^2 \delta(\eta-\eta^{\prime})\, ,\quad \text{with } M(\eta)=\frac{\mu^2}{H^2\eta^2}\, .
\end{equation}
Then, formally the solution to this equation is the inverse of the quadratic operator:
\begin{equation}
\begin{aligned}
    G^{(M)}_{BB}(\eta,\eta^{\prime};k) &= \left(D^{(0)}_{k,\eta} + M(\eta)\right)^{-1}\, \left[ -i H^2 \eta^2 \delta(\eta-\eta^\prime) \right]\\
    &= \left(M(\eta)^{-1} D^{(0)}_{k,\eta} + 1\right)^{-1} \left[ \frac{-iH^2 \eta^2}{M(\eta^\prime)} \delta(\eta-\eta^\prime) \right] ,
\end{aligned}
\end{equation}
where we defined the massless quadratic operator $D^{(0)}_{k,\eta}=\partial_{\eta}^2 - \frac{2}{\eta}\partial_{\eta} +k^2$. Under the $\delta$-shift we have $\mu^2 = \pm \delta^2$, and we can thus expand around the large $\mu$ limit to obtain:
\begin{equation}
    G^{(M)}_{BB}(\eta,\eta^{\prime};k) \to \left(1 - M^{-1}(\eta) D^{(0)}_{k,\eta}  +\cdots \right) \left[ \frac{-iH^2 \eta^2}{M(\eta^\prime)}\delta(\eta-\eta^\prime) \right],
    \label{eq:GdS_exp}
\end{equation}

Of course this expansion is only valid in the regime $M(\eta) \gg 1$, but once we put it into the time integrals, within the integration region we have $\eta \to -\infty$ which competes with $\mu \to \infty$, thus making the expansion not well defined. However, imposing the Bunch-Davies vacuum in the past precisely requires the integral to vanish in this region, and so the contribution from this region is exponentially suppressed and does not contribute to the $1/\delta$ expansion\footnote{In \cite{DuasoPueyo:2025lmq}, it was shown that in the large mass limit any non-analytic dependence on the external kinematics is exponentially suppressed in $\mu$ for the Bunch-Davies state.}.
%at   The final subtlety is related to the expansion in eq.~\eqref{eq:GdS_exp}. At very early times, $\eta \to -\infty$, the series expansion does not seem to be well defined anymore. However, here we are not interested in re-summing the series, we are always interested in picking up a specific term which functionally depends on a specific power of $1/\mu$, and only one of the terms has the correct energy dependence of the NLSM wavefunction. 

Now the goal is to express the different terms in the expansion in terms of integrals we would find in the massless theory, just like we managed to do in the flat space case -- of course, there this was trivial since the $\delta$-shift only redefined the massless momenta. To do this, we recall that the massless propagator obeys: 
\begin{equation}
    D^{(0)}_{k,\eta} G_{BB}^{(0)}(\eta,\eta_\prime;k)  = -i H^2 \eta^2 \delta(\eta-\eta^\prime)
    \label{eq:FlatBB_prop1}
\end{equation}
and so we can write: 
\begin{equation}
    G^{(M)}_{BB}(\eta,\eta^{\prime};k) \to \left(1 - M^{-1}(\eta) D^{(0)}_{k,\eta}  +\cdots \right) \left[ \frac{D^{(0)}_{k,\eta} G_{BB}^{(0)}(\eta,\eta_\prime;k)}{M(\eta^\prime)}\right], 
    \label{eq:LeadSubLead}
\end{equation}
Finally, to mimic the flat-space result and trivialize the action of $D^{(0)}_{k,\eta}$, we represent the massless bulk-to-bulk propagator in frequency space via the contour integral: 
\begin{equation}
    G_{BB}^{(0)}(\eta,\eta^{\prime};k)= \frac{H^2}{2}\int_{-\infty}^{\infty}\frac{d\omega}{\omega^2 (2 \pi i)} \frac{(\phi^{\text{dS}}_{\mathbf{\omega}}(\eta)-\phi^{\text{dS}}_{\mathbf{-\omega}}(\eta))(\phi^{\text{dS}}_{\mathbf{\omega}}(\eta^{\prime})-\phi^{\text{dS}}_{\mathbf{-\omega}}(\eta^{\prime}))}{\omega^2-k^2 + i\epsilon} \, ,
    \label{eq:FreqRep_dS}
\end{equation}
where the $\omega$-integral is to be interpreted as a residue sum, once we properly close the contour in the UHP or LHP, depending on the sign of $\eta-\eta^\prime$. Note that as $\omega \to 0$, $\Delta \phi^{\text{dS}}_\omega (\eta) \sim \mathcal{O}(\omega^3)$ and therefore the integral is completely smooth in that region. Under the action of $D_{k,\eta}^{(0)}$ we find
\begin{equation}
    D_{k,\eta}^{(0)} \Delta \phi^{\text{dS}}_\omega (\eta) = (k^2 -\omega^2) \Delta \phi^{\text{dS}}_\omega (\eta) ,
\end{equation}
which leads to a trivial transformation for $G_{BB}^{(0)}(\eta,\eta^{\prime};k)$. 
Therefore for the leading and subleading terms written explicitly in  \eqref{eq:LeadSubLead} we get  
\begin{equation}
\begin{aligned}
    G_{BB}^{(M)} \xrightarrow{M\to \infty} &- \frac{H^4 \eta^2}{ 2 \mu^2} \int \frac{d \omega}{\omega^2 (2\pi i)}  \Delta \phi^{\text{dS}}_\omega (\eta) \Delta \phi^{\text{dS}}_\omega (\eta^\prime) \\
    &- \frac{H^6 \eta^2 \eta^{\prime \, 2}}{2 \mu^4} \int \frac{d \omega}{\omega^2 (2\pi i)}  (\omega^2-k^2)\Delta \phi^{\text{dS}}_\omega (\eta) \Delta \phi^{\text{dS}}_\omega (\eta^\prime) + \mathcal{O}(1/\mu^6) \, .
\end{aligned}
\label{eq:BB_expansion}
\end{equation}

Now to build the full pion wavefunction in de Sitter we use the recursion described in section \ref{sec:RecursionFlat} -- which states that we can reconstruct the full object by ``gluing'' in frequency space the effective contact vertices $\hat{\Psi}_n$. 

To show that once again we can restrict ourselves to the terms entering in \eqref{eq:RecursionPion1}, we need to argue that once we take an extra power in $1/\delta^2$ from  a given $P_e$, and one fewer from another $P_e^\prime$; the result cancels when we sum over all triangulations of $P_e^\prime$. Recall the reason in flat-space why this was true was simply because the leading order expansion for each propagator, $1/\delta^2$, was independent of the internal $k_{i,j}$ and therefore all pure even- and odd- triangulations of $P_e^\prime$ gave the same result up to a minus sign. Looking at the leading order term in the expansion of $G_{BB}^{(M)}$ in de Sitter, we see that precisely the same is true, and with the sign also depending on $\mu^2 = \pm \delta^2$! And thus, the flat-space recursion generalizes trivially to de Sitter. 

%%%%%%%%%%
As a result, the two terms in \eqref{eq:BB_expansion} are all we need to compute the NLSM de Sitter wavefunction. Since we want to collect the piece of order $1/\delta^{2n-4}$, this comes from expanding one of the propagators to subleading order, and all $(n-4)$ remaining to leading order in $1/\delta$. So schematically we have:
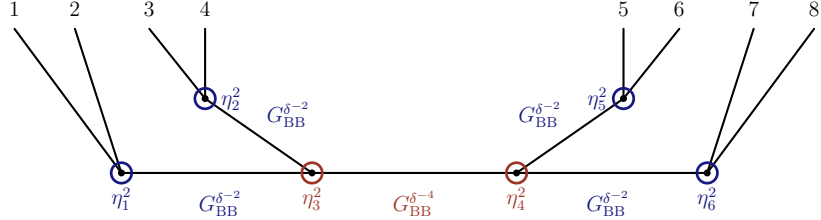
\begin{equation*}
\definecolor{vertexblue}{RGB}{22,25,128}
\definecolor{vertexred}{RGB}{155,55,38}
\resizebox{0.70\textwidth}{!}{%
\begin{tikzpicture}[
    x=0.9cm,
    y=0.9cm,
    every node/.style={font=\large},
    line cap=round,
    line join=round
]

% Vertices: exactly symmetric about x=9
\coordinate (v1) at (2.7,1.1);
\coordinate (v2) at (4.5,2.7);
\coordinate (v3) at (6.8,1.1);
\coordinate (v4) at (11.2,1.1);
\coordinate (v5) at (13.5,2.7);
\coordinate (v6) at (15.3,1.1);

% External endpoints: all aligned horizontally
\coordinate (p1) at (0.4,4.2);
\coordinate (p2) at (1.7,4.2);
\coordinate (p3) at (3.3,4.2);
\coordinate (p4) at (4.5,4.2);

\coordinate (p5) at (13.5,4.2);
\coordinate (p6) at (14.7,4.2);
\coordinate (p7) at (16.3,4.2);
\coordinate (p8) at (17.6,4.2);

% External legs
\draw[line width=1.15pt] (p1)--(v1);
\draw[line width=1.15pt] (p2)--(v1);
\draw[line width=1.15pt] (p3)--(v2);
\draw[line width=1.15pt] (p4)--(v2);

\draw[line width=1.15pt] (p5)--(v5);
\draw[line width=1.15pt] (p6)--(v5);
\draw[line width=1.15pt] (p7)--(v6);
\draw[line width=1.15pt] (p8)--(v6);

% Internal propagators
\draw[line width=1.15pt] (v1)--(v3);
\draw[line width=1.15pt] (v2)--(v3);
\draw[line width=1.15pt] (v3)--(v4);
\draw[line width=1.15pt] (v4)--(v5);
\draw[line width=1.15pt] (v4)--(v6);

% Blue vertices
\foreach \v in {v1,v2,v5,v6}{
    \draw[vertexblue,line width=1.5pt]
        (\v) circle[radius=0.22];
    \fill (\v) circle[radius=0.075];
}

% Red vertices
\foreach \v in {v3,v4}{
    \draw[vertexred,line width=1.5pt]
        (\v) circle[radius=0.22];
    \fill (\v) circle[radius=0.075];
}

% External labels
\node[above=3pt] at (p1) {$1$};
\node[above=3pt] at (p2) {$2$};
\node[above=3pt] at (p3) {$3$};
\node[above=3pt] at (p4) {$4$};

\node[above=3pt] at (p5) {$5$};
\node[above=3pt] at (p6) {$6$};
\node[above=3pt] at (p7) {$7$};
\node[above=3pt] at (p8) {$8$};

% Vertex labels
\node[vertexblue,below=4pt] at (v1) {$\eta_1^2$};
\node[vertexblue,right=5pt] at (v2) {$\eta_2^2$};

\node[vertexred,below=4pt] at (v3) {$\eta_3^2$};
\node[vertexred,below=4pt] at (v4) {$\eta_4^2$};

\node[vertexblue,left=5pt] at (v5) {$\eta_5^2$};
\node[vertexblue,below=4pt] at (v6) {$\eta_6^2$};

% Propagator labels
\node[vertexblue,below=6pt]
at ($(v1)!0.52!(v3)$)
{$G_{\mathrm{BB}}^{\delta^{-2}}$};

\node[vertexblue,above right=1pt]
at ($(v2)!0.50!(v3)$)
{$G_{\mathrm{BB}}^{\delta^{-2}}$};

\node[vertexred,below=6pt]
at ($(v3)!0.50!(v4)$)
{$G_{\mathrm{BB}}^{\delta^{-4}}$};

\node[vertexblue,above left=1pt]
at ($(v4)!0.50!(v5)$)
{$G_{\mathrm{BB}}^{\delta^{-2}}$};

\node[vertexblue,below=6pt]
at ($(v4)!0.48!(v6)$)
{$G_{\mathrm{BB}}^{\delta^{-2}}$};

\end{tikzpicture}%
}
\end{equation*}
so that effectively the powers of $\eta^2$ in the expansion of $G_{BB}^{(M)}$ nicely reorganize throughout the diagram to effectively change the power of $\eta$ at each vertex! Therefore we can recast the contribution from a given graph, $G$, to the vertex $\hat{\Psi}_n$ as follows: 
\begin{equation}
    \begin{aligned}
    \hat{\Psi}_n\vert_G={}&\left(\frac{H^2}{2}\right)^{n-3}
    \int_{\mathbb{R}}\prod_{e\in G}\frac{d\omega_e}{(2\pi i)\omega_e^2}
    \int_{-\infty}^0\prod_{v\in G}\left(\frac{d\eta_v}{H^2\eta_v^4}\,{\eta_v^2}\right)\\
    &\hspace{2cm}\times\prod_{a\in\mathrm{ext}(G)}\phi_{k_a}^{\mathrm{dS}}(\eta_{v_a})
    \prod_{e\in G}\frac{\Delta\phi^{\mathrm{dS}}_{\omega_e}(\eta_{v_e})\Delta\phi^{\mathrm{dS}}_{\omega_e}(\eta_{v_e^\prime})}{\omega_e^2-k_e^2\pm\delta^2+i\epsilon}
    \bigg\vert_{\delta^{-(2n-4)}}\, .
    \end{aligned}
\end{equation}
Here $\mathrm{ext}(G)$ denotes the set of external legs of $G$, and $v_a$ is the bulk vertex incident to external leg $a$. The sign of $\pm\delta^2$ depends on whether the chords are all even or odd, as usual.

Now just like we did in the flat-space case, we can perform the $\eta$ integrals and read off the effective $C_{\text{dS}}^{(t)}$ functions in de Sitter. We find: 
\begin{equation}
\begin{aligned}
    & \left.\begin{gathered}
\begin{tikzpicture}[
    line cap=round,
    line width=1pt,
    offshell/.style={
        draw=Maroon,
        line width=1.9pt
    },
    onshell/.style={
        draw=black,
        line width=0.9pt
    },
]
    % External legs
    \draw[onshell] (-0.7, 0.6) -- (0,0);
    \draw[onshell] (-0.7,-0.6) -- (0,0);
    \draw[offshell] (0,0) -- (0.9,0);

    % Labels
    \node at (-0.98, 0.55) {$k$};
    \node at (0.1, -0.35) {$\eta$};
    \node at (-0.98,-0.55) {$k^\prime$};
    \node at ( 1.3, 0) {$\omega_0$};

    % Black vertex
    \filldraw[black] (0,0) circle (3.2pt);
\end{tikzpicture}
    \end{gathered} \, \, \right]\,\,  \begin{aligned}  C^{(1)}_{\text{dS}}(k,k^\prime; \omega_0)& = \int_{-\infty}^0 \frac{d \eta}{\eta^{2+\alpha}} \, \phi_{k}^{\text{dS}}(\eta)\phi_{k^{\prime}}^{\text{dS}}(\eta) (\phi_{\omega_0}^{\text{dS}}(\eta)-\phi_{-\omega_0}^{\text{dS}}(\eta)) \\
    &=\frac{2 i\omega_0^3 \left(k^2+4 k k^{\prime}+{k^{\prime}}^2-\omega_0^2\right)}{((k+k^{\prime})^2-\omega_0^2)^2}\, ,\end{aligned} \\[0.8em]
&\left. \begin{gathered}
\begin{tikzpicture}[
    line cap=round,
    line width=1pt,
    offshell/.style={
        draw=Maroon,
        line width=1.9pt
    },
    onshell/.style={
        draw=black,
        line width=0.9pt
    },
]
    % External legs
    \draw[onshell] (-0.7, 0.6) -- (0,0);
    \draw[offshell] (-0.7,-0.6) -- (0,0);
    \draw[offshell] (0,0) -- (0.9,0);

    % Labels
    \node at (-0.98, 0.55) {$k$};
     \node at (0.1, -0.35) {$\eta$};
    \node at (-0.98,-0.65) {$\omega_1$};
    \node at ( 1.3, 0) {$\omega_0$};

    % Black vertex
    \filldraw[black] (0,0) circle (3.2pt);
\end{tikzpicture}
    \end{gathered} \, \, \right]\, \, \begin{aligned} C^{(2)}_{\text{dS}}(k;\omega_0,\omega_1) &=\int_{-\infty}^0 \frac{d \eta}{\eta^{2+\alpha}} \, \phi_{k}^{\text{dS}}(\eta) \prod_{j=0,1}(\phi_{\omega_j}^{\text{dS}}(\eta)-\phi_{-\omega_j}^{\text{dS}}(\eta)) \\
    &=\frac{i64 k^3 \omega_0^3 \omega_1^3}{\left((\omega_0+\omega_1)^2-k^2 + i \epsilon \right)^2\left((\omega_0-\omega_1)^2-k^2+i\epsilon \right)^2}\, ,\end{aligned} 
\\[0.8em]
&\left. \begin{gathered}
\begin{tikzpicture}[
    line cap=round,
    line width=1pt,
    offshell/.style={
        draw=Maroon,
        line width=1.9pt
    },
    onshell/.style={
        draw=black,
        line width=0.9pt
    },
]
    % External legs
    \draw[offshell] (-0.7, 0.6) -- (0,0);
    \draw[offshell] (-0.7,-0.6) -- (0,0);
    \draw[offshell] (0,0) -- (0.9,0);

    % Labels
    \node at (-0.98, 0.55) {$\omega_2$};
     \node at (0.1, -0.35) {$\eta$};
    \node at (-0.98,-0.65) {$\omega_1$};
    \node at ( 1.3, 0) {$\omega_0$};

    % Black vertex
    \filldraw[black] (0,0) circle (3.2pt);
\end{tikzpicture}  
    \end{gathered}\, \, \right] \, \,  \begin{aligned} C^{(3)}_{\text{dS}}(\omega_0,\omega_1,\omega_2) &=\int_{-\infty}^0  \frac{d \eta}{\eta^{2+\alpha}} \,  \prod_{j=0,1,2}(\phi_{\omega_j}^{\text{dS}}(\eta)-\phi_{-\omega_j}^{\text{dS}}(\eta)) \\
    &=\sum_{\sigma_{j}=\pm 1}  \sigma_0\sigma_1\sigma_2\,i\psi_3(\sigma_0 \omega_0,\sigma_1 \omega_1,\sigma_2 \omega_2)\, , \end{aligned} 
\end{aligned}
\label{eq:3ptFactors_dS}
\end{equation}
where: 
\begin{equation}
    \psi_3(k_1,k_2,k_3) = \frac{k_1 k_2 k_3}{\left(k_1+k_2+k_3-i \epsilon \right){}^2}+\frac{k_1 k_2+k_3 k_2+k_1 k_3}{k_1+k_2+k_3-i \epsilon }-k_1-k_2-k_3\, ,
\end{equation}
and the $\alpha$ is a regulator which is needed because these integrals have unphysical power divergences. From here, we can write the integral associated with each graph/triangulation, $\mathcal{T}_G$, contributing to $\hat{\Psi}_n$ as follows: 
\begin{equation}
    \hat{\Psi}_n \vert_{\mathcal{T}_G} =  \frac{1}{2^{n-3}H^2} \int_{\mathbb{R}} \prod_{(ij) \in \mathcal{T}_G} \frac{d \omega_{ij}}{(2\pi i) \omega_{ij}^2 (\omega_{ij}^2-k_{i,j}^2 \pm \delta^2)} \times \prod_{t \in \mathcal{T}_G} C^{(t)}_{\text{dS}}( \{k_i\},\{\omega_{ij}\}) \, \bigg \vert_{\delta^{-(2n-4)}},
    \label{eq:FreqRep_dS_Graph}
\end{equation}
which is precisely the analogue of \eqref{eq:FreqSpace_FlatSpace} but now for de Sitter! The main difference is that while in the flat-space case we started with such a form for each diagram and then extracted the desired power in $1/\delta$, here this representation was derived exclusively for the diagrams leading to the ``contact piece'' of $\Psi_n$ -- so the triangulations containing only $(e,e)$ or only $(o,o)$ propagators. And from these we can write the full pion wavefunction using the recursion \eqref{eq:RecursionPion1}, which in de Sitter simply reads: 
\begin{equation}
    \Psi^{\text{dS}}_n =  \sum_{\mathcal{C}} \int_{-\infty}^{+\infty} \prod_{(i,j) \in \mathcal{C}} \frac{d \omega_{i,j} \, H^2}{(2\pi i) \omega_{i,j}^2[2(w_{i,j}^2-k_{i,j}^2)]} \ \prod_{P_{e} \subset \mathcal{C}} \hat{\Psi}_{P_e} + \hat{\Psi}_{n} \, .
    \label{eq:RecursionPiondS}
\end{equation}

Now the difference between \eqref{eq:RecursionPiondS} and simply summing over all cubic diagrams, where each diagram contributes with integral of the form \eqref{eq:FreqRep_dS_Graph}, where now of course the $(e,o)$ propagators are not shifted, is that in general there are more ways to obtain the correct $1/\delta^{2n-4}$ power than those coming from the decomposition into $\hat{\Psi}_{P_e}$ -- as discussed at the end of section \ref{sec:RecursionFlat}. But it is trivial to see that the argument that showed that all these extra contributions cancel in the flat-space diagram sum, generalizes to de Sitter! And therefore we can alternatively write the full pion wavefunction as a sum over cubic graphs as follows: 
\begin{equation}
    \Psi^{\text{dS}}_n  = \sum_{\mathcal{T}} \frac{1}{2^{n-3}H^2}\int_{\mathbb{R}} \prod_{(ij) \in \mathcal{T}} \frac{d \omega_{ij}}{(2\pi i) \omega_{ij}^2 (\omega_{ij}^2-k^2_{i,j} - \delta_{i,j}^2)} \times \prod_{t \in \mathcal{T}_G} C^{(t)}_{\text{dS}}( \{k_i\},\{\omega_{ij}\}) \, \bigg \vert_{\delta^{-(2n-4)}} \, ,
    \label{eq:Flat-space analogy}
\end{equation}
with $\delta^2_{i,j}$ denoting the usual $\delta$-shift. As we will see momentarily, this representation will be particularly useful in deriving the de Sitter soft-limit, turning it into a simple analogue of what we saw in the flat-space setting. 

\subsubsection{Frequency integrals as residue sums}

Understanding that the de Sitter pion wavefunction can be extracted from the $\delta$-expansion of the frequency integral in \eqref{eq:Flat-space analogy} is the key step to be able to define it as a residue sum, exactly like in flat-space. 

Of course, the sum over all residues in the UHP/LHP from \eqref{eq:Flat-space analogy} will also include those that are $\sim \delta$, but just like we argued in the flat space case, it is simple to see also in de Sitter these lead to pure boundary contributions. Following the same logic as in flat-space, with a shifted bulk-to-bulk propagator, we have an integral of the type 
\begin{equation}
    \begin{gathered}
        \begin{tikzpicture}[scale=0.9,
    line cap=round,
    line width=1pt,
    offshell/.style={
        draw=Maroon,
        line width=1.9pt
    },
    onshell/.style={
        draw=gray,
        line width=0.9pt
    },
]
    % External legs
    \draw[onshell] (-0.8, 0.6) -- (0,0);
    \draw[onshell] (-0.4,0.6) -- (0,0);
    \draw[offshell] (0,0) -- (1,0);
    \draw[onshell] (1, 0) -- (1.4,0.6);
    \draw[onshell] (1,0) -- (1.8,0.6);
    \node at (0.5, 0.4) {$k_{i,j}$};
    \node at (-0.3, -0.2) {$\eta_{L}$};
    \node at (1.4, -0.2) {$\eta_{R}$};

    % Black vertex
    \filldraw[black] (0,0) circle (2.5pt);
    \filldraw[black] (1,0) circle (2.5pt);
\end{tikzpicture}
    \end{gathered} \quad = \,  \frac{1}{2} \int_{\mathbb{R}} \frac{d \omega_{ij}}{(2\pi i)\omega_{ij}^2} \frac{F^{\text{dS}}_L(\omega_{ij})F^{\text{dS}}_R(\omega_{ij})}{(\omega_{ij}-k_\delta)(\omega_{ij}+k_{\delta})} \, ,
\label{eq:DeltExpmth_dS}
\end{equation}
with $k_\delta = \sqrt{k_{ij}^2-\delta^2-i\epsilon}$ (assuming, without loss of generality, that $(i,j)$ are both odd). In de Sitter, $F^{\text{dS}}_L$ takes the form:
\begin{equation}
\begin{aligned}
    F^{\text{dS}}_L(\omega) &= \int_{-\infty}^0 \frac{d\eta_L}{\eta_L^{2}} f_L(\eta_L) \left( \phi^{\text{dS}}_{\omega}(\eta_L) - \phi^{\text{dS}}_{-\omega}(\eta_L) \right) \\
    &= 2i \int_{-\infty}^{0}\frac{d\eta_L}{\eta_L^{2}}f_L(\eta_L)\left(\sin(\omega\eta_L)-\omega\eta_L\cos(\omega\eta_L)\right) \, ,
\end{aligned}
\end{equation}
and similar for $F^{\text{dS}}_R$. And so integrating by parts, we can derive the asymptotic expansion for $F_L^\text{dS}$ to be
\begin{equation}
    F_{L}^{\text{dS}}(\omega) \sim -2 i\omega f_L(0) + i \pi f^{(1)}_L(0) - 2i\frac{f^{(2)}_L(0)}{\omega} + \mathcal{O}\left(\frac{1}{\omega^3}\right)\, ,  \quad \text{ (similar for }F_R^{\text{dS}})
    \label{eq:C3LdSOmExp}
\end{equation}
and so once again all the contributions come from the boundary $\eta_L/\eta_R=0$, and so do the contributions from the residue at $\omega \sim \pm i\delta$.  

In summary, just like in the flat-space case, from the frequency representation we can extract the pion wavefunction (up to boundary terms), by simply expanding the $\omega$-integrand to the correct $1/\delta$ power, and sum over the remaining finite $\omega$-residues. As we will show momentarily, in perfect analogy to the results in flat-space, this prescription to compute the wavefunction now in de Sitter will reveal an especially simple behavior under the soft limit. But let us first give an explicit example for $n=4$.  

\paragraph{Four-point example} At four-points, we must sum over the two possible triangulations of the square, corresponding to the $s$- and $t$-channels. For the $s$-channel, (triangulation with chord $(1,3)$) we get (after restoring the factor of $i^2$ from the vertex rule):
\begin{equation}
    \hat{\Psi}_4\vert_{(1,3)} = \frac{1}{2 H^2 \delta^4}\int_{\mathbb{R}}\frac{d\omega_{13}}{(2\pi i )\omega_{13}^2}(\omega_{13}^2-k_{13}^2)C_{\text{dS}}^{(1)}(k_1,k_2,\omega_{13})C_{\text{dS}}^{(1)}(k_3,k_4,\omega_{13})\, ,
    \label{eq:4pt_int_dS_13}
\end{equation}
and the $t$-channel, with chord $(2,4)$, yields
\begin{equation}
    \hat{\Psi}_4\vert_{(2,4)} = \frac{1}{2 H^2 \delta^4}\int_{\mathbb{R}}\frac{d\omega_{24}}{(2\pi i )\omega_{24}^2}(\omega_{24}^2-k_{24}^2)C_{\text{dS}}^{(1)}(k_1,k_4,\omega_{24})C_{\text{dS}}^{(1)}(k_2,k_3,\omega_{24})\, .
    \label{eq:4pt_int_dS_24}
\end{equation}
So summing over the residues in the LHP for both $\omega_{13}$ and $\omega_{24}$, we find the four-point wavefunction coefficient can be written as follows (where we suppress the coupling and factors of $H$): 
\begin{equation}
\Psi_4^{\text{dS}}
=
a_{\mathrm{dS}}
\left(
E_T-\frac{\sigma_2}{E_T}
-\frac{\sigma_3}{E_T^2}
-\frac{2\sigma_4}{E_T^3}
\right)
-\frac{\mathcal{C}}{E_T} +2 \sum_{i=1}^4 k_i^3
\label{eq:four_point_WF_dS}
\end{equation}
where 
\begin{equation}
\begin{aligned}
\sigma_2 &\equiv \sum_{i<j} k_i k_j,
\qquad
\sigma_3 \equiv \sum_{i<j<\ell} k_i k_j k_\ell,
\qquad
\sigma_4 \equiv k_1 k_2 k_3 k_4, \\
\mathcal{C}
&\equiv
E_T^2\sum_{i=1}^4 k_i^2
-k_1 k_3 (k_1+k_3)^2
-k_2 k_4 (k_2+k_4)^2
+2 k_1 k_2 k_3 k_4 \,,  \\
a_{\mathrm{dS}}
&=
\left[
(k_1+k_2)^2-k_{13}^2
\right]
+
\left[
(k_2+k_3)^2-k_{24}^2
\right]
-E_T(k_2-k_4) \, . 
\end{aligned}
\label{eq:4pt_polys_notation}
\end{equation}
To check this result, we can start by looking at the leading total-energy , and check that in the limit $E_T\to 0$ it factorizes into the flat-space amplitude as expected:
\begin{equation}
\psi_4
\underset{E_T\to 0}{\sim}
-\frac{
2 k_1 k_2 k_3 k_4
}{E_T^3}
\left(\left[
(k_1+k_2)^2-k_{13}^2
\right]
+
\left[
(k_2+k_3)^2-k_{24}^2
\right]\right)
+\mathcal{O}\!\left(E_T^{-2}\right) .
\end{equation}
In addition, one can check that the above wavefunction coefficient is de Sitter invariant by acting with the respective de Sitter group operators  (see \cite{Sun:2021thf} for a review),  namely it is invariant under special conformal transformations, and the dilatation operator. 

Following the same procedure and using the recursion in \eqref{eq:RecursionPiondS}, we can compute the pion wavefunction at higher points. We have done it explicitly up to $n=8$, and checked that our results agree up to $n=6$ with  \cite{Armstrong:2022vgl} up to local boundary terms.

\subsection{Universal soft limits}

 To end this section, let us now explain how the soft limits described in Sec.~\ref{sec:Soft_Limits_FS} trivially extend to de Sitter. Of course from the moment we derive \eqref{eq:Flat-space analogy} for de Sitter, it is clear the soft limits will extend, as long as the de Sitter $C$-factors from eq.~\eqref{eq:3ptFactors_dS} behave properly under the soft limit. 

To see this explicitly, we check the soft limit case by case, just like in Sec.~\ref{sec:Soft_Limits_FS}, starting with the case where we have an edge triangle $(1,2,3)$ and the adjacent triangle is $(1,3,n)$. Now the $\mu$-expansion has the following form:
\begin{equation}
    \begin{gathered}
     \begin{tikzpicture}[line width=1,scale=10]
    \node[regular polygon, regular polygon sides=22,  minimum size=2cm] (p) at (0,0) {};
     \draw (0,0) circle [radius=0.101cm];

    \node[scale=0.8,xshift=-10,yshift=0] at (p.corner 7) {$2$};
    \node[scale=0.8,xshift=-7,yshift=-5] at (p.corner 10) {$1$};   
    \node[scale=0.8,xshift=3,yshift=-7] at (p.corner 12) {$n$};
    \node[scale=0.8,xshift=-7,yshift=7] at (p.corner 4) {$3$};
   % \node[scale=0.8,xshift=7,yshift=7] at (p.corner 20) {$j_2$};

    \foreach \i in {7}
    {\draw[fill] (p.corner \i) circle [radius=0.1pt];}
    \foreach \i in {4}
    {\draw[fill] (p.corner \i) circle [radius=0.1pt];}
    \draw[fill] (p.corner 10) circle [radius=0.1pt];

     %\fill[Blue,opacity=0.2]  (p.corner 8) -- (p.corner 7) -- (p.corner 6) -- (p.corner 5) -- (p.corner 4) -- (p.corner 3) -- (p.corner 2) -- (p.corner 1) -- (p.corner 22)-- cycle;

     \fill[Gray,opacity=0.2]   (p.corner 10) -- (p.corner 11) -- (p.corner 12) -- (p.corner 13) -- (p.corner 14) -- (p.corner 15) -- (p.corner 16) -- (p.corner 17) -- (p.corner 18)-- (p.corner 19)-- (p.corner 20) -- (p.corner 21) -- (p.corner 22) -- (p.corner 1) -- (p.corner 2) -- (p.corner 3) -- (p.corner 4) -- cycle;

     %\node[scale=0.8,xshift=18,yshift=30] at (p.corner 8) {$\mathcal{S}_L$};
     \node[scale=0.8,xshift=30,yshift=30] at (p.corner 10) {$\tilde{\Psi}_R$};
      \draw[Maroon] (p.corner 4) -- (p.corner 10);
    \draw[Gray] (p.corner 4) -- (p.corner 12);
    \fill[Gray] (p.corner 12) circle [radius=0.15pt];
    \fill[Maroon] (p.corner 10) circle [radius=0.15pt];
     \fill[Maroon] (p.corner 7) circle [radius=0.15pt];
     \fill[Black] (p.corner 4) circle [radius=0.15pt];

    \end{tikzpicture}      
\end{gathered} =     \sum_{m=0}^{\infty} \frac{(-1)^m}{2\delta^{2m+2}}\int\frac{d\omega_{1,3}}{2\pi i\,\omega_{1,3}^2} C_{dS}^{(1)}(k_1,k_2,\omega_{1,3})\,C_{dS}^{(2)}(k_n,\omega_{1,3},\omega_{3,n}) (\omega_{1,3}^2-k_{13}^2)^m\times\tilde{\Psi}_R \, ,
\label{eq:13Soft_dS}
\end{equation}
which upon taking the soft limit, $k_1\to 0, k_{13}\to k_2 $, becomes:
\begin{equation}
    \sum_{m=0}^{\infty} \frac{(-1)^m}{2\delta^{2m+2}}\int\frac{d\omega_{1,3}}{2\pi i}\frac{128 k_n^3 \omega_{1,3}^4(\omega_{1,3}^2-k_2^2)^{m-1}\omega_{3,n}^3}{(k_n^2-(\omega_{1,3}-\omega_{3,n})^2)^2(k_n^2-(\omega_{1,3}+\omega_{3,n})^2)^2} \tilde{\Psi}_R\, ,
\end{equation}
and like before, considering the case $m=0$ separately from $m\geq1$, we find precisely like in flat-space:
\begin{equation}
\underline{m=0:} \quad \frac{i}{\delta^2} \times \Psi^R_{n-1}(2,3,\cdots,n)\,,  \quad \quad 
    \underline{m\geq1:} \quad \frac{-i}{\delta^2} \times \Psi^R_{n-1}(\hat{2},3,\cdots,n)\vert_{k_2\to \sqrt{k_2^2-\delta^2}}.
\end{equation}
%When $m=0$, we take the sum over the residues $\{k_2,k_n+\omega_{3,n},k_n-\omega_{3,n}\}$, which will precisely yield $C_{\text{dS}}^{(1)}(k_2,k_n,\omega_{3,n})\times \tilde{\Psi}_R$.
%When $m\geq 1$, the residues we sum over are $\{k_n+\omega_{3,n},k_n-\omega_{3,n}\}$, but once again we obtain the same answer as in flat-space, the lower point de Sitter wavefunction with a shifted external leg, in this case $k_2\to \sqrt{k_2^2-\delta^2}$:
%
%\begin{equation}
%    \underline{m\geq1:} \quad \frac{-i}{\delta^2} \times \Psi^R_{n-1}(\hat{2},3,\cdots,n).
%\end{equation}
%
%One may ask why is this case, since this square-root shift only really makes sense in flat-space. From the practical point of view to obtain the de Sitter wavefunction our expansion is equivalent to inserting an extra $\eta^2$ per vertex a Tr$[\phi^3]$ even point diagram and then shifting the required exchange momenta by the square root shift. Close to the soft limit $k_1\to 0$ what is really being shifted is the internal state which carries momentum $k_{13}$, since the extra $\eta^2$ measure is already incorporated in our $C_3$-factors, then the shift is precisely just the square root shift, which manifests itself as this shift of the external leg on a lower point wavefunction. Therefore the interpretation is precisely that of flat-space, the external leg of the lower point wavefunction inherits the shift of the internal leg. 

Now let's proceed to the case where instead of the internal triangle being $(1,3,n)$ we have that it is $(1,3,j)$, then $C^{(2)}_{\text{dS}}(k_n,\omega_{1,3},\omega_{1,n})$ is replaced by  $C^{(3)}_{\text{dS}}(\omega_{1,3},\omega_{1,j},\omega_{3,j})$ leading to the following:
\begin{equation}
    \begin{gathered}
    \begin{tikzpicture}[line width=1,scale=10]
    \node[regular polygon, regular polygon sides=22,  minimum size=2cm] (p) at (0,0) {};
     \draw (0,0) circle [radius=0.101cm];

    \node[scale=0.8,xshift=-10,yshift=0] at (p.corner 7) {$2$};
    \node[scale=0.8,xshift=-7,yshift=-5] at (p.corner 10) {$1$};   
    \node[scale=0.8,xshift=8,yshift=5] at (p.corner 17) {$p$};
    \node[scale=0.8,xshift=-7,yshift=7] at (p.corner 4) {$3$};
   % \node[scale=0.8,xshift=7,yshift=7] at (p.corner 20) {$j_2$};

    \foreach \i in {7}
    {\draw[fill] (p.corner \i) circle [radius=0.1pt];}
    \foreach \i in {4}
    {\draw[fill] (p.corner \i) circle [radius=0.1pt];}
    \draw[fill] (p.corner 10) circle [radius=0.1pt];

     %\fill[Blue,opacity=0.2]  (p.corner 8) -- (p.corner 7) -- (p.corner 6) -- (p.corner 5) -- (p.corner 4) -- (p.corner 3) -- (p.corner 2) -- (p.corner 1) -- (p.corner 22)-- cycle;

     \fill[Gray,opacity=0.2]   (p.corner 10) -- (p.corner 11) -- (p.corner 12) -- (p.corner 13) -- (p.corner 14) -- (p.corner 15) -- (p.corner 16) -- (p.corner 17) -- (p.corner 18)-- (p.corner 19)-- (p.corner 20) -- (p.corner 21) -- (p.corner 22) -- (p.corner 1) -- (p.corner 2) -- (p.corner 3) -- (p.corner 4) -- cycle;

     %\node[scale=0.8,xshift=18,yshift=30] at (p.corner 8) {$\mathcal{S}_L$};
     \node[scale=0.7,xshift=35,yshift=60] at (p.corner 10) {$\tilde{\Psi}_{(1)}$};
     \node[scale=0.7,xshift=40,yshift=5] at (p.corner 10) {$\tilde{\Psi}_{(2)}$};
     \node[scale=0.7,xshift=27,yshift=33] at (p.corner 10) {$C^{(3)}$};
      \draw[Maroon] (p.corner 4) -- (p.corner 10);
    \draw[Gray] (p.corner 4) -- (p.corner 18);
    \draw[Gray] (p.corner 10) -- (p.corner 18);
    \fill[Gray] (p.corner 18) circle [radius=0.15pt];
    \fill[Maroon] (p.corner 10) circle [radius=0.15pt];
     \fill[Maroon] (p.corner 7) circle [radius=0.15pt];
     \fill[Black] (p.corner 4) circle [radius=0.15pt];

    \end{tikzpicture}     
\end{gathered} = \sum_{m=0}^\infty \frac{(-1)^m}{2 \, \delta^{2m+2}} \int \frac{d \omega_{1,3}}{2\pi i \omega_{1,3}^2}  C^{(1)}_{\text{dS}}(k_1,k_2,\omega_{1,3})(\omega_{1,3}^2-k_{1,3}^2)^{m} C^{(3)}_{\text{dS}} \times \tilde{\Psi}_{(1)}\tilde{\Psi}_{(2)}\, ,
\end{equation}
which upon taking the soft limit $k_1\to0,\, k_{1,3}\to  k_2$, again we recover that only the $m=0$ case contributes, which trivially reproduces the flat-space result. And so from the piece in the Berends-Giele recursion corresponding to $j=3$, we get: 

%, because the four poles in $C^{(3)}_{\text{dS}}(\omega_{1,3},\omega_{1,j},\omega_{3,j})$ also sum up to zero, and we recover that, upon taking the soft limit and the $\omega_{1,3}=k_2$ residue, we get the $C^{(2)}_{\text{dS}}(k_2,\omega_{1,j},\omega_{3,j})$ factor. And, ultimately also shifting $k_2\to\sqrt{k_2-\delta^2}$ inside the $C^{(2)}$ factor, and then expanding in large $\delta$ produces only odd powers of $\delta$, therefore we once again show that we obtain precisely the same result as in flat-space. 
%Thus, also in de Sitter we can write that for the external triangle, $(1,2,3)$, case we obtain the soft limit:
%
\begin{equation}
    \begin{gathered}
    \begin{tikzpicture}[line width=1,scale=10]
    \node[regular polygon, regular polygon sides=22,  minimum size=2cm] (p) at (0,0) {};
     \draw (0,0) circle [radius=0.101cm];
    \node[scale=0.8,xshift=-10,yshift=0] at (p.corner 7) {$2$};
    \node[scale=0.8,xshift=-7,yshift=-5] at (p.corner 10) {$1$};   
    \node[scale=0.8,xshift=3,yshift=-7] at (p.corner 12) {$n$};
    \node[scale=0.8,xshift=-7,yshift=7] at (p.corner 4) {$3$};
   % \node[scale=0.8,xshift=7,yshift=7] at (p.corner 20) {$j_2$};

    \foreach \i in {7}
    {\draw[fill] (p.corner \i) circle [radius=0.1pt];}
    \foreach \i in {4}
    {\draw[fill] (p.corner \i) circle [radius=0.1pt];}
    \draw[fill] (p.corner 10) circle [radius=0.1pt];

     %\fill[Blue,opacity=0.2]  (p.corner 8) -- (p.corner 7) -- (p.corner 6) -- (p.corner 5) -- (p.corner 4) -- (p.corner 3) -- (p.corner 2) -- (p.corner 1) -- (p.corner 22)-- cycle;

     \fill[Gray,opacity=0.2]   (p.corner 10) -- (p.corner 11) -- (p.corner 12) -- (p.corner 13) -- (p.corner 14) -- (p.corner 15) -- (p.corner 16) -- (p.corner 17) -- (p.corner 18)-- (p.corner 19)-- (p.corner 20) -- (p.corner 21) -- (p.corner 22) -- (p.corner 1) -- (p.corner 2) -- (p.corner 3) -- (p.corner 4) -- cycle;

     %\node[scale=0.8,xshift=18,yshift=30] at (p.corner 8) {$\mathcal{S}_L$};
     \node[scale=0.8,xshift=30,yshift=30] at (p.corner 10) {$\Psi^{\text{dS}}_R$};
      \draw[Maroon] (p.corner 4) -- (p.corner 10);
    %\draw[Gray] (p.corner 4) -- (p.corner 12);
    \fill[Black] (p.corner 12) circle [radius=0.15pt];
    \fill[Maroon] (p.corner 10) circle [radius=0.15pt];
     \fill[Maroon] (p.corner 7) circle [radius=0.15pt];
     \fill[Black] (p.corner 4) circle [radius=0.15pt];
\draw[-{Stealth[length=1.5mm,width=0.9mm]}]
  (220.09:0.115cm)
  arc[
    start angle=220.09,
    end angle=185,
    radius=0.115cm
  ];

    \end{tikzpicture}     
\end{gathered} \quad \xrightarrow[k_1 \to 0]{}\quad   \frac{i}{\delta^2} \times \left[\begin{gathered}
    \begin{tikzpicture}[line width=1,scale=10]
    \node[regular polygon, regular polygon sides=22,  minimum size=2cm] (p) at (0,0) {};
     \draw (0,0) circle [radius=0.101cm];
   % \draw[Blue!80!White,dashed,line width=2]
  %({90-180/22+3*360/22}:0.101cm)
  %arc[
  %  start angle={90-180/22+3*360/22},
  %  end angle={90-180/22+6*360/22},
   % radius=0.101cm
  %];

    \node[scale=0.8,xshift=-9,yshift=6] at (p.corner 7) {$2$};
    \node[scale=0.8,xshift=-7,yshift=-5] at (p.corner 7) {$1_{\text{\tiny{R}}}$};   
    \node[scale=0.8,xshift=-6,yshift=-5] at (p.corner 10) {$n$};
    \node[scale=0.8,xshift=-7,yshift=7] at (p.corner 4) {$3$};
   % \node[scale=0.8,xshift=7,yshift=7] at (p.corner 20) {$j_2$};
\draw[Maroon]
  ({90-180/22+3*360/22}:0.101cm)
  arc[
    start angle={90-180/22+3*360/22},
    end angle={90-180/22+6*360/22},
    radius=0.101cm
  ];

    \foreach \i in {7}
    {\draw[Maroon,fill] (p.corner \i) circle [radius=0.1pt];}
    \foreach \i in {4}
    {\draw[fill] (p.corner \i) circle [radius=0.1pt];}
    \draw[fill] (p.corner 10) circle [radius=0.1pt];

     \fill[Gray,opacity=0.2]   (p.corner 10) -- (p.corner 11) -- (p.corner 12) -- (p.corner 13) -- (p.corner 14) -- (p.corner 15) -- (p.corner 16) -- (p.corner 17) -- (p.corner 18)-- (p.corner 19)-- (p.corner 20) -- (p.corner 21) -- (p.corner 22) -- (p.corner 1) -- (p.corner 2) -- (p.corner 3) -- (p.corner 4) -- (p.corner 5) -- (p.corner 6) -- (p.corner 7) -- (p.corner 8) -- (p.corner 9)-- cycle;

     %\node[scale=0.8,xshift=18,yshift=30] at (p.corner 8) {$\mathcal{S}_L$};
     \node[scale=0.8,xshift=25,yshift=30] at (p.corner 10) {$\Psi^{R,\text{dS}}_{n-1}$};
      %\draw[Gray] (p.corner 4) -- (p.corner 10);

    \fill[Black] (p.corner 10) circle [radius=0.15pt];
     \fill[Black] (p.corner 4) circle [radius=0.15pt];

    \end{tikzpicture}     
\end{gathered} \quad - \quad  \begin{gathered}
    \begin{tikzpicture}[line width=1,scale=10]
    \node[regular polygon, regular polygon sides=22,  minimum size=2cm] (p) at (0,0) {};
     \draw (0,0) circle [radius=0.101cm];
    \draw[Blue!80!White,dashed,line width=2]
  ({90-180/22+3*360/22}:0.101cm)
  arc[
    start angle={90-180/22+3*360/22},
    end angle={90-180/22+6*360/22},
    radius=0.101cm
  ];

    \node[scale=0.8,xshift=-9,yshift=6] at (p.corner 7) {$2$};
    \node[scale=0.8,xshift=-7,yshift=-5] at (p.corner 7) {$1_{\text{\tiny{R}}}$};   
    \node[scale=0.8,xshift=-6,yshift=-5] at (p.corner 10) {$n$};
    \node[scale=0.8,xshift=-7,yshift=7] at (p.corner 4) {$3$};
   % \node[scale=0.8,xshift=7,yshift=7] at (p.corner 20) {$j_2$};

    \foreach \i in {7}
    {\draw[Maroon,fill] (p.corner \i) circle [radius=0.1pt];}
    \foreach \i in {4}
    {\draw[fill] (p.corner \i) circle [radius=0.1pt];}
    \draw[fill] (p.corner 10) circle [radius=0.1pt];

     \fill[Gray,opacity=0.2]   (p.corner 10) -- (p.corner 11) -- (p.corner 12) -- (p.corner 13) -- (p.corner 14) -- (p.corner 15) -- (p.corner 16) -- (p.corner 17) -- (p.corner 18)-- (p.corner 19)-- (p.corner 20) -- (p.corner 21) -- (p.corner 22) -- (p.corner 1) -- (p.corner 2) -- (p.corner 3) -- (p.corner 4) -- (p.corner 5) -- (p.corner 6) -- (p.corner 7) -- (p.corner 8) -- (p.corner 9)-- cycle;

     %\node[scale=0.8,xshift=18,yshift=30] at (p.corner 8) {$\mathcal{S}_L$};
     \node[scale=0.8,xshift=25,yshift=30] at (p.corner 10) {$\Psi^{R,\text{dS}}_{n-1}$};
      %\draw[Gray] (p.corner 4) -- (p.corner 10);

    \fill[Black] (p.corner 10) circle [radius=0.15pt];
     \fill[Black] (p.corner 4) circle [radius=0.15pt];

    \end{tikzpicture}     
\end{gathered} \right]\,,
\label{eq:Softj3_dS}
\end{equation}
and similarly for the $(1,2,n)$ triangle case, we obtain the equivalent soft limit to eq.~\eqref{eq:Softjn}. Finally just like in flat-space, it is worth emphasizing that also in de Sitter we can interpret these $(n-1)$-point wavefunctions, as the mixed theory wavefunctions: 
\begin{equation}
\begin{aligned}
\Psi^{R,\text{dS}}_{n-1} &\equiv \Psi^{\text{NLSM}\oplus \phi^3,\text{dS}}_{n-1}(2^{\phi},3^{\phi},4^\pi,\cdots,(n-1)^\pi,n^\phi) \,, \\
     \Psi^{L,\text{dS}}_{n-1} &\equiv \Psi^{\text{NLSM}\oplus \phi^3,\text{dS}}_{n-1}(2^{\phi},3^{\pi},\cdots,(n-2)^\pi,(n-1)^\phi,n^\phi) \,.  
\end{aligned}
\end{equation}

Finally, for the remaining terms in the Berends-Giele recursion, so for an internal triangle $(1,2,j)$, we have:
\begin{equation}
    \begin{gathered}
    \begin{tikzpicture}[line width=1,scale=10]
    \node[regular polygon, regular polygon sides=22,  minimum size=2cm] (p) at (0,0) {};
     \draw (0,0) circle [radius=0.101cm];

    \node[scale=0.8,xshift=-10,yshift=0] at (p.corner 7) {$2$};
    \node[scale=0.8,xshift=-7,yshift=-5] at (p.corner 10) {$1$};    
    \node[scale=0.8,xshift=7,yshift=7] at (p.corner 19) {$j$};
   % \node[scale=0.8,xshift=7,yshift=7] at (p.corner 20) {$j_2$};

    \foreach \i in {7}
    {\draw[fill] (p.corner \i) circle [radius=0.1pt];}
    \foreach \i in {19}
    {\draw[fill] (p.corner \i) circle [radius=0.1pt];}
    \draw[fill] (p.corner 10) circle [radius=0.1pt];

     %\fill[Blue,opacity=0.2]  (p.corner 8) -- (p.corner 7) -- (p.corner 6) -- (p.corner 5) -- (p.corner 4) -- (p.corner 3) -- (p.corner 2) -- (p.corner 1) -- (p.corner 22)-- cycle;

     \fill[Gray,opacity=0.2]   (p.corner 10) --  (p.corner 19)--(p.corner 18) -- (p.corner 17) -- (p.corner 16) -- (p.corner 15) -- (p.corner 14) -- (p.corner 13) -- (p.corner 12)-- (p.corner 11)-- cycle;
    
    \fill[Gray,opacity=0.2]   (p.corner 7) --  (p.corner 6)--(p.corner 5) -- (p.corner 4) -- (p.corner 3) -- (p.corner 2) -- (p.corner 1) -- (p.corner 21) -- (p.corner 20)-- (p.corner 19)-- cycle;

     %\node[scale=0.8,xshift=18,yshift=30] at (p.corner 8) {$\mathcal{S}_L$};
     \node[scale=0.8,xshift=38,yshift=10] at (p.corner 10) {$\Psi_R$};
     \node[scale=0.8,xshift=20,yshift=48] at (p.corner 10) {$\Psi_L$};
      \draw[Maroon] (p.corner 7) -- (p.corner 19);
      \draw[Maroon] (p.corner 10) -- (p.corner 19);
    \fill[Maroon] (p.corner 10) circle [radius=0.15pt];
     \fill[Maroon] (p.corner 7) circle [radius=0.15pt];
     \fill[Black] (p.corner 19) circle [radius=0.15pt];

    \end{tikzpicture}     
\end{gathered} = \sum_{m=0}^\infty\frac{(-1)}{4 \, \delta^{2m+2}} \int \frac{d \omega_{2,j} d \omega_{1,j}}{(2\pi i)^2 \omega_{1,j}^2 \omega_{2,j}^2} \Psi_L \times \frac{(\omega_{2,j}^2-k_{2,j}^2)^m C^{(2)}_{\text{dS}}(k_1;\omega_{1,j},\omega_{2,j})}{(\omega_{1,j}^2-k_{1,j}^2)}  \times \Psi_R \, ,
\end{equation}
also here in the soft limit, $k_{1,j} \to k_{2,j}$, from the form of the $C^{(2)}$ we only get contributions from the $\omega$ poles at $\omega_{1,j}=\pm \omega_{2,j}$ -- which lead to exactly same soft limit as in eq.~\eqref{eq:softLim2}. This means, the de Sitter wavefunction for the NLSM follows exactly the same soft limit as the flat-space NLSM wavefunction:
\begin{equation}
    \Psi_n^{\text{dS}} \quad \xrightarrow{k_1 \to 0} \quad \frac{i}{\delta^2} \sum_{j=3}^n (-1)^{j+1} \left[ \begin{gathered}
    \begin{tikzpicture}[line width=1,scale=10]
    \node[regular polygon, regular polygon sides=22,  minimum size=2cm] (p) at (0,0) {};
     \draw (0,0) circle [radius=0.101cm];

    \node[scale=0.8,xshift=-10,yshift=5] at (p.corner 7) {$2_{\text{\tiny{L}}}$};
    \node[scale=0.8,xshift=-10,yshift=-5] at (p.corner 7) {$1_{\text{\tiny{R}}}$};    
    \node[scale=0.8,xshift=7,yshift=7] at (p.corner 19) {$j$};
   % \node[scale=0.8,xshift=7,yshift=7] at (p.corner 20) {$j_2$};

    \foreach \i in {7}
    {\draw[fill] (p.corner \i) circle [radius=0.1pt];}
    \foreach \i in {19}
    {\draw[fill] (p.corner \i) circle [radius=0.1pt];}
    %\draw[fill] (p.corner 10) circle [radius=0.1pt];

     %\fill[Blue,opacity=0.2]  (p.corner 8) -- (p.corner 7) -- (p.corner 6) -- (p.corner 5) -- (p.corner 4) -- (p.corner 3) -- (p.corner 2) -- (p.corner 1) -- (p.corner 22)-- cycle;

     \fill[Gray,opacity=0.2]   (p.corner 10) -- (p.corner 11) -- (p.corner 12) -- (p.corner 13) -- (p.corner 14) -- (p.corner 15) -- (p.corner 16) -- (p.corner 17) -- (p.corner 18)-- (p.corner 19)-- (p.corner 20) -- (p.corner 21) -- (p.corner 22) -- (p.corner 1) -- (p.corner 2) -- (p.corner 3) -- (p.corner 4) -- (p.corner 5) -- (p.corner 6) -- (p.corner 7) -- (p.corner 8) -- (p.corner 9)-- cycle;

     %\node[scale=0.8,xshift=18,yshift=30] at (p.corner 8) {$\mathcal{S}_L$};
     \node[scale=0.8,xshift=25,yshift=15] at (p.corner 10) {$\Psi^{\text{dS}}_R$};
     \node[scale=0.8,xshift=20,yshift=48] at (p.corner 10) {$\Psi^{\text{dS}}_L$};
      \draw[Maroon] (p.corner 7) -- (p.corner 19);
      %\draw[Maroon] (p.corner 10) -- (p.corner 19);
    %\fill[Maroon] (p.corner 10) circle [radius=0.15pt];
     \fill[Maroon] (p.corner 7) circle [radius=0.15pt];
     %\fill[Black] (p.corner 19) circle [radius=0.15pt];

    \end{tikzpicture}     
\end{gathered} \quad - \quad \begin{gathered}
    \begin{tikzpicture}[line width=1,scale=10]
    \node[regular polygon, regular polygon sides=22,  minimum size=2cm] (p) at (0,0) {};
     \draw (0,0) circle [radius=0.101cm];

    \node[scale=0.8,xshift=-10,yshift=5] at (p.corner 7) {$2_{\text{\tiny{L}}}$};
    \node[scale=0.8,xshift=-10,yshift=-5] at (p.corner 7) {$1_{\text{\tiny{R}}}$};    
    \node[scale=0.8,xshift=7,yshift=7] at (p.corner 19) {$j$};
   % \node[scale=0.8,xshift=7,yshift=7] at (p.corner 20) {$j_2$};

    \foreach \i in {7}
    {\draw[fill] (p.corner \i) circle [radius=0.1pt];}
    \foreach \i in {19}
    {\draw[fill] (p.corner \i) circle [radius=0.1pt];}
    %\draw[fill] (p.corner 10) circle [radius=0.1pt];

     %\fill[Blue,opacity=0.2]  (p.corner 8) -- (p.corner 7) -- (p.corner 6) -- (p.corner 5) -- (p.corner 4) -- (p.corner 3) -- (p.corner 2) -- (p.corner 1) -- (p.corner 22)-- cycle;

     \fill[Gray,opacity=0.2]   (p.corner 10) -- (p.corner 11) -- (p.corner 12) -- (p.corner 13) -- (p.corner 14) -- (p.corner 15) -- (p.corner 16) -- (p.corner 17) -- (p.corner 18)-- (p.corner 19)-- (p.corner 20) -- (p.corner 21) -- (p.corner 22) -- (p.corner 1) -- (p.corner 2) -- (p.corner 3) -- (p.corner 4) -- (p.corner 5) -- (p.corner 6) -- (p.corner 7) -- (p.corner 8) -- (p.corner 9)-- cycle;

     %\node[scale=0.8,xshift=18,yshift=30] at (p.corner 8) {$\mathcal{S}_L$};
     \node[scale=0.8,xshift=25,yshift=15] at (p.corner 10) {$\Psi^{\text{dS}}_R$};
     \node[scale=0.8,xshift=20,yshift=48] at (p.corner 10) {$\Psi^{\text{dS}}_L$};
      \draw[Maroon] (p.corner 7) -- (p.corner 19);
      \draw[Blue!80!White,dashed,line width=2] (p.corner 7) -- (p.corner 19);
      %\draw[Maroon] (p.corner 10) -- (p.corner 19);
    %\fill[Maroon] (p.corner 10) circle [radius=0.15pt];
     \fill[Maroon] (p.corner 7) circle [radius=0.15pt];
     %\fill[Black] (p.corner 19) circle [radius=0.15pt];

    \end{tikzpicture}     
\end{gathered} \right] \, . 
\label{eq:FullSoftLimit_dS}
\end{equation}
Finally, since the soft-limits for each case are the precisely the same, the cancellation argument between all terms follows automatically leaving us with the de Sitter analogue of ~\eqref{eq:FullSoftLimit_final}: 
\begin{equation}
\begin{aligned}
    \Psi^{\text{dS}}_n  \to & -\frac{i}{\delta^2} \left[ \Psi^{\text{NLSM}\oplus \phi, \text{dS}}_{n-1}(2^{\phi},3^{\phi},4^\pi,\cdots,(n-1)^\pi,n^\phi) \vert_{\sqrt{k_2^2-\delta^2}} \right. \\
    & \left. \quad \quad \quad \quad \quad \quad \quad -  \Psi^{\text{NLSM}\oplus \phi, \text{dS}}_{n-1}(2^{\phi},3^{\pi},\cdots,(n-2)^\pi,(n-1)^\phi,n^\phi) \vert_{\sqrt{k_n^2+\delta^2}} \right] 
\end{aligned} \, . 
\label{eq:SoftLimitdS}
\end{equation}

This is an interesting observation, even though our $C_{\text{dS}}^{(i)}$ factors look quite different from the corresponding factors in flat-space, the wavefunctions have exactly the same soft limits. Of course, it is not entirely unexpected since the form of the Ward identity for the non-linear symmetry of this theory will be the same irrespective of the background, as long as we are computing the wavefunction in a spatially homogeneous slice. Of course, using the Ward identity one can derive the soft theorems at each multiplicity  (see \cite{Bittermann:2022nfh} for example), but the remarkable point of eq.~\eqref{eq:FullSoftLimit_final} is that it tells us a universal formula for the soft limit, at any $n$, and which is valid both in flat-space and de Sitter. 

We can check our statements from eq.~\eqref{eq:four_point_WF_dS} when we set $k_1\to 0, k_{13}\to k_2, k_{24}\to k_4$, we obtain the same as we would by employing eq.~\eqref{eq:FullSoftLimit_final} directly:
\begin{equation}
    \psi_4^{\text{dS}}\vert_{k_1\to 0} = k_2^3 +2 k_3^3 + k_4^3\, .
\end{equation}

Another interesting consequence of the form of the soft limit in \eqref{eq:SoftLimitdS}, it shows the wavefunction is much ``softer'' than expected. This is, the vanishing of the amplitude in the soft limit ensures that the coefficient of the leading total energy pole must vanish in the soft limit; but from \eqref{eq:SoftLimitdS} we learn that not only the leading power in $1/E_t$ vanishes, but the $\delta$-shift in $k_2$ ensures \textit{any} multiplicity of the total energy pole vanishes in the soft limit! 

\section{Pion correlators in flat-space and de Sitter}
\label{sec:Correlators}

Now that we have understood how to compute the NLSM wavefunction in flat-space and de Sitter, we can compute the correlator directly using the mod-square of the wavefunction as follows:
\begin{align}
    \langle \Phi^N \rangle &= \int \mathcal{D}\Phi\:\prod_{i}^{N} \Phi(0,\vec{x}_i)\: |\Psi[\Phi]|^2 \, ,
\end{align}
which in practice leads to the following expansion for the $n$-point correlator at tree-level, with only connected parts, and stripping of the external legs, in terms of $\Psi_k$ for $k\leq n$ \cite{Cosmohedron,Figueiredo:2025daa}:
\begin{equation}
 \langle \Phi_1 \Phi_2 \cdots \Phi_n \rangle = 2\text{Re}(\Psi_n) + \sum_{\mathcal{C} \neq  \emptyset} \prod_{(i,j)\in\mathcal{C}} \frac{1}{2\text{Re}[\psi_2(k_{i,j})]} \prod_{P \in \mathcal{P}_\mathcal{C}} 2\text{Re}(\Psi_P) \, ,
    \label{eq:Corr}
\end{equation}
where we  sum over all possible collections of non-overlapping chords $\mathcal{C}$ -- partial and full triangulations -- which divide the original $n$-gon into a collection of subpolygons, $\mathcal{P}_\mathcal{C}$, from which we get the product of the respective lower-point wavefunctions, $\Psi_{P}$, for all $P \in \mathcal{P}_\mathcal{C}$. 

This decomposition of the correlator immediately tells us that the correlator will not obey the same simple form of soft limits found for the wavefunction. This is due to the presence of additional poles, $1/k_{i,j}$, which will also, in general, be shifted. We leave the exploration of the behavior under soft-limit for future work and here we focus on explaining the different ways to compute NLSM correlators in flat-space and de Sitter.

\subsection{Flat-space correlators}
\label{sec:Corrs_FS}

Naturally, one way to obtain the NLSM correlator is simply by applying the $\delta$-shift to the Tr$[\phi^3]$ correlator. It obviously yields the correct answer for the same reason we argued in Sec.~\ref{sec:NLSM_WF_FS}. It is of course also possible to build the correlator using \eqref{eq:Corr} but inserting the NLSM wavefunction which we learned how to compute in the previous sections. However, as it turns out, the frequency representation for the correlator is just as simple as for the wavefunction. 

To understand this, we recall the recent works pointing out how, in some aspects, the correlator, both in flat-space and de Sitter, is vastly simpler than the wavefunction \cite{Arkani-Hamed:2025mce,Chowdhury:2023arc,chowdhury_simplicity_2026,Donath:2024utn,Glew:2025mry}. For example, the flat-space correlator at tree-level has a time integral representation which can be obtained from the wavefunction by simply swapping the minus sign in the bulk-to-bulk propagator to a plus, thus we can effectively write that (at tree-level):
\begin{equation}
    \langle\Phi^n\rangle_G = 2\int_{-\infty(1-i \epsilon)}^{0}\prod_{v\in \mathcal{V}(G)} d\eta_i \; \phi_{k_v}(\eta_v)\prod_{e\in \mathcal{E}(G)} \tilde{G}_{BB}(\eta_{v_e},\eta_{v_e^{\prime}}; k_e) \; ,
    \label{eq:time_int_rep_Corr}
\end{equation}
where the tilde above the propagator simply stands for the change in sign:
\begin{equation}
    \tilde{G}_{BB}(\eta,\eta^\prime; k) = \frac{1}{2\,k}\left(e^{-i k |\eta-\eta^\prime|} +e^{i k (\eta+\eta^\prime)}\right)\, .
    \label{eq:BBProp_FS_corr}
\end{equation}
We refer the reader to \cite{Arkani-Hamed:2025mce,Glew:2025mry,chowdhury_simplicity_2026} in order to learn about these remarkable simplifications of the correlator. As for \eqref{eq:time_int_rep_Corr}-\eqref{eq:BBProp_FS_corr}, these are simply a result of the wavefunction being real for the theory under consideration. When the wavefunction is real, Re$[\Psi_k] =\Psi_k$, picking the boundary term in the Green's function is equivalent to taking the product of the lower point wavefunctions on either side of the $G_{BB}$. This is exactly what the second term in eq.~\eqref{eq:Corr} is doing as well. Therefore, there will be cancellations of terms when computing the correlator, in the end these cancellations amount exactly into computing the correlator as in \eqref{eq:time_int_rep_Corr}-\eqref{eq:BBProp_FS_corr}. Since the NLSM wavefunction in de Sitter is also real, this will also hold in for de Sitter, which we discuss in the next section. 

Considering that the tree-level correlator is simply obtained by turning the minus signs in the bulk-to-bulk propagators into plus signs, then the frequency representation for the bulk-to-bulk propagator is simply:
\begin{equation}
    \tilde{G}_{BB}(\eta,\eta^\prime; k) = -\frac{1}{2}\int_{-\infty}^{\infty}\frac{d\omega}{2 \pi i} \,\frac{(e^{i \omega \eta}+e^{-i \omega \eta})(e^{i \omega \eta^\prime}+e^{-i \omega \eta^\prime})}{\omega^2 - k^2 + i \epsilon}\, .
    \label{eq:freq_rep_FS_Corr}
\end{equation}
The factored time integrals are just obtained by swapping the minus signs in eqn.\eqref{eq:3ptFactors}, to obtain:
\begin{equation}
\begin{aligned}
    & \left.\begin{gathered}
\begin{tikzpicture}[
    line cap=round,
    line width=1pt,
    offshell/.style={
        draw=Maroon,
        line width=1.9pt
    },
    onshell/.style={
        draw=black,
        line width=0.9pt
    },
]
    % External legs
    \draw[onshell] (-0.7, 0.6) -- (0,0);
    \draw[onshell] (-0.7,-0.6) -- (0,0);
    \draw[offshell] (0,0) -- (0.9,0);

    % Labels
    \node at (-0.98, 0.55) {$k$};
    \node at (0.1, -0.35) {$\eta$};
    \node at (-0.98,-0.55) {$k^\prime$};
    \node at ( 1.3, 0) {$\omega_0$};

    % Black vertex
    \filldraw[black] (0,0) circle (3.2pt);
\end{tikzpicture}
    \end{gathered} \, \, \right]\,\,  \begin{aligned}  C^{(1)}_{\text{Corr}}(k,k^\prime; \omega_0)& = \int_{-\infty}^0 d \eta \, e^{i(k+k^\prime)\eta} (e^{i\omega_0\eta}+e^{-i\omega_0\eta}) \\
    &=\frac{2 i (k+k^\prime)}{\omega_0^2-(k+k^\prime)^2}\end{aligned} \\[0.8em]
&\left. \begin{gathered}
\begin{tikzpicture}[
    line cap=round,
    line width=1pt,
    offshell/.style={
        draw=Maroon,
        line width=1.9pt
    },
    onshell/.style={
        draw=black,
        line width=0.9pt
    },
]
    % External legs
    \draw[onshell] (-0.7, 0.6) -- (0,0);
    \draw[offshell] (-0.7,-0.6) -- (0,0);
    \draw[offshell] (0,0) -- (0.9,0);

    % Labels
    \node at (-0.98, 0.55) {$k$};
     \node at (0.1, -0.35) {$\eta$};
    \node at (-0.98,-0.65) {$\omega_1$};
    \node at ( 1.3, 0) {$\omega_0$};

    % Black vertex
    \filldraw[black] (0,0) circle (3.2pt);
\end{tikzpicture}
    \end{gathered} \, \, \right]\, \, \begin{aligned} C^{(2)}_{\text{Corr}}(k;\omega_0,\omega_1) &=\int_{-\infty}^0 d \eta \, e^{ik\eta} \prod_{j=0,1}(e^{i\omega_j\eta}+e^{-i\omega_j\eta}) \\
    &=\frac{4 i k \left(-k^2+\omega _0^2+\omega _1^2\right)}{\left((\omega_0+\omega_1)^2-k^2 + i \epsilon \right)\left((\omega_0-\omega_1)^2-k^2+i\epsilon \right)} \end{aligned} 
\\[0.8em]
&\left. \begin{gathered}
\begin{tikzpicture}[
    line cap=round,
    line width=1pt,
    offshell/.style={
        draw=Maroon,
        line width=1.9pt
    },
    onshell/.style={
        draw=black,
        line width=0.9pt
    },
]
    % External legs
    \draw[offshell] (-0.7, 0.6) -- (0,0);
    \draw[offshell] (-0.7,-0.6) -- (0,0);
    \draw[offshell] (0,0) -- (0.9,0);

    % Labels
    \node at (-0.98, 0.55) {$\omega_2$};
     \node at (0.1, -0.35) {$\eta$};
    \node at (-0.98,-0.65) {$\omega_1$};
    \node at ( 1.3, 0) {$\omega_0$};

    % Black vertex
    \filldraw[black] (0,0) circle (3.2pt);
\end{tikzpicture}  
    \end{gathered}\, \, \right] \, \,  \begin{aligned} C^{(3)}_{\text{Corr}}(\omega_0,\omega_1,\omega_2) &=\int_{-\infty}^0 d \eta \,  \prod_{j=0,1,2}(e^{i\omega_j\eta}+e^{-i\omega_j\eta}) \\
    &=\sum_{\sigma_{j}=\pm 1}  \frac{-i}{(\sigma_0 \omega_0 + \sigma_1 \omega_1 +\sigma_2 \omega_2) - i \epsilon} \end{aligned} 
\end{aligned}
\label{eq:three_point_funcs_Corr}
\end{equation}
Naturally, the frequency representation for the correlator will also have the same subtleties as in the wavefunction, which were discussed in Sec.~\ref{sec:Freq_Rep_FS}. The resolution of these subtleties is the same, given that the contributions are boundary local, then the correlators with versus without the pole at infinity differ by boundary field redefinitions. 

Having understood the wavefunction computation, then there is no mystery about the correlators, either through the frequency representation or using \eqref{eq:Corr}. It is nevertheless interesting to note that the simplifications observed in \cite{Arkani-Hamed:2025mce} are also observed for the NLSM correlator in flat-space. For example, take the correlator contribution from the graph:
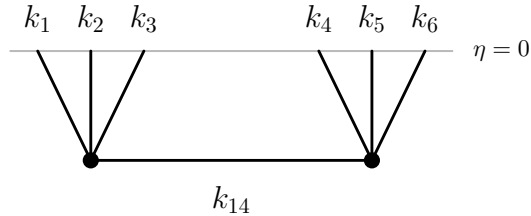
\begin{figure}[H]
\centering

\resizebox{0.45\textwidth}{!}{%
\begin{tikzpicture}[
    propagator/.style={
        draw=black,
        line width=1.3pt,
        line cap=round
    },
    boundary/.style={
        draw=gray!55,
        line width=0.9pt
    },
    vertex/.style={
        circle,
        fill=black,
        inner sep=2.6pt
    },
    momentum/.style={
        font=\Large
    }
]

% Bulk vertices
\coordinate (XL) at (-2.25,0);
\coordinate (XR) at ( 2.25,0);

% Boundary endpoints
\coordinate (K1) at (-3.10,1.75);
\coordinate (K2) at (-2.25,1.75);
\coordinate (K3) at (-1.40,1.75);

\coordinate (K4) at ( 1.40,1.75);
\coordinate (K5) at ( 2.25,1.75);
\coordinate (K6) at ( 3.10,1.75);

% eta = 0 boundary
\draw[boundary]
    (-3.55,1.75) -- (3.55,1.75)
    node[right=5pt,black] {$\eta=0$};

% External lines
\draw[propagator] (K1)--(XL);
\draw[propagator] (K2)--(XL);
\draw[propagator] (K3)--(XL);

\draw[propagator] (K4)--(XR);
\draw[propagator] (K5)--(XR);
\draw[propagator] (K6)--(XR);

% Internal line
\draw[propagator] (XL)--(XR);

% Vertices
\node[vertex] at (XL) {};
\node[vertex] at (XR) {};

% External momentum labels
\node[momentum,above=5pt] at (K1) {$k_1$};
\node[momentum,above=5pt] at (K2) {$k_2$};
\node[momentum,above=5pt] at (K3) {$k_3$};

\node[momentum,above=5pt] at (K4) {$k_4$};
\node[momentum,above=5pt] at (K5) {$k_5$};
\node[momentum,above=5pt] at (K6) {$k_6$};

% Internal momentum label
\node[momentum,below=8pt] at (0,0) {$k_{14}$};

\end{tikzpicture}%
}

\end{figure}
If we perform the re-scaling every external energy by $k_i \to \epsilon \, k_i$, and then expand around $\epsilon=0$ we obtain:
\begin{equation}
\langle\phi^6\rangle_{k_{14}}\sim \frac{2(k_{13}^2 + k_{24}^2) (k_{15}^2 + k_{46}^2)}{k_{14}^2(k_1+k_2+k_3+k_4+k_5+k_6)\epsilon^1} + \mathcal{O}(\epsilon)\, ,
\label{eq:ET_rescale}
\end{equation}
where we can see that we obtain the corresponding six-point scattering amplitude, and the term $\epsilon^0$ vanishes. Furthermore, if we do the same re-scaling for a different pole, $k_1\to\epsilon k_1,\,k_2\to\epsilon k_2,\,k_3\to\epsilon k_3,\, k_{14}\to\epsilon k_{14}$, we obtain:
\begin{equation}
    \langle\phi^6\rangle_{k_{14}}\sim \frac{2(k_{13}^2 + k_{24}^2) (k_{15}^2 + k_{46}^2 + k_{4} k_{5} + 2 k_{4} k_{6} + k_{5} k_{6})}{k_{14} (k_1 + k_{14} + k_2 + k_3) (k_4 + k_5 + k_6)\epsilon^2} + \mathcal{O}(\epsilon^{0})\, ,
    \label{eq:4pt_WF_Eps_Exp}
\end{equation}
which is precisely the factorization of the four-point amplitude and the four-point correlator with the internal edge set to zero, as predicted in \cite{Arkani-Hamed:2025mce}.
Finally, there is one last simplification we expect to see, if we set all but one external energy (any single one), $k_i$, to zero, then we obtain:
\begin{equation}
    \langle\phi^6\rangle_{k_{14}}\sim \frac{2(k_{13}^2 + k_{24}^2) (k_{15}^2 + k_{46}^2)}{k_{14}^2\,k_i}\, ,
\end{equation}
which is precisely the scattering amplitude. This discussion shows that there is yet another way to compute the NLSM correlator: we can simply bootstrap it by imposing the sub-leading zeros in the pole expansions, which is dramatically simpler than imposing factorization structure for each pole. 

\subsection{De Sitter correlators}

Now, let us consider correlators in de Sitter. The formula \eqref{eq:Corr} is always valid, so once we compute the required wavefunctions we can simply obtain the correlator. However, as stated in Sec.~\ref{sec:DSSpacePsi} the de Sitter tree-level wavefunction for the NLSM is real\footnote{Assuming we are regulating the boundary power divergences appropriately, which with a cut-off regulator are imaginary parts in the wavefunction.}\cite{Goodhew:2024eup}. Then our assumptions of the previous section continue to hold, and we can compute correlators directly by swapping the minus signs to plus signs in the frequency representation as discussed above. Therefore, our three-point building blocks for the correlator in de Sitter are simply :
\begin{equation}
\begin{aligned}
    & \left.\begin{gathered}
\begin{tikzpicture}[
    line cap=round,
    line width=1pt,
    offshell/.style={
        draw=Maroon,
        line width=1.9pt
    },
    onshell/.style={
        draw=black,
        line width=0.9pt
    },
]
    % External legs
    \draw[onshell] (-0.7, 0.6) -- (0,0);
    \draw[onshell] (-0.7,-0.6) -- (0,0);
    \draw[offshell] (0,0) -- (0.9,0);

    % Labels
    \node at (-0.98, 0.55) {$k$};
    \node at (0.1, -0.35) {$\eta$};
    \node at (-0.98,-0.55) {$k^\prime$};
    \node at ( 1.3, 0) {$\omega_0$};

    % Black vertex
    \filldraw[black] (0,0) circle (3.2pt);
\end{tikzpicture}
    \end{gathered} \, \, \right]\,\,  \begin{aligned}  C^{(1), dS}_{\text{Corr}}(k,k^\prime; \omega_0)& = \int_{-\infty}^0 \frac{d \eta}{\eta^{2+\alpha}} \, \phi_{k}^{\text{dS}}(\eta)\phi_{k^{\prime}}(\eta) (\phi_{\omega_0}^{\text{dS}}(\eta)+\phi_{-\omega_0}^{\text{dS}}(\eta)) \\
    &=\frac{2 i\left(k^3+{k^{\prime}}^3\right) \omega _0^2-2 i \left(k+k^{\prime}\right){}^3 \left(k^2+k^{\prime} k+{k^{\prime}}^2\right)}{((k+k^{\prime})^2-\omega_0^2)^2}\, ,\end{aligned} \\[0.8em]
&\left. \begin{gathered}
\begin{tikzpicture}[
    line cap=round,
    line width=1pt,
    offshell/.style={
        draw=Maroon,
        line width=1.9pt
    },
    onshell/.style={
        draw=black,
        line width=0.9pt
    },
]
    % External legs
    \draw[onshell] (-0.7, 0.6) -- (0,0);
    \draw[offshell] (-0.7,-0.6) -- (0,0);
    \draw[offshell] (0,0) -- (0.9,0);

    % Labels
    \node at (-0.98, 0.55) {$k$};
     \node at (0.1, -0.35) {$\eta$};
    \node at (-0.98,-0.65) {$\omega_1$};
    \node at ( 1.3, 0) {$\omega_0$};

    % Black vertex
    \filldraw[black] (0,0) circle (3.2pt);
\end{tikzpicture}
    \end{gathered} \, \, \right]\, \, \begin{aligned} C^{(2), dS}_{\text{Corr}}(k;\omega_0,\omega_1) &=\int_{-\infty}^0 \frac{d \eta}{\eta^{2+\alpha}} \, \phi_{k}^{\text{dS}}(\eta) \prod_{j=0,1}(\phi_{\omega_j}^{\text{dS}}(\eta)+\phi_{-\omega_j}^{\text{dS}}(\eta)) \\
    &=\frac{-4 i k^3 \left(k^2-\omega _0^2-\omega _1^2\right) \left(-2 \omega _1^2 \left(k^2+5 \omega _0^2\right)+\left(k^2-\omega _0^2\right){}^2+\omega_1^4\right)}{\left((\omega_0+\omega_1)^2-k^2 + i \epsilon \right)^2\left((\omega_0-\omega_1)^2-k^2+i\epsilon \right)^2}\, ,\end{aligned} 
\\[0.8em]
&\left. \begin{gathered}
\begin{tikzpicture}[
    line cap=round,
    line width=1pt,
    offshell/.style={
        draw=Maroon,
        line width=1.9pt
    },
    onshell/.style={
        draw=black,
        line width=0.9pt
    },
]
    % External legs
    \draw[offshell] (-0.7, 0.6) -- (0,0);
    \draw[offshell] (-0.7,-0.6) -- (0,0);
    \draw[offshell] (0,0) -- (0.9,0);

    % Labels
    \node at (-0.98, 0.55) {$\omega_2$};
     \node at (0.1, -0.35) {$\eta$};
    \node at (-0.98,-0.65) {$\omega_1$};
    \node at ( 1.3, 0) {$\omega_0$};

    % Black vertex
    \filldraw[black] (0,0) circle (3.2pt);
\end{tikzpicture}  
    \end{gathered}\, \, \right] \, \,  \begin{aligned} C^{(3), dS}_{\text{Corr}}(\omega_0,\omega_1,\omega_2) &=\int_{-\infty}^0 \frac{d \eta}{\eta^{2+\alpha}} \,  \prod_{j=0,1,2}(\phi_{\omega_j}^{\text{dS}}(\eta)+\phi_{-\omega_j}^{\text{dS}}(\eta)) \\
    &=\sum_{\sigma_{j}=\pm 1}  i \psi_3(\sigma_0 \omega_0,\sigma_1 \omega_1,\sigma_2 \omega_2)\, , \end{aligned}
\end{aligned}
\label{eq:3ptFactors_dS_Corr}
\end{equation}
where: 
\begin{equation}
    \psi_3(k_1,k_2,k_3) = \frac{k_1 k_2 k_3}{\left(k_1+k_2+k_3-i \epsilon \right){}^2}+\frac{k_1 k_2+k_3 k_2+k_1 k_3}{k_1+k_2+k_3-i \epsilon }-k_1-k_2-k_3\, .
\end{equation}
Also, for de Sitter we are interested in checking whether the simplifications observed in the flat-space correlator hold. Therefore, we start by re-scaling the external energies of the four-point de Sitter, which for four points can be read off eq.~\eqref{eq:four_point_WF_dS}, given that the correlator is simply twice the real part of the wavefunction in this case. The re-scaling is the same as in eq.~\eqref{eq:ET_rescale}, which is $k_i \to \epsilon \, k_i$, and what we obtain is:
\begin{equation}
\langle\phi^4\rangle
\sim
-2\epsilon \left(
k_{13}^2
+
k_{24}^2
\right)
\left(
E_T-\frac{\sigma_2}{E_T}
-\frac{\sigma_3}{E_T^2}
-\frac{2\sigma_4}{E_T^3}
\right)
+ \mathcal{O}(\epsilon^3) \, ,
\label{eq:four_point_Corr_Eps_Exp}
\end{equation}
where the notation was defined in eq.~\eqref{eq:4pt_polys_notation}. Obviously, the amplitude emerges as a factor in the leading term, but once again the sub-leading term in $\epsilon$ vanishes! One can check that the same is true for the six-point correlator; we leave a more systematic investigation for future work.  

\section{Outlook}
In this note, we showed that the simple kinematic shift that relates Tr$[\phi^3]$ amplitudes and NLSM amplitudes extends to the vacuum wavefunction, and equal time correlators both in a Minkowski background and a de Sitter background. 

In flat-space, we showed how the square root shift, $k_{ij}\to\sqrt{k_{ij}^2\pm\delta^2}$, on the Tr$[\phi^3]$ wavefunction reproduces the correct NLSM wavefunction, and, using the frequency representation, we derived a simple universal behavior for the wavefunction under the soft limit. Of course, one naturally expects the soft limit to be ruled by lower-point wavefunctions, which can be derived for each multiplicity from the Ward identity. Nevertheless, our result holds for any multiplicity and gives a simple interpretation for the functions appearing in the soft limit: as related to wavefunctions in the mixed theory, Tr$(\phi^3)\oplus$NLSM. 

In de Sitter, remarkably, we find that using the frequency representation, we can fully recast the $\delta$-shifted problem in a completely analogous way to that of flat-space. Even though the  $\delta$-shift ought to produce the correct NLSM wavefunction in de Sitter, the technical complexity of de Sitter computations made it naively challenging to implement this limit. After all, the massless Tr$[\phi^3]$ wavefunction at tree-level is a transcendental function with physical logarithmic divergences. Irrespective of this, once we found the correct way to represent it via the frequency representation, we automatically see that all the properties (recursion and soft-behavior) automatically generalize to de Sitter! 

This work leaves several avenues of research in the open, some of which we discuss here. First and foremost is the generalization of this construction to loop-level. In \cite{Arkani-Hamed:2024nhp}, the authors generalized the $\delta$-shift for all loop order amplitudes in the NLSM, and it is clear that the flat-space loop wavefunction, and correlators, can be obtained in a similar way. One interesting aspect, already present at one-loop, is that using the generalization of kinematics associated to curves on surfaces --\textit{ surface kinematics} \cite{SurfaceKin} -- plays a crucial role in seeing the simple behavior on the soft limit at integrand level for both pion and gluon amplitudes  \cite{Splits,Backus:2025njt}. It would be interesting to understand whether the same generalization of kinematics for the wavefunction lets us extend the soft-limits found at tree-level to the loop-integrands. Finally, it would be very interesting to understand how the loop-level $\delta$-shift generalizes to de Sitter.

Then, in Sec.~\ref{sec:SqrtShift_FS}, we discussed a suggestive way to write the wavefunction as an object that manifestly becomes the amplitude in the $E_T\to 0$ limit, plus an $E_T$ proportional deformation, in the form of $Q_{L/R}$ factors. We found that this deformation's complexity keeps increasing at higher points, and were not able to identify an especially simple pattern. There are of course many different ways to parametrize the kinematics -- in terms of perimeters, $k_{i,j}$, $\dots$ -- so it would be interesting to find one which most transparently encodes the structure of this numerators, as this constitutes a possible pathway to better understanding the combinatorics of the numerators of derivative interactions (in this case $\phi \phi (\partial\phi)^2$) -- which, in general, have received substantially less attention, when compared to polynomial interactions. 

Finally, here we only discussed massless Goldstone bosons, however in inflation we are usually interested in having a potential which generates a nearly constant energy density that drives the expansion, for example slow-roll inflation. By breaking the symmetry (adding a mass term) we can have a non-zero potential, therefore an important extension of this work would be to consider light Goldstone bosons. 
\\ \\
\textbf{Acknowledgments ---} {We thank Nima Arkani-Hamed, Ross Glew, Austin Joyce, Enrico Pajer, Guilherme Pimentel,  Kostas Skenderis and ChatGPT for discussions. C.F. is supported by the Society of Fellows at Harvard University. J.M. is supported by the European Union (ERC, UNIVERSE PLUS, 101118787). Y.M. is supported by the National Natural Science Foundation of China under Grant Nos.~12225510 and 12447101, and by the New Cornerstone Science Foundation. F.V. is supported by the Institute for Advanced Study and an anonymous donor. F.V also thanks the organizers of the meeting ``Celestial meets de Sitter Holography II" for the opportunity to present a preliminary version of this work. Views and opinions expressed are, however, those of the authors only and do not necessarily reflect those of the European Union or the European Research Council Executive Agency. Neither the European Union nor the granting authority can be held responsible for them.}

\bibliography{refs.bib} 

\providecommand{\href}[2]{#2}\begingroup\raggedright\begin{thebibliography}{10}

\bibitem{Arkani-Hamed:2017fdk}
N.~Arkani-Hamed, P.~Benincasa and A.~Postnikov, \emph{{Cosmological Polytopes and the Wavefunction of the Universe}},  \href{https://arxiv.org/abs/1709.02813}{{\ttfamily 1709.02813}}.

\bibitem{Cosmohedron}
N.~Arkani-Hamed, C.~Figueiredo and F.~Vaz{\~a}o, \emph{{Cosmohedra}}, \href{https://doi.org/10.1007/JHEP11(2025)029}{\emph{JHEP} {\bfseries 11} (2025) 029} [\href{https://arxiv.org/abs/2412.19881}{{\ttfamily 2412.19881}}].

\bibitem{Figueiredo:2025daa}
C.~Figueiredo and F.~Vaz{\~a}o, \emph{{Correlator Polytopes}},  \href{https://arxiv.org/abs/2506.19907}{{\ttfamily 2506.19907}}.

\bibitem{KinFlow}
N.~Arkani-Hamed, D.~Baumann, A.~Hillman, A.~Joyce, H.~Lee and G.L.~Pimentel, \emph{{Kinematic Flow and the Emergence of Time}}, \href{https://doi.org/10.1103/dsjm-tckw}{\emph{Phys. Rev. Lett.} {\bfseries 135} (2025) 031602} [\href{https://arxiv.org/abs/2312.05300}{{\ttfamily 2312.05300}}].

\bibitem{GeometryCosmoCorr}
P.~Benincasa and G.~Dian, \emph{{The Geometry of Cosmological Correlators}}, \href{https://doi.org/10.21468/SciPostPhys.18.3.105}{\emph{SciPost Phys.} {\bfseries 18} (2025) 105} [\href{https://arxiv.org/abs/2401.05207}{{\ttfamily 2401.05207}}].

\bibitem{Benincasa:2024lxe}
P.~Benincasa and F.~Vaz{\~a}o, \emph{{The asymptotic structure of cosmological integrals}}, \href{https://doi.org/10.21468/SciPostPhys.19.2.029}{\emph{SciPost Phys.} {\bfseries 19} (2025) 029} [\href{https://arxiv.org/abs/2402.06558}{{\ttfamily 2402.06558}}].

\bibitem{Arkani-Hamed:2018kmz}
N.~Arkani-Hamed, D.~Baumann, H.~Lee and G.L.~Pimentel, \emph{{The Cosmological Bootstrap: Inflationary Correlators from Symmetries and Singularities}}, \href{https://doi.org/10.1007/JHEP04(2020)105}{\emph{JHEP} {\bfseries 04} (2020) 105} [\href{https://arxiv.org/abs/1811.00024}{{\ttfamily 1811.00024}}].

\bibitem{Baumann:2019oyu}
D.~Baumann, C.~Duaso~Pueyo, A.~Joyce, H.~Lee and G.L.~Pimentel, \emph{{The cosmological bootstrap: weight-shifting operators and scalar seeds}}, \href{https://doi.org/10.1007/JHEP12(2020)204}{\emph{JHEP} {\bfseries 12} (2020) 204} [\href{https://arxiv.org/abs/1910.14051}{{\ttfamily 1910.14051}}].

\bibitem{Goodhew:2020hob}
H.~Goodhew, S.~Jazayeri and E.~Pajer, \emph{{The Cosmological Optical Theorem}}, \href{https://doi.org/10.1088/1475-7516/2021/04/021}{\emph{JCAP} {\bfseries 04} (2021) 021} [\href{https://arxiv.org/abs/2009.02898}{{\ttfamily 2009.02898}}].

\bibitem{Melville:2021lst}
S.~Melville and E.~Pajer, \emph{{Cosmological Cutting Rules}}, \href{https://doi.org/10.1007/JHEP05(2021)249}{\emph{JHEP} {\bfseries 05} (2021) 249} [\href{https://arxiv.org/abs/2103.09832}{{\ttfamily 2103.09832}}].

\bibitem{Baumann:2024ttn}
D.~Baumann, G.~Mathys, G.L.~Pimentel and F.~Rost, \emph{{A New Twist on Spinning (A)dS Correlators}},  \href{https://arxiv.org/abs/2408.02727}{{\ttfamily 2408.02727}}.

\bibitem{DiffEq_CosmoCorr}
N.~Arkani-Hamed, D.~Baumann, A.~Hillman, A.~Joyce, H.~Lee and G.L.~Pimentel, \emph{{Differential equations for cosmological correlators}}, \href{https://doi.org/10.1007/JHEP09(2025)009}{\emph{JHEP} {\bfseries 09} (2025) 009} [\href{https://arxiv.org/abs/2312.05303}{{\ttfamily 2312.05303}}].

\bibitem{De:2024zic}
S.~De and A.~Pokraka, \emph{{A physical basis for cosmological correlators from cuts}}, \href{https://doi.org/10.1007/JHEP03(2025)040}{\emph{JHEP} {\bfseries 03} (2025) 040} [\href{https://arxiv.org/abs/2411.09695}{{\ttfamily 2411.09695}}].

\bibitem{Glew:2025ypb}
R.~Glew and A.~Pokraka, \emph{{Kinematic flow from the flow of cuts}},  \href{https://arxiv.org/abs/2508.11568}{{\ttfamily 2508.11568}}.

\bibitem{Cespedes:2023aal}
S.~C\'espedes, A.-C.~Davis and D.-G.~Wang, \emph{{On the IR Divergences in de Sitter Space: loops, resummation and the semi-classical wavefunction}},  \href{https://arxiv.org/abs/2311.17990}{{\ttfamily 2311.17990}}.

\bibitem{korte_cosmological_2026}
C.d.~Korte, H.~Goodhew, K.S.~Vaziri and N.~Weiss, \emph{Cosmological {Weight}-{Shifting} {Matrices}},  May, 2026.
\newblock 10.48550/arXiv.2605.30475.

\bibitem{baumann_differential_2026}
D.~Baumann, A.~Joyce, H.~Lee and K.S.~Vaziri, \emph{Differential {Equations} for {Massive} {Correlators}},  2026.
\newblock 10.48550/ARXIV.2604.08658.

\bibitem{LoopInts}
P.~Benincasa, G.~Brunello, M.K.~Mandal, P.~Mastrolia and F.~Vaz{\~a}o, \emph{{One-loop corrections to the Bunch-Davies wave function of the universe}}, \href{https://doi.org/10.1103/PhysRevD.111.085016}{\emph{Phys. Rev. D} {\bfseries 111} (2025) 085016} [\href{https://arxiv.org/abs/2408.16386}{{\ttfamily 2408.16386}}].

\bibitem{Zeros}
N.~Arkani-Hamed, Q.~Cao, J.~Dong, C.~Figueiredo and S.~He, \emph{{Hidden zeros for particle/string amplitudes and the unity of colored scalars, pions and gluons}},  \href{https://arxiv.org/abs/2312.16282}{{\ttfamily 2312.16282}}.

\bibitem{Arkani-Hamed:2024nhp}
N.~Arkani-Hamed, Q.~Cao, J.~Dong, C.~Figueiredo and S.~He, \emph{{Nonlinear Sigma model amplitudes to all loop orders are contained in the Tr({\ensuremath{\Phi}}3) theory}}, \href{https://doi.org/10.1103/PhysRevD.110.065018}{\emph{Phys. Rev. D} {\bfseries 110} (2024) 065018} [\href{https://arxiv.org/abs/2401.05483}{{\ttfamily 2401.05483}}].

\bibitem{Cachazo:2016njl}
F.~Cachazo, P.~Cha and S.~Mizera, \emph{{Extensions of Theories from Soft Limits}}, \href{https://doi.org/10.1007/JHEP06(2016)170}{\emph{JHEP} {\bfseries 06} (2016) 170} [\href{https://arxiv.org/abs/1604.03893}{{\ttfamily 1604.03893}}].

\bibitem{Splits}
N.~Arkani-Hamed and C.~Figueiredo, \emph{{All-order splits and multi-soft limits for particle and string amplitudes}},  \href{https://arxiv.org/abs/2405.09608}{{\ttfamily 2405.09608}}.

\bibitem{Anninos:2014lwa}
D.~Anninos, T.~Anous, D.Z.~Freedman and G.~Konstantinidis, \emph{{Late-time Structure of the Bunch-Davies De Sitter Wavefunction}}, \href{https://doi.org/10.1088/1475-7516/2015/11/048}{\emph{JCAP} {\bfseries 11} (2015) 048} [\href{https://arxiv.org/abs/1406.5490}{{\ttfamily 1406.5490}}].

\bibitem{KinFlowLoop}
D.~Baumann, H.~Goodhew and H.~Lee, \emph{{Kinematic flow for cosmological loop integrands}}, \href{https://doi.org/10.1007/JHEP07(2025)131}{\emph{JHEP} {\bfseries 07} (2025) 131} [\href{https://arxiv.org/abs/2410.17994}{{\ttfamily 2410.17994}}].

\bibitem{CosmoReview}
P.~Benincasa, \emph{{Amplitudes meet Cosmology: A (Scalar) Primer}},  \href{https://arxiv.org/abs/2203.15330}{{\ttfamily 2203.15330}}.

\bibitem{PajerUnitarityLocality}
S.~Jazayeri, E.~Pajer and D.~Stefanyszyn, \emph{{From locality and unitarity to cosmological correlators}}, \href{https://doi.org/10.1007/JHEP10(2021)065}{\emph{JHEP} {\bfseries 10} (2021) 065} [\href{https://arxiv.org/abs/2103.08649}{{\ttfamily 2103.08649}}].

\bibitem{GoodhewCutCosmoCorr}
H.~Goodhew, S.~Jazayeri, M.H.G.~Lee and E.~Pajer, \emph{{Cutting cosmological correlators}}, \href{https://doi.org/10.1088/1475-7516/2021/08/003}{\emph{JCAP} {\bfseries 08} (2021) 003} [\href{https://arxiv.org/abs/2104.06587}{{\ttfamily 2104.06587}}].

\bibitem{Benincasa:2025uie}
P.~Benincasa and F.~Vaz{\~a}o, \emph{{Observables in Expanding Universes}}, \href{https://doi.org/10.1007/978-3-031-90352-6_4}{\emph{Lect. Notes Phys.} {\bfseries 1041} (2025) 131}.

\bibitem{CosmoSMatrix}
S.~Melville and G.L.~Pimentel, \emph{{A de Sitter S-matrix from amputated cosmological correlators}}, \href{https://doi.org/10.1007/JHEP08(2024)211}{\emph{JHEP} {\bfseries 08} (2024) 211} [\href{https://arxiv.org/abs/2404.05712}{{\ttfamily 2404.05712}}].

\bibitem{Sachs2}
T.~Heckelbacher, I.~Sachs, E.~Skvortsov and P.~Vanhove, \emph{{Analytical evaluation of cosmological correlation functions}}, \href{https://doi.org/10.1007/JHEP08(2022)139}{\emph{JHEP} {\bfseries 08} (2022) 139} [\href{https://arxiv.org/abs/2204.07217}{{\ttfamily 2204.07217}}].

\bibitem{CosmoTreeTheo}
S.~Agui~Salcedo and S.~Melville, \emph{{The cosmological tree theorem}}, \href{https://doi.org/10.1007/JHEP12(2023)076}{\emph{JHEP} {\bfseries 12} (2023) 076} [\href{https://arxiv.org/abs/2308.00680}{{\ttfamily 2308.00680}}].

\bibitem{Meltzer}
D.~Meltzer, \emph{{The inflationary wavefunction from analyticity and factorization}}, \href{https://doi.org/10.1088/1475-7516/2021/12/018}{\emph{JCAP} {\bfseries 12} (2021) 018} [\href{https://arxiv.org/abs/2107.10266}{{\ttfamily 2107.10266}}].

\bibitem{Bittermann:2022nfh}
N.~Bittermann and A.~Joyce, \emph{{Soft limits of the wavefunction in exceptional scalar theories}}, \href{https://doi.org/10.1007/JHEP03(2023)092}{\emph{JHEP} {\bfseries 03} (2023) 092} [\href{https://arxiv.org/abs/2203.05576}{{\ttfamily 2203.05576}}].

\bibitem{CespedesScott}
S.~C\'espedes, A.-C.~Davis and S.~Melville, \emph{{On the time evolution of cosmological correlators}}, \href{https://doi.org/10.1007/JHEP02(2021)012}{\emph{JHEP} {\bfseries 02} (2021) 012} [\href{https://arxiv.org/abs/2009.07874}{{\ttfamily 2009.07874}}].

\bibitem{Sachs3}
C.~Chowdhury, A.~Lipstein, J.~Marshall, J.~Mei and I.~Sachs, \emph{{Cosmological Dressing Rules}},  \href{https://arxiv.org/abs/2503.10598}{{\ttfamily 2503.10598}}.

\bibitem{HiddenZerosWF}
S.~De, S.~Paranjape, A.~Pokraka, M.~Spradlin and A.~Volovich, \emph{{Hidden zeros of the cosmological wavefunction}}, \href{https://doi.org/10.1007/JHEP07(2025)174}{\emph{JHEP} {\bfseries 07} (2025) 174} [\href{https://arxiv.org/abs/2503.23579}{{\ttfamily 2503.23579}}].

\bibitem{Baumann:2025qjx}
D.~Baumann, H.~Goodhew, A.~Joyce, H.~Lee, G.L.~Pimentel and T.~Westerdijk, \emph{{Geometry of Kinematic Flow}},  \href{https://arxiv.org/abs/2504.14890}{{\ttfamily 2504.14890}}.

\bibitem{Glew:2025otn}
R.~Glew and T.~Lukowski, \emph{{Amplitubes: graph cosmohedra}}, \href{https://doi.org/10.1007/JHEP09(2025)074}{\emph{JHEP} {\bfseries 09} (2025) 074} [\href{https://arxiv.org/abs/2502.17564}{{\ttfamily 2502.17564}}].

\bibitem{Glew:2025ugf}
R.~Glew, \emph{{Wavefunction coefficients from amplitubes}}, \href{https://doi.org/10.1007/JHEP07(2025)064}{\emph{JHEP} {\bfseries 07} (2025) 064} [\href{https://arxiv.org/abs/2503.13596}{{\ttfamily 2503.13596}}].

\bibitem{Benincasa:2019vqr}
P.~Benincasa, \emph{{Cosmological Polytopes and the Wavefuncton of the Universe for Light States}},  \href{https://arxiv.org/abs/1909.02517}{{\ttfamily 1909.02517}}.

\bibitem{Fan:2024iek}
B.~Fan and Z.-Z.~Xianyu, \emph{{Cosmological amplitudes in power-law FRW universe}}, \href{https://doi.org/10.1007/JHEP12(2024)042}{\emph{JHEP} {\bfseries 12} (2024) 042} [\href{https://arxiv.org/abs/2403.07050}{{\ttfamily 2403.07050}}].

\bibitem{Arkani-Hamed:2025mce}
N.~Arkani-Hamed, R.~Glew and F.~Vaz{\~a}o, \emph{{Correlators are simpler than wavefunctions}},  \href{https://arxiv.org/abs/2512.23795}{{\ttfamily 2512.23795}}.

\bibitem{Ardila-Mantilla:2026cbo}
F.~Ardila{\nobreakdash-}Mantilla, N.~Arkani{\nobreakdash-}Hamed, C.~Figueiredo and F.~Vaz{\~a}o, \emph{{Combinatorics of the Cosmohedron}},  \href{https://arxiv.org/abs/2603.03425}{{\ttfamily 2603.03425}}.

\bibitem{Bunch:1978yq}
T.S.~Bunch and P.C.W.~Davies, \emph{{Quantum Field Theory in de Sitter Space: Renormalization by Point Splitting}}, \href{https://doi.org/10.1098/rspa.1978.0060}{\emph{Proc. Roy. Soc. Lond. A} {\bfseries 360} (1978) 117}.

\bibitem{Allen:1985ux}
B.~Allen, \emph{{Vacuum States in de Sitter Space}}, \href{https://doi.org/10.1103/PhysRevD.32.3136}{\emph{Phys. Rev. D} {\bfseries 32} (1985) 3136}.

\bibitem{Ford:1984hs}
L.H.~Ford, \emph{{Quantum Instability of De Sitter Space-time}}, \href{https://doi.org/10.1103/PhysRevD.31.710}{\emph{Phys. Rev. D} {\bfseries 31} (1985) 710}.

\bibitem{Gorbenko:2019rza}
V.~Gorbenko and L.~Senatore, \emph{{$\lambda \phi^4$ in dS}},  \href{https://arxiv.org/abs/1911.00022}{{\ttfamily 1911.00022}}.

\bibitem{Burgess:2015ajz}
C.P.~Burgess, R.~Holman and G.~Tasinato, \emph{{Open EFTs, IR effects \textbackslash{}\& late-time resummations: systematic corrections in stochastic inflation}}, \href{https://doi.org/10.1007/JHEP01(2016)153}{\emph{JHEP} {\bfseries 01} (2016) 153} [\href{https://arxiv.org/abs/1512.00169}{{\ttfamily 1512.00169}}].

\bibitem{Rajaraman:2010xd}
A.~Rajaraman, \emph{{On the proper treatment of massless fields in Euclidean de Sitter space}}, \href{https://doi.org/10.1103/PhysRevD.82.123522}{\emph{Phys. Rev. D} {\bfseries 82} (2010) 123522} [\href{https://arxiv.org/abs/1008.1271}{{\ttfamily 1008.1271}}].

\bibitem{Beneke:2012kn}
M.~Beneke and P.~Moch, \emph{{On \textquotedblleft{}dynamical mass\textquotedblright{} generation in Euclidean de Sitter space}}, \href{https://doi.org/10.1103/PhysRevD.87.064018}{\emph{Phys. Rev. D} {\bfseries 87} (2013) 064018} [\href{https://arxiv.org/abs/1212.3058}{{\ttfamily 1212.3058}}].

\bibitem{Mirbabayi:2019qtx}
M.~Mirbabayi, \emph{{Infrared dynamics of a light scalar field in de Sitter}}, \href{https://doi.org/10.1088/1475-7516/2020/12/006}{\emph{JCAP} {\bfseries 12} (2020) 006} [\href{https://arxiv.org/abs/1911.00564}{{\ttfamily 1911.00564}}].

\bibitem{Baumgart:2020oby}
M.~Baumgart and R.~Sundrum, \emph{{Manifestly Causal In-In Perturbation Theory about the Interacting Vacuum}}, \href{https://doi.org/10.1007/JHEP03(2021)080}{\emph{JHEP} {\bfseries 03} (2021) 080} [\href{https://arxiv.org/abs/2010.10785}{{\ttfamily 2010.10785}}].

\bibitem{Benincasa:2024rfw}
P.~Benincasa and F.~Vaz{\~a}o, \emph{{Cosmological infrared subtractions {\&} infrared-safe computables}}, \href{https://doi.org/10.21468/SciPostPhys.18.6.176}{\emph{SciPost Phys.} {\bfseries 18} (2025) 176} [\href{https://arxiv.org/abs/2405.19979}{{\ttfamily 2405.19979}}].

\bibitem{Beneke:2023wmt}
M.~Beneke, P.~Hager and A.F.~Sanfilippo, \emph{{Cosmological Correlators in massless ${\phi}^4$-theory and the Method of Regions}},  \href{https://arxiv.org/abs/2312.06766}{{\ttfamily 2312.06766}}.

\bibitem{Cohen:2020php}
T.~Cohen and D.~Green, \emph{{Soft de Sitter Effective Theory}}, \href{https://doi.org/10.1007/JHEP12(2020)041}{\emph{JHEP} {\bfseries 12} (2020) 041} [\href{https://arxiv.org/abs/2007.03693}{{\ttfamily 2007.03693}}].

\bibitem{Kitamoto:2010et}
H.~Kitamoto and Y.~Kitazawa, \emph{{Non-linear sigma model in de Sitter space}}, \href{https://doi.org/10.1103/PhysRevD.83.104043}{\emph{Phys. Rev. D} {\bfseries 83} (2011) 104043} [\href{https://arxiv.org/abs/1012.5930}{{\ttfamily 1012.5930}}].

\bibitem{Armstrong:2022csc}
C.~Armstrong, H.~Gomez, R.~Lipinski~Jusinskas, A.~Lipstein and J.~Mei, \emph{{Effective field theories and cosmological scattering equations}}, \href{https://doi.org/10.1007/JHEP08(2022)054}{\emph{JHEP} {\bfseries 08} (2022) 054} [\href{https://arxiv.org/abs/2204.08931}{{\ttfamily 2204.08931}}].

\bibitem{Goodhew:2022ayb}
H.~Goodhew, \emph{{Rational wavefunctions in de Sitter spacetime}}, \href{https://doi.org/10.1088/1475-7516/2023/03/036}{\emph{JCAP} {\bfseries 03} (2023) 036} [\href{https://arxiv.org/abs/2210.09977}{{\ttfamily 2210.09977}}].

\bibitem{Skenderis:2002wp}
K.~Skenderis, \emph{{Lecture notes on holographic renormalization}}, \href{https://doi.org/10.1088/0264-9381/19/22/306}{\emph{Class. Quant. Grav.} {\bfseries 19} (2002) 5849} [\href{https://arxiv.org/abs/hep-th/0209067}{{\ttfamily hep-th/0209067}}].

\bibitem{Goodhew:2024eup}
H.~Goodhew, A.~Thavanesan and A.C.~Wall, \emph{{The Cosmological CPT Theorem}},  \href{https://arxiv.org/abs/2408.17406}{{\ttfamily 2408.17406}}.

\bibitem{Werth:2024mjg}
D.~Werth, \emph{{Spectral Representation of Cosmological Correlators}},  \href{https://arxiv.org/abs/2409.02072}{{\ttfamily 2409.02072}}.

\bibitem{Raju:2011mp}
S.~Raju, \emph{{Recursion Relations for AdS/CFT Correlators}}, \href{https://doi.org/10.1103/PhysRevD.83.126002}{\emph{Phys. Rev. D} {\bfseries 83} (2011) 126002} [\href{https://arxiv.org/abs/1102.4724}{{\ttfamily 1102.4724}}].

\bibitem{Albayrak:2020isk}
S.~Albayrak, C.~Chowdhury and S.~Kharel, \emph{{Study of momentum space scalar amplitudes in AdS spacetime}}, \href{https://doi.org/10.1103/PhysRevD.101.124043}{\emph{Phys. Rev. D} {\bfseries 101} (2020) 124043} [\href{https://arxiv.org/abs/2001.06777}{{\ttfamily 2001.06777}}].

\bibitem{Mei:2024abu}
J.~Mei and Y.~Mo, \emph{{On-shell Bootstrap for n-gluons and gravitons scattering in (A)dS, Unitarity and Soft limit}},  \href{https://arxiv.org/abs/2402.09111}{{\ttfamily 2402.09111}}.

\bibitem{Mei:2024sqz}
J.~Mei and Y.~Mo, \emph{{From on-shell amplitude in AdS to cosmological correlators: Gluons and gravitons}}, \href{https://doi.org/10.1103/fh8s-rlxf}{\emph{Phys. Rev. D} {\bfseries 113} (2026) 105008} [\href{https://arxiv.org/abs/2410.04875}{{\ttfamily 2410.04875}}].

\bibitem{circles}
N.~Arkani-Hamed and C.~Figueiredo, \emph{{Circles and Triangles, the NLSM and Tr($\Phi^3$)}},  \href{https://arxiv.org/abs/2403.04826}{{\ttfamily 2403.04826}}.

\bibitem{DuasoPueyo:2025lmq}
C.~Duaso~Pueyo, H.~Goodhew, C.~McCulloch and E.~Pajer, \emph{{On the asymptotic nature of cosmological effective theories}}, \href{https://doi.org/10.1007/JHEP01(2026)009}{\emph{JHEP} {\bfseries 01} (2026) 009} [\href{https://arxiv.org/abs/2505.17820}{{\ttfamily 2505.17820}}].

\bibitem{Sun:2021thf}
Z.~Sun, \emph{{A note on the representations of SO(1,d + 1)}}, \href{https://doi.org/10.1142/S0129055X24300073}{\emph{Rev. Math. Phys.} {\bfseries 37} (2025) 2430007} [\href{https://arxiv.org/abs/2111.04591}{{\ttfamily 2111.04591}}].

\bibitem{Armstrong:2022vgl}
C.~Armstrong, A.~Lipstein and J.~Mei, \emph{{Enhanced soft limits in de Sitter space}}, \href{https://doi.org/10.1007/JHEP12(2022)064}{\emph{JHEP} {\bfseries 12} (2022) 064} [\href{https://arxiv.org/abs/2210.02285}{{\ttfamily 2210.02285}}].

\bibitem{Chowdhury:2023arc}
C.~Chowdhury, A.~Lipstein, J.~Mei, I.~Sachs and P.~Vanhove, \emph{{The subtle simplicity of cosmological correlators}}, \href{https://doi.org/10.1007/JHEP03(2025)007}{\emph{JHEP} {\bfseries 03} (2025) 007} [\href{https://arxiv.org/abs/2312.13803}{{\ttfamily 2312.13803}}].

\bibitem{chowdhury_simplicity_2026}
C.~Chowdhury, S.~He, Y.-X.~Su and D.~Yang, \emph{On the simplicity of de {Sitter} correlators},  Apr., 2026.
\newblock 10.48550/arXiv.2604.26421.

\bibitem{Donath:2024utn}
Y.~Donath and E.~Pajer, \emph{{The in-out formalism for in-in correlators}}, \href{https://doi.org/10.1007/JHEP07(2024)064}{\emph{JHEP} {\bfseries 07} (2024) 064} [\href{https://arxiv.org/abs/2402.05999}{{\ttfamily 2402.05999}}].

\bibitem{Glew:2025mry}
R.~Glew, \emph{{Correlators from graphical amplitudes}}, \href{https://doi.org/10.1103/cg2r-gtkt}{\emph{Phys. Rev. D} {\bfseries 112} (2025) L061302}.

\bibitem{SurfaceKin}
N.~Arkani-Hamed, Q.~Cao, J.~Dong, C.~Figueiredo and S.~He, \emph{{Surface Kinematics and the Canonical Yang-Mills All-Loop Integrand}}, \href{https://doi.org/10.1103/PhysRevLett.134.171601}{\emph{Phys. Rev. Lett.} {\bfseries 134} (2025) 171601} [\href{https://arxiv.org/abs/2408.11891}{{\ttfamily 2408.11891}}].

\bibitem{Backus:2025njt}
J.V.~Backus and C.~Figueiredo, \emph{{Surface gauge invariance, soft limits and the transmutation of gluons into scalars}}, \href{https://doi.org/10.1007/JHEP09(2025)069}{\emph{JHEP} {\bfseries 09} (2025) 069} [\href{https://arxiv.org/abs/2505.17179}{{\ttfamily 2505.17179}}].

\end{thebibliography}\endgroup
\bibliographystyle{JHEP}

\end{document}